\documentclass[%
 reprint,
 amsmath,amssymb,
 aps,
]{revtex4-2}

\usepackage{graphicx}% Include figure files
\usepackage{dcolumn}% Align table columns on decimal point
\usepackage{bm}% bold math
\usepackage[dvipsnames]{xcolor}
\usepackage{float}
\usepackage{tikz}
\usepackage{tkz-graph}
\usepackage{standalone}
\usetikzlibrary{graphs, graphs.standard, math}
\usepackage{graphicx} % Required for inserting images

\usepackage[shortlabels]{enumitem}
\usepackage{hyperref}

\newcommand{\dd}[2]{\frac{d #1}{d #2}} % Makes a derivative

\newcommand{\set}[1]{\left\{ #1 \right\}}
\newcommand{\react}{\rightleftharpoons}

\definecolor{DB}{RGB}{31,120,180}
\definecolor{LB}{RGB}{166,206,227}
\definecolor{G}{RGB}{178,223,138}

\usetikzlibrary{cd}
\usetikzlibrary{automata}
\usetikzlibrary{positioning}
\usetikzlibrary{arrows.meta}
\usetikzlibrary{decorations.fractals}
\usetikzlibrary{decorations.markings}
\usetikzlibrary{decorations.pathreplacing}
\usetikzlibrary{calligraphy}
\usetikzlibrary{shapes.geometric}
\usetikzlibrary{matrix, patterns, hobby}
\usetikzlibrary{intersections}

\usepackage[new]{old-arrows}
\usepackage[outline]{contour}
\contourlength{.6pt}

\definecolor{americanrose}{rgb}{1.0, 0.01, 0.24}
\definecolor{applegreen}{rgb}{0.55, 0.71, 0.0}
\definecolor{azure}{rgb}{0.0, 0.5, 1.0}
\definecolor{corn}{rgb}{0.98, 0.93, 0.36}

\newcommand{\cx}{americanrose}
\newcommand{\cy}{applegreen}
\newcommand{\cz}{azure}
\newcommand{\cw}{corn}

\definecolor{myblue}{rgb}{0.2, 0.5, 0.91}  % my blue
\definecolor{purpleheart}{rgb}{0.41, 0.21, 0.61} % purpleheart
\definecolor{mygreen}{rgb}{0.2, 0.7, 0.2} % my green
\definecolor{carnelian}{rgb}{0.7, 0.11, 0.11} % carnelian
\definecolor{amber}{rgb}{1.0, 0.75, 0.0} % amber
\definecolor{beige}{rgb}{0.96, 0.96, 0.86}
\definecolor{darktangerine}{rgb}{1.0, 0.66, 0.07}

\definecolor{arylideyellow}{rgb}{0.91, 0.84, 0.42} %arylideyellow
\definecolor{cosmiclatte}{rgb}{1.0, 0.97, 0.91}

\definecolor{almond}{rgb}{0.94, 0.87, 0.8}  % almond
\definecolor{antiquewhite}{rgb}{0.98, 0.92, 0.84}
\definecolor{floralwhite}{rgb}{1.0, 0.98, 0.94}

\tikzset{
pics/square2/.style args={#1/#2}{
 code = {
   \draw[black,fill=#2] (0,0) rectangle (1,1) node[midway] {\Large #1}; 
   }
 }
}
\tikzset{
    vertex/.style={circle, draw, fill=black!30, inner sep=0pt, minimum width=12pt}
}
\tikzset{
    vertexs/.style={circle, draw, fill=black!30, inner sep=0pt, minimum width=9pt}
}
\tikzset{
    vxs/.style={circle, draw, fill, inner sep=0pt, minimum width=5pt}
}

\tikzset{
    edge/.style={ultra thick}
}
\tikzset{
    strong/.style={line width = 1.5mm, color=black}
}
\tikzset{
    weak/.style={line width = 0.5mm, color=black!17}
}
\tikzset{
    medium/.style={line width = 1.0mm, color=black!30}
}
\tikzset{
    zero/.style={dashed, line width = 0.5mm, color=black!17, draw opacity=0}
}

\begin{document}

\preprint{APS/123-QED}

\title{Optimal interactions for addressable self-assembly}% Force line breaks with \\

\author{Tighe McAsey}
 \affiliation{Department of Mathematics, University of British Columbia.}%Lines break automatically or can be forced with \\
 
\author{Sushrut Tadwalkar}%
\affiliation{Department of Mathematics, University of British Columbia.}

\author{Ali Fele-Paranj}%
\affiliation{Department of Mathematics, University of British Columbia.}

\author{Miranda Holmes-Cerfon}%
 \email{holmescerfon@math.ubc.ca}
\affiliation{Department of Mathematics, University of British Columbia.}

\date{\today}% It is always \today, today,

\begin{abstract}
Addressable self-assembly asks that each building block assemble into a particular location in a target structure. Although particles may all be distinct, achieving high yield is a challenge because of monomer depletion: more target structures can nucleate than there are building blocks for, so they form partial fragments which cannot complete growth. We ask how to design the interactions between building blocks to achieve the highest yield in a given time. Using reaction equations describing all the intermediate steps of assembly, combined with numerical optimization, we show that the optimal interactions are such that (i) all bonds are either very strong or very weak, and (ii) the strong bonds form a spanning tree of the target structure. We then prove that when interactions form a spanning tree, monomer depletion cannot occur: assembly can always proceed downhill in energy space, with no kinetic traps. This result is a combinatorial property of the underlying interaction graph, and does not depend on the particular model for the kinetics. It suggests a robust design principle: create a network of strong interactions that has no loops, and make all other interactions much weaker. We validate this principle in numerical simulations of larger structures, and we further show that spanning trees that are topologically more compact have typically better yield. Our results suggest a new framework for understanding monomer depletion and addressable self-assembly, which may be applied to DNA nanotechnology and which may give insight into the assembly pathways of certain multiprotein complexes. 
\end{abstract}

\maketitle

\section{Introduction}

Addressable self-assembly is a flavour of self-assembly whereby each particle in a target structure has a unique location, or ``address'',  where it must end up upon assembly \cite{Rothemund.2012,Rothemund.2006,Sacanna.2013,Wang.2012,Rogers.2015,Rogers.2016,Hong.2017,murugan2015multifarious}. 
It is studied both as a model for biological processes which assemble from many distinct components, such as the ribosome \cite{de2015functions,klinge2019ribosome} and other multiprotein complexes \cite{kuhner2009proteome,Sartori.2020zom}, and for its potential technological applications, including in drug delivery, biosensing, vaccine development, data storage, nanoscale computing and robotics \cite{green2017complex,woo2024molecular,fu2026programmable,madhanagopal2026light,zeng2025dna}. Achieving addressable self-assembly generally requires each particle to be distinct, with distinct interactions with its neighbours, a feat which has been demonstrated using short DNA strands to control interactions \cite{Wei.2012,Ke.2012,Ong.2017,evans2024pattern}. Yet despite the possibility of precise control over particle interactions, achieving addressable self-assembly is not simple, because of the tradeoff that generically occurs between the thermodynamic stability of the system, and its kinetic accessibility -- making the ground state very low-energy, often creates many kinetic traps along the assembly pathway \cite{Chatterjee.2025}. This tradeoff was brought to the spotlight by the so-called Levinthal's paradox of protein folding \cite{Shakhnovich.2006}, and is ubiquitous in self-assembling systems.

In addressable self-assembly, the thermodynamic-kinetic tradeoff manifests as the phenomenon of monomer depletion: when interactions are strong enough to thermodynamically favour the target structure, then many structures simultaneously nucleate and grow, using up the monomers and leaving none to complete the growth. The system thus becomes kinetically trapped in a state with many partially-formed fragments, for which the timescale to break apart and reassemble into the target structure can be exceedingly long. To overcome the kinetic barriers, experiments use temperature cycling to control the nucleation behaviour and help break up structures with defects \cite{Reinhardt.2014,Jacobs.2015}. 
 Annealing protocols in both simulations \cite{bupathy2022temperature} and experiments \cite{evans2024pattern} have also been successful at achieving multifarious assembly,  whereby one of several target structures can be retreived via careful seeding.  %and some attention has been devoted to understanding when and why this leads to successful assembly .

The question still remains of whether there are assembly strategies which overcome the thermodynamic-kinetic tradeoff in a setting which is not externally controlled -- as for natural proteins, which have evolved a variety of strategies to create a ``funnelled'' energy landscape where most kinetic pathways lead efficiently to the ground state. 
Early attempts to answer this question for addressable assembly focused on the role of interactions parameters, with some studies claiming that approximately equal strengths are best \cite{Hormoz.2011,hedges2014growth}, while others 
showing that increased variability can aid assembly \cite{Jacobs.2015,Madge.2018}. 
Murugan et al \cite{murugan2015undesired} proposed to use the initial concentrations of monomers as a design parameter, and showed that using nonequal concentrations can dramatically reduce the impact of undesired interactions.
Recently Holmes-Cerfon et al \cite{holmes2025hierarchical} proposed choosing interactions with differing scales to promote ``hierarchical'' self-assembly, where smaller units form first, which assemble into larger units, and so on.  While these strategies have been shown to help the assembly process, sometimes significantly, it is not known if they are optimal  -- could there be an even better strategy yet to be discovered?

%Most experiments and theoretical studies operate in the regime where the interactions between monomers have approximately the same strength. 
%, although some heterogeneity is expected physically \mhc{ref-arvind?} and has been shown to aid assembly \mhc{ref-jacobs}. 
%What has been  less explored in the context of addressable self-assembly is the possibility of choosing interaction strengths or other design parameters specifically to aid assembly. Some studies have claimed that approximately equal strengths are optimal \cite{Hormoz.2011,hedges2014growth}, while others have shown that increased variability can aid assembly \cite{Jacobs.2015,Madge.2018}. Murugan et al \cite{murugan2015undesired} proposed to use the initial concentrations of monomers as a design parameter, and showed that using nonequal concentrations can dramatically reduce the impact of undesired interactions. %Carefully designed time-dependent temperature protocols have been shown to aid retrieval of addressable systems with multiple target structures  \cite{bupathy2022temperature}. Recently \cite{holmes2025hierarchical} proposed choosing interactions with differing scales to promote ``hierarchical'' self-assembly, where smaller units form first, which assemble into larger units, and so on.  While these strategies have been shown to help the assembly process, sometimes significantly, it is not known if they are optimal  -- could there be an even better strategy yet to be discovered? 

It has recently become possible to ask about optimizing loss functions which depend on the kinetics of a system, thanks to algorithmic and software advances which have made it feasible to directly optimize through dynamic simulations \cite{rackauckas2017differentialequations,schoenholz2020jax}. 
%A promising approach to finding optimal parameters, which has been recently enabled by the advent of machine learning with its associated algorithms and software packages, is to directly optimize through dynamic simulations. 
This approach has been used in a handful of self-assembling systems, including in designing transition rates between small sphere clusters \cite{goodrich2021designing}, in optimizing time-dependent potentials \cite{engel2023optimal}, and in designing patchy particles %which assemble into open lattices and self-limiting rings 
\cite{king2024programming}. 
%\cite{Hubl.2026}
%\cite{jhaveri2024discovering}

In this paper we take advantage of these methods to ask: what are the optimal interaction energies for finite-time addressable self-assembly? We focus on the case of constant temperature, so particles assemble in equilibrium conditions, and we further wish to minimize waste: we want as many of our starting monomers as possible to assemble into a target structure. Thus, we ask to maximize the yield of perfectly-formed target structures at some finite time $T$. 
We start with small target structures by enumerating all the intermediate sub-species of the assembly process, and writing down reaction equations governing the evolution of all of their concentrations. The yield at time $T$ is then found by solving this nonlinear system of ODEs, and we use numerical optimisation to maximize it. 
This approach is similar to the ones followed  in \cite{jhaveri2024discovering}, \cite{Hubl.2026}, though our focus is on learning design principles from our optimal parameter sets. 

Indeed, we find 
the optimal strengths have a striking property: they contain a network of very strong bonds which form a spanning tree of the target structure. This suggests there a deeper property of self-assembling systems, arising from the combinatorial structure of their underlying interaction graphs. We then prove a theorem which says that when strong interactions form a spanning tree, then monomer depletion cannot occur: assembly can always proceed downhill in energy space, with no kinetic traps. This ``Spanning Tree Theorem'' suggests a design principle: create a network of very strong bonds which form a spanning tree of the target structure, and make all other bonds much weaker. We verify that this principle gives excellent assembly properties in simulations of larger systems where we cannot optimise the interactions, and we further provide some guidance about how to choose among the many different spanning trees. Overall, our design principle leads to a practical method to achieve high-yield addressable self-assembly, which doesn't require precise tuning of parameters. It also beats the thermodynamic-kinetic tradeoff -- it appears that having an underlying spanning tree of strong bonds creates a funnelled energy landscape for addressable self-assembly.

%, which suggests a design principle we propose that is easy to implement, robust, and leads to extremely fast, high-yield assembly, a fact we test in simulations of larger systems. 
%Overall, our design principle leads to a practical method to achieve high-yield addressable self-assembly, which doesn't require precise tuning of parameters. 

%The optimal strengths have a striking property: they contain a network of very strong bonds which form a spanning tree of the target structure. This suggests there a deeper property of self-assembling systems, arising from the combinatorial structure of their underlying interaction graphs. We then prove a theorem which says that when strong interactions form a spanning tree, then monomer depletion cannot occur: assembly can always proceed downhill in energy space, with no kinetic traps. 
%This ``Spanning Tree Theorem'' suggests a design principle: create a network of very strong bonds which form a spanning tree of the target structure, and make all other bonds much weaker. We verify that this principle gives excellent assembly properties in simulations of larger systems where we cannot optimise the interactions, and we further provide some guidance about how to choose among the many different spanning trees. Overall, our design principle leads to a practical method to achieve high-yield addressable self-assembly, which doesn't require precise tuning of parameters. 

\section{Finding the optimal interactions by numerical optimisation}\label{sec:optimisation}

\subsection{Example: a 2x2 square}\label{sec:square}

\begin{figure*}
\includegraphics{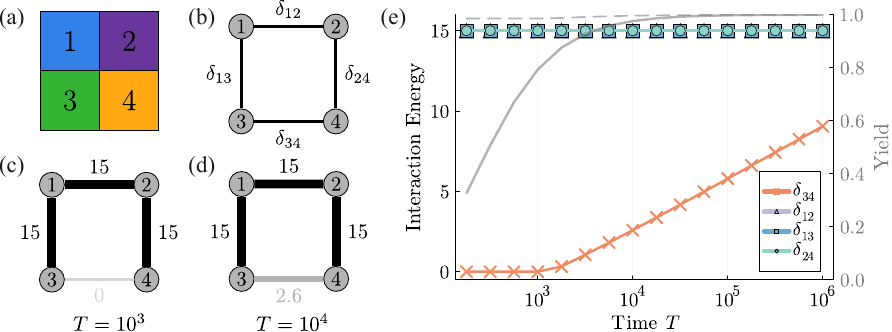}
%--------------------------------------------------------------%
\caption{Optimising the interaction energies for a 2x2 square. (a) Target structure, a square consisting of 4 distinguishable monomers. We wish to create multiple copies of this target structure. (b) The target structure can be represented a graph, with vertices representing particles, and edges representing interactions. The strength of each interaction is a given parameter shown on each edge. (c,d) Interaction energies that maximize the concentration of the target at times  (c) $T=10^3$ (d) $T=10^4$. 
(e) Optimal interaction parameters as a function of time $T$ (left axis), and the corresponding optimal yield (right axis, solid grey line). Dashed grey line shows the equilibrium yield for the optimal interaction energies.
}\label{fig:square}
\end{figure*}

We illustrate our optimization procedure on a square formed from 4 distinct particles  as in Fig. \ref{fig:square}(a).   The particles have pairwise interactions, with $\delta_{ij}$ denoting the interaction energy between particles $i$ and $j$, measured in units of $k_BT$. We only allow native interaction energies to be nonzero, so there are four interaction parameters, $\delta_{12}, \delta_{13},\delta_{24},\delta_{34}$, whose values we wish to choose to assemble many copies of the square. We assume the directionality of the interactions is controlled such that these interactions may form with no frustration. 
The embedding space of the particles does not need to be specified at this stage, although one possibility is a lattice model where particles are squares moving on a two-dimensional square lattice, as was studied in \cite{holmes2025hierarchical}. 
%\mhc{might need more explanation here about what graph represents, assumptions on interactions, directionality, etc.}

The target structure and the allowable interactions may be conveniently represented as a graph, Fig. \ref{fig:square} (b), whose vertices correspond to distinct particles, and whose edges represent the set of possible pairwise interactions. 

As each square assembles it moves through a sequence of intermediate states, each corresponding to a connected subgraph of the interaction graph. 
For this small example there are 13 such connected subgraphs,  which we call ``clusters'':
$$\{1,2,3,4,12,13,24,34,123,134,124,234,1234\}.$$
Two clusters may combine reversibly to make a larger cluster, such as $1+2\leftrightarrow 12$, $13+24\leftrightarrow 1234$, etc. There are 18 total such reactions, shown in the Appendix, Sec. \ref{Asec:square}. 

When the number of copies of the target is large, and the system is well-mixed, we may use the law of mass action  to write down a system of ODEs describing the assembly kinetics. 
Let $c_i$ be the concentration of a cluster $i$. For each pair of clusters $A,B$ forming another cluster $AB$, the product forms with rate $k_{AB}^{\rm on}$ and it breaks apart with rate $k_{AB}^{\rm off}$. Assuming  detailed balance, the rate parameters satisfy 
\begin{equation}\label{konoff}
\frac{k_{A,B}^{\rm on}}{k_{A,B}^{\rm off}} = e^{\delta_{A,B}},
\end{equation} 
where $\delta_{A,B}$ is the nondimensional energy change in forming the product, equal to the sum of the energies of the additional bonds. For a system of $A$s and $B$s in isolation this gives ODEs
\begin{align*}
\dd{c_A}{t} &= k_{A,B}^{\rm off}c_{AB} - k_{A,B}^{\rm on}c_Ac_B, \\
\dd{c_B}{t} &= k_{A,B}^{\rm off}c_{AB} - k_{A,B}^{\rm on}c_Ac_B,\\
\dd{c_{AB}}{t} &= k_{A,B}^{\rm on}c_Ac_B - k_{A,B}^{\rm off}c_{AB}. 
\end{align*}
To describe the full assembly kinetics we sum the rates associated with each of the 18 reactions, to obtain a system of 13 ODEs (Appendix \ref{Asec:square}). 
These ODEs are equipped with initial condition corresponding to an equal concentration of the starting monomers: $c_1(0){=}c_2(0){=}c_3(0){=}c_4(0){=}\phi/4$ where $\phi$ is the volume fraction of the system, and $c_i(0) = 0$ for $i\neq\{1,2,3,4\}$. Most of our computations use $\phi=0.05$. For the rate parameter we start by setting $k_{i}^{\rm on}=1$ for all $i$ and set $k_i^{\rm off}$ using \eqref{konoff}. This is the simplest possible model that allows us to focus on the underlying combinatorics without making assumptions about the physics. 
%We call this model for the \emph{standard rate model}. 
%We briefly explore another model for the rate parameters in Section \ref{sec:},\mhc{REF} which accounts for the fact that larger fragments may diffuse more slowly than smaller fragments. 
(Our later simulations incorporate more physical diffusion by requiring larger fragments to diffuse more slowly than smaller ones.)

We wish to find values of $\delta_{12}, \delta_{13},\delta_{24},\delta_{34}$ which maximize the concentration of target structures at some finite time $T$. 
%yield of the target structure at some finite time $T$, where the yield is the total fraction of fully completed target structures: $c_{1234}(T)/(\phi/4)$. 
To this end, we use numerical optimisation to maximize $c_{1234}(T)$ for different values of $T$, under the constraint that the parameters lie in the range $[0,\delta_{\rm max}]$, with $\delta_{\rm max} = 15$. Specifically, our computations are performed in the Julia programming language, using the SciML ecosystem of packages to solve the ODEs numerically, compute the gradient of $c_{1234}(T)$ using automatic differentiation (which we use in forward-mode because the number of parameters is smaller than the number of equations), and implement a limited memory version of the Broyden–Fletcher–Goldfarb–Shanno optimisation algorithm (L-BFGS) starting with random initial conditions for the optimisation \cite{rackauckas2017differentialequations}. %\mhc{ref to github where codes lie}

The result of our optimization is always qualitatively the same, for all times $T$: there are always three interaction energies that are equally strong, and one which is much weaker (Fig. \ref{fig:square}(c,d)). 
The strong bonds are always the maximum allowed, which by symmetry we may take to be $\delta_{12}{=}\delta_{13}{=}\delta_{24}{=}\delta_{\rm max}$. The value of the weak bond, $\delta_{34}$, equals 0 for times smaller than around $T\approx 10^3$, % about 1360
but it subsequently increases, as \(\approx 1.4\log T\) (Fig. \ref{fig:square}(e)). The overall yield of the target structure, defined as the fraction of fully completed squares $c_{1234}(T)/(\phi/4)$, similarly increases with $T$. 
Fig. \ref{fig:square}(e) shows the equilibrium yield for the optimal interaction parameters, computed by solving for the steady-state of the reaction equations, to show when the system reaches equilibrium.
%Notably, the yield is always higher than the maximum yield obtainable with equal energy parameters, $\delta_{12}{=}\delta_{13}{=}\delta_{24}{=}\delta_{34}=\delta$, when we maximized this yield over the single parameter $\delta$  (Fig. \ref{fig:square} (e)).
%, the equal energy parameters are chosen similarly using L-BFGS, by optimizing with respect to a single parameter \(\delta = \delta_{12} = \delta_{13} = \delta_{24} = \delta_{34}\) (Fig. \ref{fig:square2}). 

\subsection{Optimising a general target structure}

    \begin{figure*}
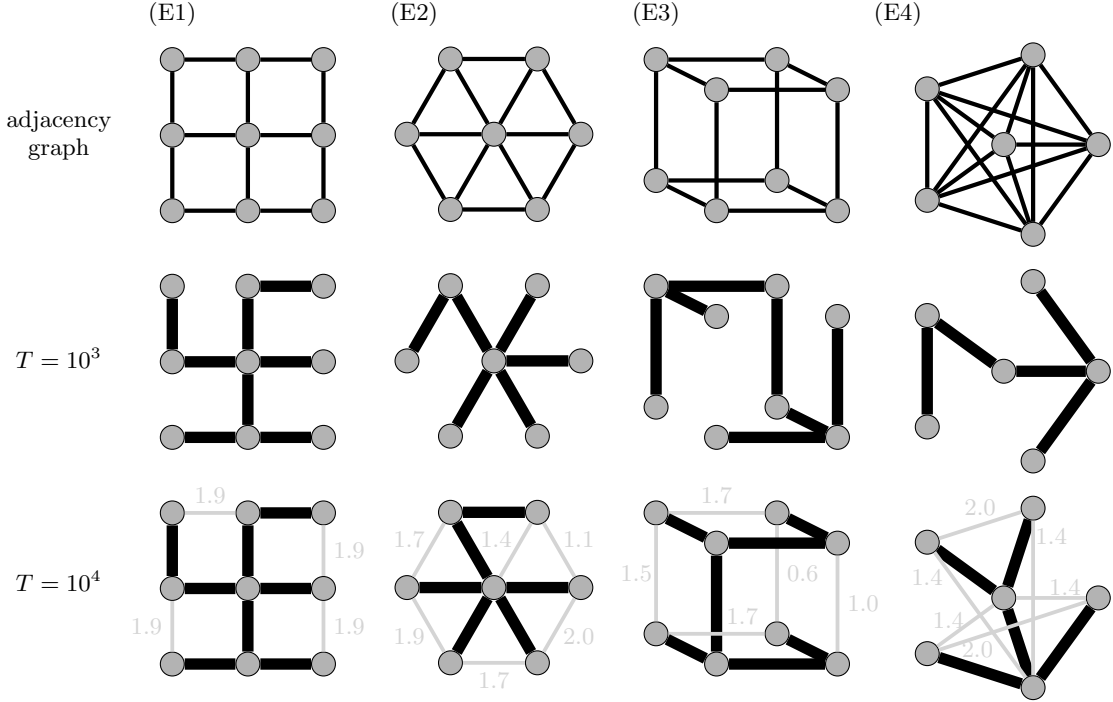

        \centering
            \include{Include/tikzexamples_nolabels.tex}
        \caption{Optimal interaction energies for various target structures, labelled (E1), (E2), (E3), (E4), for the standard model of reaction rates. 
        Top row shows the input interaction graph, middle row shows the optimal interactions at  time $T=10^3$, and bottom row shows the optimal interactions at time $T=10^4$. Thick black lines represent interactions with the maximum strength ($\delta_{\rm max}{=}15$), thin grey lines with labels are weaker interactions, and no line indicates an interaction energy that was numerically 0. 
} 
        \label{fig:examples}
        \end{figure*}

We repeated this optimisation procedure on several other examples shown in Fig.  \ref{fig:examples}: (E1) a 3x3 square with 9 particles, which could be formed from particles on a square lattice; (E2) a hexagon with 7 particles, which could be formed from particles on a triangular lattice; (E3) a 2x2x2 cube with 8 particles, which could be formed on a three-dimensional cubic lattice; (E4) a complete graph with 6 particles. % (note the use of standard notation \(K_n\) for the complete graph on \(n\)-vertices). 
The latter example probably doesn't correspond to any physical embedding of particles, but we include it anyways as this graph add richness to our examples. 

We developed an algorithm that takes as input an adjacency graph, then computes a list of all connected subgraphs (clusters), identifies the set of all reactions between clusters, and writes a text file corresponding to Julia code which evaluates the right-hand side of the reaction equations for that system (see Appendix Sec. \ref{Asec:alg} for details). For examples E1, E2 which can be embedded in a two-dimensional lattice, the algorithm also (optionally) removes reactions that require particles to physically move through each other, as illustrated in the Appendix, Fig. \ref{Afig:blockages}. We then use this code as input to the optimisation procedure described in Sec. \ref{sec:square}. 

For each example, we ran the optimisation with 6 different random starting conditions for each of two different end times, $T=10^3,10^4$, with volume fraction $\phi= 0.05$. %The list of optimal parameters for each run are shown in the Appendix, Section \ref{sec:data}. 
Figure \ref{fig:examples} shows the graphs corresponding to the run with the best yield over all 6 runs, for each $T$, for each example. (The full set of optima are reported in the Appendix, Section \ref{sec:data}.) %(In the case of a tie, we chose the run which we performed earlier in time.) 

These optima have two striking properties. The first is that \emph{the interaction energies come in two types}: either they are as strong as possible, equal to $\delta_{\rm max}=15$, or, they are much weaker. For the short time $T=10^3$ the weak interactions are all numerically 0. For the longer time $T=10^4$, the weak interactions vary in strength, but are all $\leq 2$. 
The second striking observation, is that \emph{the network of strong bonds forms a spanning tree of the input adjacency graph} -- it links together all the vertices without forming any loops. 
This observation also held for the 2x2 square.

These observations hold for almost all of the other local minima, which are similarly visualized as graphs in 
 the Appendix, Section \ref{sec:data}.  Almost every optimisation run produced a different optimum, which are almost all distinct even when considering graph isomorphisms. The optimisation landscape is therefore very rugged. Yet, virtually all of the local minima show the properties above: some interaction parameters are as strong as possible and form a spanning tree of the target structure, while the remaining interaction parameters are much smaller -- numerically 0 for $T=10^3$ in all cases, and  less than 2 for $T=10^4$ in almost all cases. One small exception to this statement is example E3, where in %runs 10,11,12 
three runs the strong bonds are not fully connected, but rather form 2-3 disjoint pieces that are glued together with medium-strength bonds, with parameters in the range 3-5. These runs had slightly lower yield than the others.  
 
The yields are uniformly high by $T{=}10^4$, at 87\%, 92\%, 89\%, 94\% for examples E1-E4 respectively. The yields for most runs are very similar, typically varying by less than 0.1\%. %An exception is E1 run 6, which produced a ``blocked'' spanning tree and for which the yield was about 4\% lower than the optimal, and E3 runs 10-12 which are not connected by strong bonds, and for which the yield was about 0.3\% lower. 

%We compared the optimal yields to those obtained by requiring all interactions to be equal, and found... \tm{description here when you remake fig. \ref{fig:yields}.}

One might wonder how sensitive are the yields to the particular values of the weak bonds which produce a local optimum in yield, especially since fabricating interactions with precise values is experimentally challenging. To address this question, we started with the optimal interactions in Fig. \ref{fig:examples} for $T=10^4$, and set all the weak bonds to the same strength $\delta_w$. We then used the reaction equations to compute the yield at times $T=10^3,10^4,10^5$, over a range of values of $\delta_w$ (Fig. \ref{Afig:weakvary}). For all $T$, the yield is nearly constant, up to a cutoff (which increases with $T$), beyond which the yield drops quickly.  For time $T=10^4$,  the yield is within 0.5\% its optimal value, up to the value of $\delta_w$ where it declines. 
For time $T=10^5$, the yield reaches 99\% for all examples, even though we have set the interaction parameters approximately, rather than using the optimal parameters. 
Hence, the yield is not sensitive to the precise value of $\delta_w$ for these examples. 

These experiments show another useful property of the optimal solutions -- for a given value of $\delta_w$, and for large enough $\delta_s$, the yield increases as time $T$ increases, since the equilibrium yield is very high. This is a useful property as it means that even if one didn't choose the optimal parameter values, one can increase the yield simply by waiting longer.   This is notably not the case when all interaction parameters have the same strength $\delta_w$, a case labelled ``Equal'' in Fig.  \ref{Afig:weakvary}. For Equal interactions, the yield displays a peak at an intermediate value of $\delta_w$. As $T$ increases, the yield to the left of this peak stays the same -- indicating that the yield is set by equilibrium considerations -- while only the yield to the right of the peak increases. %Therefore, the yield with equal interactions does not always increase uniformly with time $T$.  \mhc{rephrase -- it does increase, but sometimes it has an upper bound.} \tighe{The spanning trees also have an upper bound given by equilibrium concentrations. This equillibrium concentration is just higher in case of spanning trees.}

\medskip

We remark that computational limitations prevent us from exploring larger systems with this method of exhaustively enumerating reactions. Each single optimisation took anywhere from 10 minutes to 27 hours, with a typical run taking around 9 hours. 
Table \ref{tbl:complexity} in the Appendix shows how various measures of the complexity of the reaction equations
%, and the  file size of the resulting code, 
changes with the example. 
\section{Why spanning trees?}

We now ask: why should the strong bonds form a spanning tree? We will argue, by stating and proving a theorem regarding the combinatorics of graphs, that if there are only strong bonds, and if they form a spanning tree, then the system has no kinetic traps: assembly can proceed downhill in energy space until all target structures are formed. Then, we will compare spanning trees, and show empirically that spanning trees that are more compact seem to give the highest finite-time yield, when the weak bonds have nonzero energies.

%\madd{In sections 3.1, 3.2 and 3.3, non-lattice interactions are allowed in each of the examples. This is to eliminate the impact of lattice interactions on yields achieved by different instances of spanning trees, in order to try to understand the contribution of more nuanced factors. The effects of lattice interactions are quite straightforward; spanning tree structures should avoid 'blockages' as shown in Figure \ref{Afig:blockages}, the contribution of blockages seems to be the primary factor influencing spanning tree assembly on lattices, as explored in section 3.4.}

%%%%%%%%%%%%%%%%%%%%%%%%%%
%%%%%  Spanning trees avoid kinetic traps  %%%%%
%%%%%%%%%%%%%%%%%%%%%%%%%

\subsection{Spanning trees avoid kinetic traps}

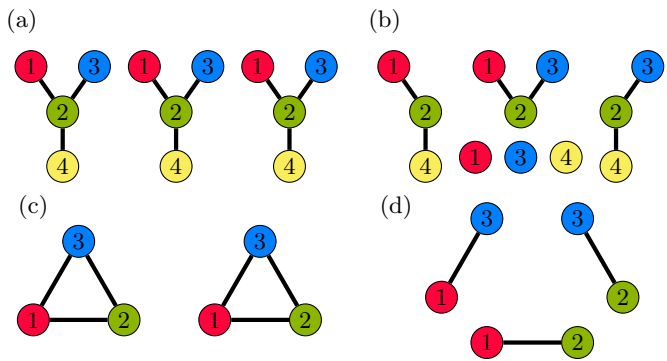
\begin{figure}
    \centering
        % !TEX root = main.tex
\begin{tikzpicture}[scale=0.6]
\begin{scope}[xshift=0cm,yshift=0.5cm]
\node at (-0.9,2) {(a)};
    \node[vertex,fill=\cx] (1a) at (-0.7,1) {1};
    \node[vertex,fill=\cy] (2a) at (0,0) {2};
    \node[vertex,fill=\cz] (3a) at (0.7,1) {3};
    \node[vertex,fill=\cw] (4a) at (0,-1.2) {4};
    \draw[edge] (1a)--(2a);
    \draw[edge] (2a)--(3a);
    \draw[edge] (2a)--(4a);
\end{scope}
\begin{scope}[xshift=2.5cm,yshift=0.5cm]
    \node[vertex,fill=\cx] (1a) at (-0.7,1) {1};
    \node[vertex,fill=\cy] (2a) at (0,0) {2};
    \node[vertex,fill=\cz] (3a) at (0.7,1) {3};
    \node[vertex,fill=\cw] (4a) at (0,-1.2) {4};
    \draw[edge] (1a)--(2a);
    \draw[edge] (2a)--(3a);
    \draw[edge] (2a)--(4a);
\end{scope}
\begin{scope}[xshift=5cm,yshift=0.5cm]
    \node[vertex,fill=\cx] (1a) at (-0.7,1) {1};
    \node[vertex,fill=\cy] (2a) at (0,0) {2};
    \node[vertex,fill=\cz] (3a) at (0.7,1) {3};
    \node[vertex,fill=\cw] (4a) at (0,-1.2) {4};
    \draw[edge] (1a)--(2a);
    \draw[edge] (2a)--(3a);
    \draw[edge] (2a)--(4a);
\end{scope}
% \begin{scope}[xshift=7cm,yshift=0.5cm]
%     \node[vertex,fill=\cx] (1a) at (-1,0) {1};
%     \node[vertex,fill=\cy] (2a) at (1,0) {2};
%     \node[vertex,fill=\cz] (3a) at (0,1.7321) {3};
%     \node[vertex,fill=\cw] (4a) at (0,-1.7321) {4};
%     \draw[edge] (1a)--(2a);
%     \draw[edge] (2a)--(3a);
%     \draw[edge] (2a)--(4a);
% \end{scope}

\begin{scope}[xshift=8cm,yshift=0.5cm]
\node at (-0.9,2) {(b)};
    % 1-2-4
    \node[vertex,fill=\cx] (1a) at (-0.7,1){1};
    \node[vertex,fill=\cy] (2a) at (0,0) {2};
    \node[vertex,fill=\cw] (4a) at (0,-1.2) {4};
    \draw[edge] (1a)--(2a);
    \draw[edge] (2a)--(4a);
    \begin{scope}[xshift=2.1cm]
    % 1-2-3 
    \node[vertex,fill=\cx] (1b) at (-0.7,1) {1};
    \node[vertex,fill=\cy] (2b) at (0,0) {2};
    \node[vertex,fill=\cz] (3b) at (0.7,1) {3};
    \draw[edge] (1b)--(2b);
    \draw[edge] (3b)--(2b);
    % extra 3,4,1
    \node[vertex,fill=\cx] (1c) at (-1,-1){1};
    \node[vertex,fill=\cz] (3a) at (0,-1) {3};
    \node[vertex,fill=\cw] (4b) at (1,-1) {4};
    \end{scope}
    \begin{scope}[xshift=4.2cm]
    % 3-2-4
    \node[vertex,fill=\cy] (2c) at (0,0) {2};
    \node[vertex,fill=\cz] (3c) at (0.7,1)  {3};
    \node[vertex,fill=\cw] (4c) at (0,-1.2) {4};
    \draw[edge] (2c)--(3c);
    \draw[edge] (2c)--(4c);
    \end{scope}
\end{scope}
\end{tikzpicture}
        % !TEX root = main.tex
\begin{tikzpicture}[scale=0.6]
\begin{scope}[xshift=0cm,yshift=0.5cm]
\node at (-1,2.5) {(c)};
    \node[vertex,fill=\cx] (1a) at (-1,0) {1};
    \node[vertex,fill=\cy] (2a) at (1,0) {2};
    \node[vertex,fill=\cz] (3a) at (0,1.7321) {3};
    \draw[edge] (1a)--(2a);
    \draw[edge] (2a)--(3a);
    \draw[edge] (3a)--(1a);
\end{scope}
\begin{scope}[xshift=4cm,yshift=0.5cm]
    \node[vertex,fill=\cx] (1a) at (-1,0) {1};
    \node[vertex,fill=\cy] (2a) at (1,0) {2};
    \node[vertex,fill=\cz] (3a) at (0,1.7321) {3};
    \draw[edge] (1a)--(2a);
    \draw[edge] (2a)--(3a);
    \draw[edge] (3a)--(1a);
\end{scope}

% \begin{scope}[xshift=9cm]
% \node at (-2,3) {(a)};
%     \node[vertex,fill=\cx] (a) at (0,0) {1};
%     \node[vertex,fill=\cy] (b) at (2,0) {2};
%     \node[vertex,fill=\cx] (c) at (-1,1) {1};
%     \node[vertex,fill=\cz] (d) at (0,2.7321) {3};
%     \node[vertex,fill=\cy] (e) at (3,1) {2};
%     \node[vertex,fill=\cz] (f) at (2,2.7321) {3};
% \end{scope}

\begin{scope}[xshift=9cm]
\node at (-2,3) {(d)};
    % 1-2
    \node[vertex,fill=\cx] (a) at (0,0) {1};
    \node[vertex,fill=\cy] (b) at (2,0) {2};
    \draw[edge] (a)--(b);
    % 1-3
    \node[vertex,fill=\cx] (c) at (-1,1) {1};
    \node[vertex,fill=\cz] (d) at (0,2.7321) {3};
    \draw[edge] (c)--(d);
    % 2-3
    \node[vertex,fill=\cy] (e) at (3,1) {2};
    \node[vertex,fill=\cz] (f) at (2,2.7321) {3};
    \draw[edge] (e)--(f);
\end{scope}
\end{tikzpicture}
    \caption{Illustrations of the Spanning Tree Theorem. Top:  an example where the Spanning Tree Theorem hold. (a) We wish to form $N_c=3$ copies of the tree shown. (b) Monomers from the 3 trees have been partially assembled using edges in the trees. No matter how this is done, we can always continue adding edges to complete the assembly. Bottom row: an example where the Spanning Tree Theorem doesn't hold. (c) We wish to form $N_c=2$ copies of a triangle. (d) The vertices can be assembled into three incomplete fragments, which can't be assembled into  triangles without breaking an edge.
    }
    \label{fig:thm}
\end{figure}

Our main result is the Spanning Tree Theorem, which is written in words below, and formulated more precisely in the Appendix, Section \ref{Asec:proof}.  
Consider a target structure, represented as a graph $G$, and suppose it has a subgraph $H\subset G$ which forms a spanning tree: it is a connected subgraph of $G$ which includes all the vertices and has no loops. %Let's call the sub-graph corresponding to this spanning tree $H_s$, and let's refer to the associated interactions as the \emph{strong bonds}. 
Suppose we start with a collection of disconnected monomers, or vertices, which can form $N_c$ copies of $G$. Suppose we let the system assemble, by allowing the monomers to form bonds in $H$. Then: 

\begin{quote} \emph{\underline{Spanning Tree theorem:}\label{theorem} When $H$ is a spanning tree of $G$, then no matter which order the bonds are formed, it is always possible to continue adding bonds from $H$ to form the full set of $N_c$ target structures.}
\end{quote}

In particular, when the subgraph $H$ represents strong bonds, this means the system has no kinetic traps -- assembly can \textit{always} proceed downhill on the energy landscape, by adding bonds in any order until all target structures are completed. 
Notably, this theorem is independent of the model used for binding rates -- it depends only on the underlying graph. Hence, it applies even for rate models that include different physics from what we have considered.  

Fig. \ref{fig:thm}(top) shows an illustration of this theorem. We start with vertices that can form $N_c=3$ copies of the spanning tree $H$, and we add bonds arbitrarily and then stop. One can see that it is possible to glue together the fragments using bonds in $H$ to form the full set of 3 target structures. 

Fig. \ref{fig:thm}(bottom) shows a minimal example where the conditions of the Spanning tree theorem don't hold. Here $H$ is a triangle, which contains a loop. Starting with disconnected vertices that can form $N_c=2$ copies of $H$, we form 3 edges, each contained in $H$. There is no way to assemble these fragments together into two triangles without first removing an edge. The system is kinetically trapped.

We prove the Spanning Tree theorem in more detail in the Appendix, Section \ref{Asec:proof}, but here we give a brief argument for why it is true. A key fact is that for a connected graph with no loops, there is exactly one path between every pair of vertices.  
Now, suppose we have a connected subgraph $G_1\subset H$, and which does not equal $H$. Choose a vertex $v$ which is not in $G_1$, but such that it is connected by an edge $(u,v)$ to $G_1$, where $u$ is a vertex in $G_1$. There can only be one such edge, by the fact we just presented. 
Now, we claim there must exist a graph $G_2$, with vertices disjoint from $G_1$, containing $v$ but not $u$. If there isn't, then every graph that contains $v$ also contains $u$. But there are $N_c$ copies of $v$, so this makes $N_c+1$ copies of $u$ ($N_c$ copies in the graphs containing $v$, and one additional copy in $G_1$). This is impossible since we started with only $N_c$ copies of each vertex. Therefore, it must be possible to find a graph $G_2$ to glue to $G_1$. Repeat, until we form $H$. Then continue in the same way with the remaining subgraphs to form the remaining copies of $H$.

\bigskip

We remark that other studies have pointed out the negative effects of loops on assembly. Deeds et al \cite{Deeds:2012} considered ring-like structures in large multiprotein complexes, and found  that structures containing one weak bond assembled more efficiently and robustly in simulations than structures where all bonds had the same strength. Furthermore, they found that most known naturally-occurring ring-like complexes contained a bond that was much weaker than the others. 
Madge et al \cite{Madge.2018} developed a ``completability index'' to count how many complete target structures are possible to form from fragments in simulations of patchy particles, and observed that when the bonding network has no loops, the index always predicts full completion. Jhaveri et al \cite{jhaveri2024discovering} used automatic differentiation to optimize the binding rates in models of small multiprotein complexes. They found that the optimal rates involved one monomer binding rapidly to all other monomers, while all other reactions were much slower. This binding strategy is equivalent (at certain timescales) to a ``star-shaped'' graph of strong interactions, which we discuss below is the optimal spanning tree.

%%%%%%%%%%%%%%%%%%%%%%%%%%
%%%%%  Weak bonds  %%%%%
%%%%%%%%%%%%%%%%%%%%%%%%%

\subsection{Which spanning tree is best?}

\begin{figure*}
\centering
\includegraphics[width=1\linewidth]{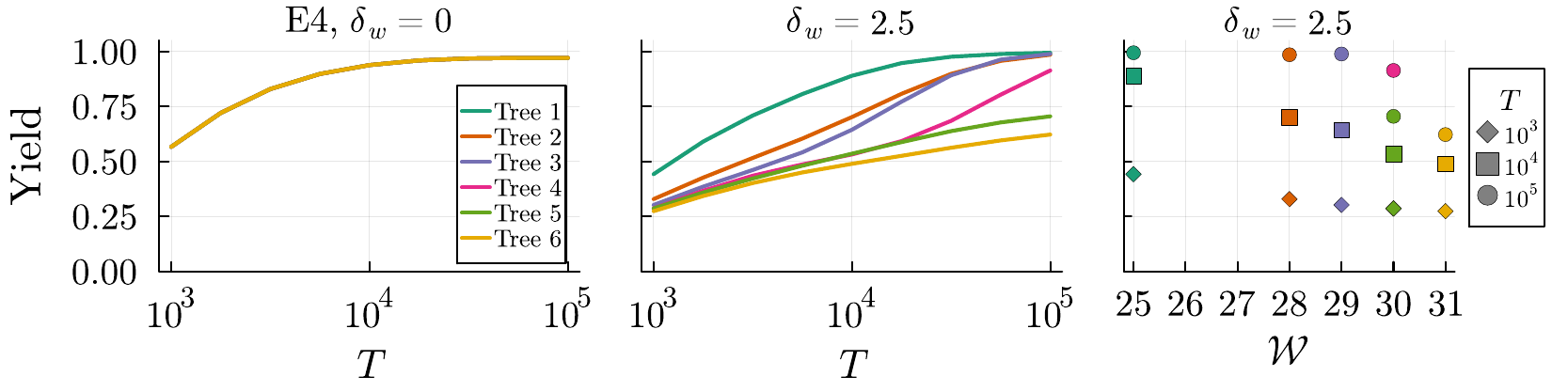}
 % !TEX root = main.tex
\definecolor{cta}{rgb}{0.106, 0.62, 0.467}
\definecolor{ctb}{rgb}{0.851, 0.373, 0.008}
\definecolor{ctc}{rgb}{0.459, 0.439, 0.702}
\definecolor{ctd}{rgb}{0.906, 0.161, 0.541}
\definecolor{cte}{rgb}{0.4, 0.651, 0.118}
\definecolor{ctf}{rgb}{0.902, 0.671, 0.008}
\begin{tikzpicture}[scale=0.82]
\begin{scope}[draw=cta,fill=cta,shift={(0,1.2)}]
    \coordinate (a) at (0,0);   % center
    \coordinate (b) at (0.31,0.95);  %b = diagonal top, then go clockwise
    \coordinate (c) at (1,0);
    \coordinate (d) at (0.31,-0.95);
    \coordinate (e) at (-0.81,-0.59);
     \coordinate (f) at (-0.81,0.59);

    \node[vxs] at (a) {}; 
    \node[vxs] at (b) {};
    \node[vxs] at (c) {};
    \node[vxs] at (d) {};
    \node[vxs] at (e) {};
    \node[vxs] at (f) {};

    \draw[edge] (a)--(b);
    \draw[edge] (a)--(c);
    \draw[edge] (a)--(d);
    \draw[edge] (a)--(e);
    \draw[edge] (a)--(f);

    \node[below] at (0,-1.2) {Tree 1};
\end{scope}
                
    \begin{scope}[draw=ctb,fill=ctb,shift={(2.75,1.2)}]
    \coordinate (a) at (0,0);
        \coordinate (b) at (0.31,0.95);
        \coordinate (c) at (1,0);
        \coordinate (d) at (0.31,-0.95);
        \coordinate (e) at (-0.81,-0.59);
        \coordinate (f) at (-0.81,0.59);

        \node[vxs] at (a) {}; 
        \node[vxs] at (b) {};
        \node[vxs] at (c) {};
        \node[vxs] at (d) {};
        \node[vxs] at (e) {};
        \node[vxs] at (f) {};

        \draw[edge] (a)--(b);
        \draw[edge] (a)--(c);
        \draw[edge] (a)--(d);
        \draw[edge] (a)--(f);
        \draw[edge] (f)--(e);

        \node[below] at (0,-1.2) {Tree 2};
        
    \end{scope}

    \begin{scope}[draw=ctc,fill=ctc,shift={(5.5,1.2)}]
        \coordinate (a) at (0,0);
        \coordinate (b) at (0.31,0.95);
        \coordinate (c) at (1,0);
        \coordinate (d) at (0.31,-0.95);
        \coordinate (e) at (-0.81,-0.59);
        \coordinate (f) at (-0.81,0.59);

        \node[vxs] at (a) {}; 
        \node[vxs] at (b) {};
        \node[vxs] at (c) {};
        \node[vxs] at (d) {};
        \node[vxs] at (e) {};
        \node[vxs] at (f) {};

        \draw[edge] (a)--(b);
        \draw[edge] (a)--(c);
        \draw[edge] (a)--(e);
        \draw[edge] (e)--(d);
        \draw[edge] (e)--(f);

        \node[below] at (0,-1.2) {Tree 3};
    \end{scope}

    \begin{scope}[draw=ctd,fill=ctd,shift={(8.25,1.2)}]
        \coordinate (a) at (0,0);
        \coordinate (b) at (0.31,0.95);
        \coordinate (c) at (1,0);
        \coordinate (d) at (0.31,-0.95);
        \coordinate (e) at (-0.81,-0.59);
        \coordinate (f) at (-0.81,0.59);

        \node[vxs] at (a) {}; 
        \node[vxs] at (b) {};
        \node[vxs] at (c) {};
        \node[vxs] at (d) {};
        \node[vxs] at (e) {};
        \node[vxs] at (f) {};

        \draw[edge] (a)--(b);
        \draw[edge] (b)--(c);
        \draw[edge] (a)--(f);
        \draw[edge] (f)--(e);
        \draw[edge] (a)--(d);
        \node[below] at (0,-1.2) {Tree 4};
    \end{scope}

    \begin{scope}[draw=cte,fill=cte,shift={(11,1.2)}]
     \coordinate (a) at (0,0);
        \coordinate (b) at (0.31,0.95);
        \coordinate (c) at (1,0);
        \coordinate (d) at (0.31,-0.95);
        \coordinate (e) at (-0.81,-0.59);
        \coordinate (f) at (-0.81,0.59);

        \node[vxs] at (a) {}; 
        \node[vxs] at (b) {};
        \node[vxs] at (c) {};
        \node[vxs] at (d) {};
        \node[vxs] at (e) {};
        \node[vxs] at (f) {};

        \draw[edge] (a)--(b);
        \draw[edge] (a)--(c);
        \draw[edge] (a)--(d);
        \draw[edge] (d)--(e);
        \draw[edge] (e)--(f);

        \node[below] at (0,-1.2) {Tree 5};
    \end{scope}

    \begin{scope}[draw=ctf,fill=ctf,shift={(13.75,1.2)}]
    \coordinate (a) at (0,0);
    \coordinate (b) at (0.31,0.95);
    \coordinate (c) at (1,0);
    \coordinate (d) at (0.31,-0.95);
    \coordinate (e) at (-0.81,-0.59);
    \coordinate (f) at (-0.81,0.59);

    \node[vxs] at (a) {}; 
    \node[vxs] at (b) {};
    \node[vxs] at (c) {};
    \node[vxs] at (d) {};
    \node[vxs] at (e) {};
    \node[vxs] at (f) {};

    \draw[edge] (a)--(b);
    \draw[edge] (b)--(c);
    \draw[edge] (c)--(d);
    \draw[edge] (d)--(e);
    \draw[edge] (e)--(f);

    \node[below] at (0,-1.2) {\(\scalebox{1.05}{Tree 6}\)};
    \end{scope}
 \end{tikzpicture}
\caption{Comparing spanning trees for the complete graph of 6 vertices, (E4), with $\delta_s=15$. Top left: yield versus time for all spanning trees, with $\delta_w=0$. Top middle: same, but with $\delta_w=2.5$. Top right: yield versus Weiner index $\mathcal W$ for all spanning trees, at times $T=10^3,10^4,10^5$ as shown in the legend. Bottom row shows the complete set of spanning trees.
}\label{fig:E4}
\end{figure*}
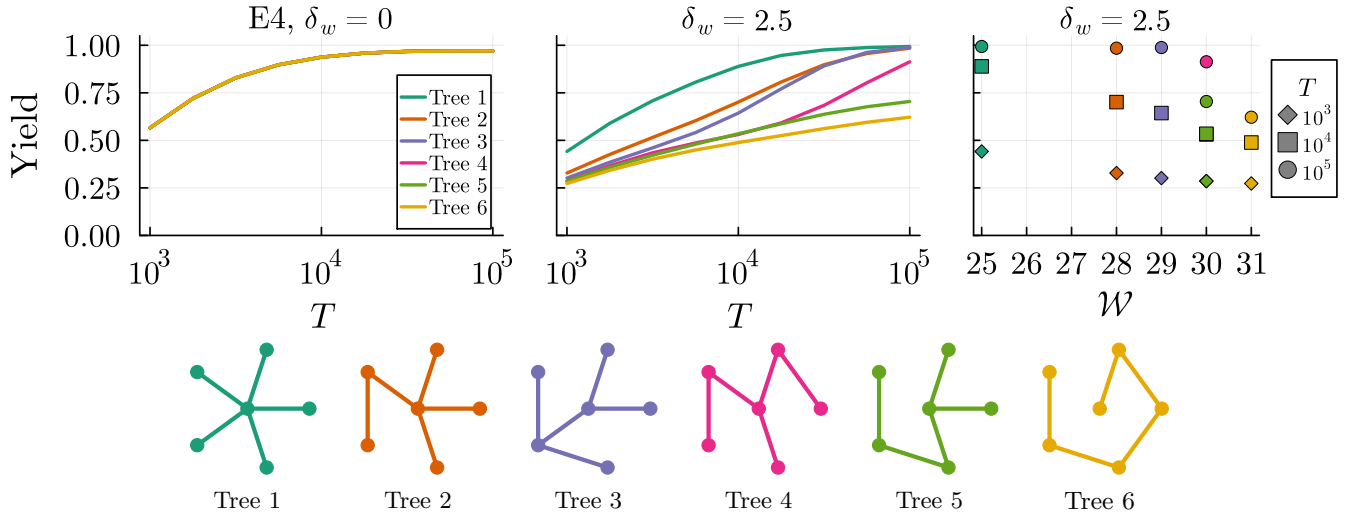

Now that we know that spanning trees have good kinetic properties, we may ask: are some spanning trees better than others? 
Consider the complete graph E4, which has six nonisomorphic spanning trees, shown in  Fig. \ref{fig:E4}. We computed the yield as a function of time with strong bonds $\delta_s=15$ and different strengths for the weak bonds, $\delta_w=0$. Fig.~\ref{fig:E4}  shows that when $\delta_w=0$, \emph{all spanning trees give the same yield!} When the weak bonds are increased, however, $\delta_w>0$, the yields are no longer the same (Fig.~\ref{fig:E4}(middle)). 

Inspecting the different spanning trees, which have been ordered in Fig.~\ref{fig:E4} by their yield at $T=10^4$, it appears that trees with the highest yield have lots of branches and are quite compact, while trees with lower yield are more string-like and spread out. We may measure the compactness of a tree by the Wiener Index \cite{wiener1947structural,Dobrynin:2001}, defined for a graph $G$  as the sum of the distances between every pair of vertices:
\[
\mathcal{W}(G) := \sum_{\set{u,v} \subset V(G)} d(u,v).
\]
Here $d(u,v)$ is the graph-distance between vertices $u,v$, namely the length of the shortest path between $u,v$. The Wiener index has been frequently been invoked in chemical graph theory to explain how topological properties of molecules affect their chemical properties \cite{Bonchev:1980,rouvray1987modeling}. 
For the complete graph, it is minimized by a star-shaped graph \cite{xu2014survey,lu2018sharp} -- Tree 1 in Fig.\ref{fig:E4}, and maximized by the path graph -- Tree 6 in Fig.\ref{fig:E4}. %\tm{tighe -- is this true? if so, could you put a proof or a reference somewhere in the appendix? \tighe{It is stated in the paper \href{https://doi.org/10.1016/j.disc.2017.11.009}{here}, it can also be proven by the greedy algorithm in 8.6.1 but replacing the invariant \(N\) with \(\mathcal{W}\)}}
We plotted the yield with $\delta_w=2.5$ versus the Wiener index for the 6 different spanning trees (Fig.~\ref{fig:E4}(right)), and find excellent anticorrelation, especially at early and intermediate times (correlation coefficient -0.99 at $T=10^4$) -- trees with lower Wiener indices, hence which are more compact, have higher yields. 

We now turn to generalize these observations, and to provide a qualitative explanation for them, giving insight into the key features that cause this anticorrelation. %We leave a more complete and more quantitative theory for future work, recognizing that this might depend on the detailed kinetic model and on the embedding geometry. 
Our theory is presented in a few key steps; first starting with understanding the role of the weak bonds.

\paragraph{When $\delta_w=0$, all trees give the same yield.}
We repeated our calculations of the yield for all spanning trees for examples E1-E3, and found that when $\delta_w=0$, all trees give the same yield (Fig.\ref{Afig:yields}, left column). 
There is a simple explanation for this. Since $k_{\rm on}=1$ for all reactions, our kinetic model when $\delta_w=0$ and $\delta_s\gg 1$ is equivalent to adding edges randomly at a constant rate. Since each spanning tree has the same number of edges, spanning trees will be completed at the same rate, independent of their topology. %Hence, the yield curves as a function of time will be identical. 

\paragraph{Weak bond strength $\delta_w>0$ aids equilibrium yields over longer timescales for optimal spanning trees.}

We next ask: what is the role of the weak bonds when they are nonzero? 
To answer this we computed, for each optimum at $T=10^4$ in Fig. \ref{fig:examples} and for different values of $\delta_s$,  the  weak bond strength $\delta_w^*(T)$, which optimizes the yield at different times $T$ (Fig. \ref{Afig:dwopt}). For a given $\delta_s$, $\delta_w^*(T)=0$ for small $T$, and then beyond a critical value $T^*$ it begins to increase with $T$. The threshhold value $T^*$ decreases as $\delta_s$ increases, and for large enough $\delta_s$, we have $\delta_w^*(T) = 0 $ for all $T$.
%The increase is logarithmic, as roughly \mhc{insert, when have data}. The threshhold value $T^*$ decreases with $\delta_s$.
%\mhc{add logarithmic comment?}

These observations suggest that the role of the weak bonds is to increase the equilibrium yield, in the case when $\delta_s$ is small enough and $T$ is large enough that the strong bonds can sometimes break.  
This observation is supported by our yield calculations in Fig.~\ref{fig:E4} (also Fig.~\ref{Afig:yields}), middle column, where it is seen that $\delta_w>0$ causes the yields to be smaller at short times compared to $\delta_w=0$, but to eventually asymptote to larger values for the best trees at long times. 

Note that an alternative explanation could be that the weak bonds alter the kinetics, by increasing the number of pathways for forming the target structure. We do not believe this is the explanation here, because the weak bonds do not increase the yield over short times, only over longer timescales.

\paragraph{Trees with fewer off-pathway clusters have higher yield at intermediate times.}

We have shown that $\delta_w>0$ aids the yield at long times for some spanning trees (notably the optima in Fig.~\ref{fig:examples}),  but it can hurt it at intermediate times and for other spanning trees. One reason why is that weak bonds create the possibility of kinetic traps. A kinetic trap may only occur when there is a cluster whose vertices are not connected by edges in the spanning tree. We call such a cluster an ``off-pathway'' cluster. 

We tested this hypothesis by comparing the yield at times $T=10^3,10^4,10^5$, to the volume (fraction) of the system in off-pathway clusters, for all spanning trees and for each example, Fig.~\ref{Afig:offpathway}. We used $\delta_s=15$, and for each example we chose a value of $\delta_w$ that was slightly larger than the optimal value in Fig.~\ref{Afig:weakvary}, but such that the yield had only decreased by a couple of percentage points. (Our results are not sensitive to the precise value of $\delta_w$.) The yield shows excellent anti-correlation with the off-pathway volume, confirming that it is these kinetic traps which serve to decrease the yield. 

We now wish to relate the off-pathway volume to a combinatorial property of the graphs. 
A simple relation comes from counting 
the number of off-pathway clusters. 
For a graph $G$, define $\rho(G)$ to be the number of connected subgraphs of $G$:
\begin{equation*}
\rho(G) := \#\set{J \subset G \mid J \text{ is a connected subgraph of $G$}}.
\end{equation*}
This is sometimes called the $\rho$-index in graph theory \cite{wagner2007correlation}.
%Both $S(G), S(H)$ can be computed in \(\mathcal{O}(n)\) time using a Depth First Search (DFS) algorithm. \tm{what is DFS? can you spell it out? is it obvious, or should we include a reference? intermediate would be to reference your code on github \tighe{DFS is Depth First Search, this algorithm is very well known and has runtime \(\mathcal{O}(n)\) for trees. I also have the code I used to compute this on Github if we want to reference it}}
Given a target structure $G$ with spanning tree $H\subset G$, the number of off-pathway clusters is therefore 
\[
\rho(G)-\rho(H).
\] 
If $H, H'$ are two spanning trees of the same graph $G$, then $\rho(H) > \rho(H')$ implies that $H$ has fewer off-pathway clusters than $H'$, so we expect $H$ to have a better yield than $H'$. Fig.~\ref{Afig:rho} shows that $\rho(H)$ and the measured off-pathway volumes have a good anticorrelation. Thus, we expect $\rho(H)$ to be a good predictor of the yields. 

\paragraph{Trees with fewer off-pathway clusters have lower Wiener indices.}

Finally, we turn to the relationship between the more commonly-used graph invariant $\mathcal{W}(H)$, and the invariant $\rho(H)$ introduced above. 
It has been shown theoretically that $\mathcal{W}(H), \rho(H)$ are strongly anticorrelated, in the sense that if a tree is chosen uniformly at random from the set of rooted, 
ordered trees of $n$ vertices, then its Wiener index $W_n$ and its $\rho$-index $\rho_n$ have the following correlation, asymptotically in $n$ \cite{wagner2007correlation}: 
\[
\mbox{corr}(\mathcal{W}_n, \rho_n) %= \frac{\mathbb E \mathcal{W}_n\rho_n - \mathbb E \mathcal W_n \mathbb E \rho_n}{\mbox{std}(\mathcal W_n)\mbox{std}(\rho_n)}
\sim (-1.78357)\cdot(0.98209)^n.
\]
Empirically, when $n$ is small the correlation is particularly strong \cite{wagner2007correlation}. 
We computed $\mathcal W,\rho$ for all spanning trees for all examples, and found excellent anti-correlation (Fig.~\ref{Afig:Wrho}).

No explanation is offered for this correlation; indeed, according to the authors of \cite{wagner2007correlation}: ``It seems to be a challenging graph-theoretical problem to explain this phenomenon.'' 
However, we may observe that the star-shaped graph of $n$ vertices, which has one vertex is connected to all others,  has the smallest possible value of $\mathcal W$ \cite{xu2014survey,lu2018sharp}, and the largest possible value of $\rho$ -- the proof of the latter is a simple application of a greedy algorithm (Sect.\ref{Asec:greedy}).
Similarly, $\mathcal W$ is maximized and $\rho$ is minimized for the path graph, where all vertices are in one line. Hence, these two invariants share the same extremal graphs.

\paragraph{Summary -- trees with lower $\mathcal W$ have higher yields.}

Putting our arguments together, show that  trees with lower Wiener indices $\mathcal W$ have higher $\rho$ and hence lower off-pathway volumes, hence, when $\delta_w>0$, they should have higher yields. This should hold particularly at intermediate times when the system has not had time to equilibrate. We compare the yields at different times with the Wiener index for our remaining examples E1-E3 in Fig.~\ref{Afig:yields}(right column). There is good anti-correlation between $\mathcal W$ and the yield, particularly at short and intermediate times, though the correlation is not as clean as for E4, and notably not as linear. We also compared the yields at different times with the $\rho$-index (Fig.~\ref{Afig:rhoYield}), which exhibits even stronger correlations -- perhaps because it is the $\rho$-index which captures the essential features of the kinetics.

One reason why the relationship between the yield and the topological indices  is more complicated for E1-E3 is that the complete graph has the special property that the embeddings of a given spanning tree are all equivalent, up to graph isomorphism. Therefore,  for a spanning tree with a given topology, the set of off-pathway clusters is identical no matter how the tree is embedded. 
This property does not hold for the other examples, where there are spanning trees with the same topology but with a different embedding into the ambient graph $G$, leading to a different set of off-pathway clusters. A graph invariant that depends only on the strong bond topology cannot distinguish between embeddings, and explaining the yield is more subtle than simply counting the number of off-pathway clusters. 

We remark that the star-shaped graph (Tree 1 in Fig~\ref{fig:E4}), which is optimal for E4 (and also E2), has another nice property, that every off-pathway cluster contains no strong bonds, a fact we prove in Section \ref{Asec:starenergetic}. 
This is consistent with the results of \cite{jhaveri2024discovering}, which also found that a star-shaped graph of fast rates led to optimal assembly of small multiprotein complexes.

%\medskip \mhc{update starting here -- show optimal trees -- highest yield = lowest W, always. check. }
%
%\mhc{move this comment somewhere -- to paragraph below?} Star-shaped graphs have another nice property, in that every off-pathway cluster contains no strong bonds. This is proven in Section \ref{Asec:starenergetic}. 
%
%\mhc{might want to change this -- haven't mentioned optimal graphs yet. should introduce optimal graphs?}
%We found in our optimization in Figs.  \ref{Afig:best} that star-shaped graphs were optimal for the examples E2 and E4 where such graphs were subgraphs of the target graph, partially validating our analysis. This fact was also observed in a different study, which \mhc{... describe Maggie's study, and result of a star-shaped graph}

%%%%%%%%%%%%%%%%%%%%%%%%%
%%%%%  Theory       %%%%%
%%%%%%%%%%%%%%%%%%%%%%%%%
%\section{A general design principle for addressable self-assembly and its illustration in larger systems}
\section{Spanning trees work effectively in larger systems}

\begin{figure*}
    \centering
    \includegraphics{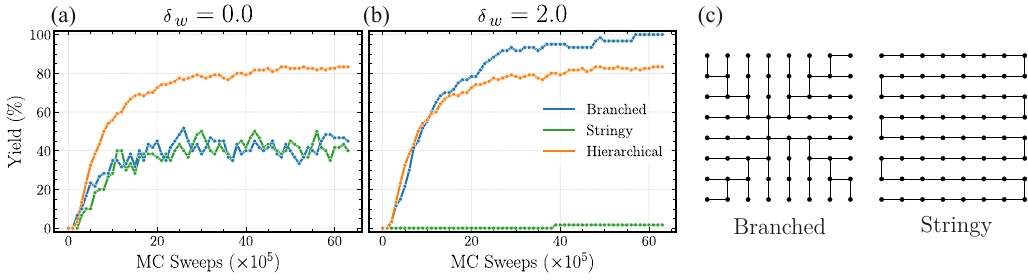}
    %\includegraphics[width=0.35\linewidth]{SimulationFigs/d_8x8_12c_b_vs_s_5.pdf}
    %\includegraphics[width=0.35\linewidth]{SimulationFigs/d_w_8x8_12b_vs_s_5.pdf}
    %
    %\input{Simulations appendix material/branched_st}\hspace{0.5cm}
    %input{Simulations appendix material/stringy_st}
    %
    \caption{Yield (\% completed target structures) versus time for different choices of interaction energies: a highly branched spanning tree, a stringy spanning tree, and a hierarchical strategy  as proposed in \cite{holmes2025hierarchical}. All spanning trees used strong bond energy $\delta_s=15$, and weak bond energy (a) $\delta_w=0$ and (b) $\delta_w=2$. The hierarchical energies were chosen to have strongest bond $\delta_1=8.5$ with remaining energies chosen according to the hierarchical scheme described in \cite{holmes2025hierarchical}. 
    All simulations used $N_c=60$ copies at volume fraction $0.05$. %The hierarchical simulations were repeated twice to form an averaged yield. 
    The spanning trees are shown in (c), with strong bonds drawn as lines, and with weak bonds corresponding to the remaining edges of the square lattice. 
    %First plot is comparing the average \% completed fragments of 12 copies of an \(8\times 8\)-square with the branched spanning tree, stringy spanning tree, and hierarchical assembly. The energies \(\delta_s=15.0\), \(\delta_w=0.0\), and \(\delta_h=8.75\). The second plot is the same but with \(\delta_w=2.0\) now. The Hierarchical bond energy is the optimum mentioned in Miranda's paper \sushrut{As a note, we expect  the spannig tree to do better than the hierarchical assembly even in the zero weak bonds case, only the strong energy will have to be much stronger. This is something that we can explore if we feel like including it in the paper as well. I have the code for this, but for now I'm not running it again since we did not discuss including it, and it seems like Ali requires a lot of server space.}
  }\label{fig:sims}
\end{figure*}

Our results suggest a general design principle for choosing interaction energies for addressable self-assembly: 

\begin{quote} 
\textit{Choose a spanning tree of the target structure, and make those bonds very strong. Make all other bonds much weaker.  }
%\textit{Make bonds of two different strengths, strong and weak, and choose the strong bonds to form a spanning tree of the target structure. }
\end{quote}

The spanning tree should be as compact as possible, i.e. with many branches, and the strong bonds should be as strong as possible. The weak bonds should be weak enough to break and form quickly, but strong enough to stabilize the yield. 

We tested these principles in simulations, to see whether they hold in larger systems for which we cannot write the reaction equations. Our simulations considered sticky squares on a lattice, simulated using the Virtual Move Monte Carlo algorithm, which is a Monte Carlo algorithm that also approximates dynamics on a lattice, in the sense that it moves clusters as a unit, with diffusivities given by an approximation to the Stokes-Einstein diffusivities, and it breaks clusters apart with Arrhenius rates \cite{Whitelam.2007}.  Details of the system and simulations are identical to those in \cite{holmes2025hierarchical}. 
Our simulations consider $N_c$ copies of a target structure, with the initial condition chosen by placing the monomers required to form those target structures randomly on the lattice. 
All our simulations are run at a volume fraction of 
 \(\phi\approx 0.05\) (the approximation arises because of the discreteness of the lattice). We do not allow orientational changes, which speeds up our code without changing the results qualitatively. 
 We measure the yield of a target structure at time \(t\) by the total number of fully-formed target structures $N_f(t)$, divided by the total number that were available to form: \(N_f(t)/N_c\).
A natural timescale for the dynamics is the number of MC sweeps. %; in these units the cluster diffusivities are all unitary: ($1$ lattice vector $^2$)/(time unit).

We considered a target structure consisting of 64 monomers arranged in an $8\times 8$ square. We compared two different spanning trees: a tree constructed to be highly branched, which should have low Weiner index, and the path tree, which has the highest Weiner index. We also considered a hierarchical assembly strategy which was shown in \cite{holmes2025hierarchical} to be highly effective, using interaction energies that were shown to be optimal for that example. %In this scheme, bonds are chosen with different strengths to promote assembly of a hierarchical nature, with smaller units forming then combing to form medium-size units, which further combine to form the largest scale of target structure. 
Fig.~\ref{fig:sims} shows the yield versus time for these three strategies, with strong bonds $\delta_s{=}15$ for the spanning trees. When the weak bonds are $\delta_w=0$, the spanning tree designs asymptote to approximately 40\% yield, while the hierarchical strategy increases initially rapidly  then plateaus and gradually approaches high yield. When the weak bonds are nonzero, $\delta_w{=}2$, the branched spanning tree rapidly approaches near-100\% yield; it reaches higher yield than the hierarchical strategy over the times considered, without any plateaus. Meanwhile the yield of the path tree is nearly zero; visual inspection shows this is because of the presence of vacancy defects. We remark that as well as validating our design strategy, this example further illustrates the role of weak bonds in improving equilibrium concentrations for certain spanning trees. 

We verified that our strategy is robust to the presence of non-native interactions, also called crosstalk. Some amount of crosstalk is inevitable in physical systems \cite{Huntley.2016} and it can significantly degrade the yield of designed structures \cite{murugan2015undesired}. Fig.~\ref{Afig:crosstalk} shows that, in the pessimistic case when all nonnative interactions are given the same energy $c$, the yield of the branched spanning tree (with $\delta_w{=}2$) remains high for small values of $c$, and then rapidly decreases beyond a critical value $c^*\approx 0.8$. The decrease occurs at the same critical value $c^*$ for the hierarchical strategy. Hence, our strategy appears to be as robust to crosstalk as the hierarchical strategy.

\begin{figure*}
    \centering
\includegraphics{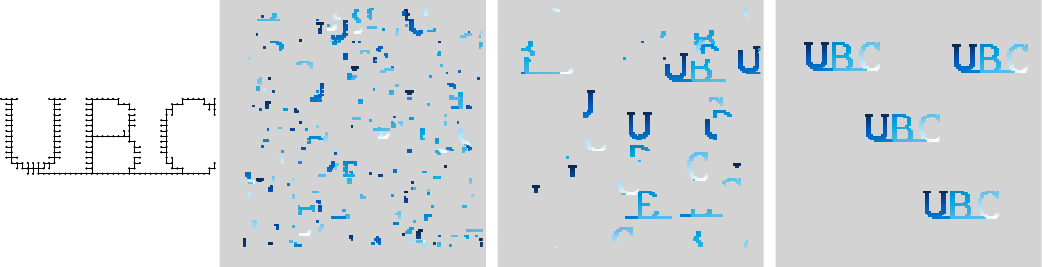}
\caption{Simulating our institution's acronym. The target structure has 213 monomers and we used the spanning tree shown on the left with $\delta_s=20$, $\delta_w=2$. Snapshots are at times $T=4,32,194\times 10^4$ MC sweeps respectively left to right. By the final time, all four copies of the target structure have assembled perfectly.}\label{fig:ubc}
\end{figure*}

As a final proof of principle we made our institution's acronym, a target structure of 213 monomers, formed from a subset of the square lattice using a branched spanning tree. Fig.~\ref{fig:ubc} shows snapshots of the simulations, and shows that all four copies of the target structure assemble perfectly.  Notably, and unlike our optimizations, all of these simulations incorporate the fact that larger fragments diffuse more slowly than smaller ones -- partially validating that the spanning tree design principle is robust to assumptions about the physical nature of diffusion.

\section{Conclusion} 

We considered a system of ``addressable'' particles, which are all distinct with distinct interactions, and asked what are the interaction energies which optimize the yield of the target structure at some finite time $T$. We found that optimal interactions are such that bonds fall into two categories: either very strong, or much weaker; furthermore the strong bonds form a spanning tree of the target structure. We then proved that when interactions form a spanning tree, then the target structure can always assemble perfectly, without kinetic traps. This result is a combinatorial property of graphs, and is independent of any physical considerations.  We further compared different spanning trees, and found that more topologically compact spanning trees, as measured by the Weiner and $\rho$-indices, have higher yield when strong bonds are  breakable and weak bonds are nonzero. A qualitative explanation for this is that compact spanning trees have fewer ways to create ``off-pathway'' clusters, i.e. which are connected subgraphs of the target graph but not of the spanning tree, and these off-pathway clusters create kinetic traps and hence slow down assembly. The number of off-pathway clusters is measured by the $\rho$-index, which correlates well with assembly yields. The more commonly-used Weiner index is strongly correlated with the $\rho$-index, hence it also correlates well with assembly yields, though the reasons for this are not understood. 
%correlated with the number of off-pathway clusters is not understood, though the $\rho$-index, which is more fundamentally related to the existence of kinetic traps, could alternatively be used as a measure of the effectiveness of a spanning tree. 
Both the $\rho$-index and the Weiner index can be computed efficiently in $O(n)$ time for trees, and the Weiner index can be tractably computed for arbitrary graphs \cite{mohar1988compute}.

Our model treats the assembly process as a set of reaction equations, which ignores the spatial embedding of particles and hence the constraints that these impose on assembly pathways. We tested our design principle in simulations which do account for the spatial embedding, and which have a physical model of diffusion, and found it works very well to form essentially 100\% yield of target structures containing hundreds of monomers, provided the spanning tree is chosen appropriately. %We note however that our model, as an idealized one, contains limitations including that it does not account for geometric frustration, does not include as an idealized model, 

Thus, we propose the following design principle for addressable self-assembly: choose a  spanning tree of strong interactions, where the spanning tree is highly branched, and the interactions are as strong as is physically possible; make all other interactions much weaker. This principle is simple, and it is robust to the precise values of the interaction energies. We anticipate this strategy will lead to faster, more robust, and higher-yield assembly than current strategies, which typically require a precise annealing protocol and do not typically lead to high yield. We note however that further work is needed to develop this principle for larger systems, as the spatial embeddings of particles or geometrical frustration effects not captured by our model might lead to further constraints on the spanning trees. This principle could be used in combination with other assembly strategies including control of monomer concentrations \cite{murugan2015undesired} to further enhance assembly. %We hope this principle can be tested experimentally, e.g. using DNA bricks \cite{Wei.2012,Ke.2012}, to see if this principle can lead to faster, more robust, and higher-yield assembly than currently strategies, which typically require a precise annealing protocol and do not typically lead to high yield. 

Our findings raise the question of whether certain biological systems such as multiprotein complexes might use the principle of spanning trees to assemble efficiently. 
%Finally, we suggest it would be fruitful to search for some manifestation of spanning trees  in biological assembly systems. 
Some hints have been found in ring-like protein complexes, where it was found that the majority of evolved heteromeric three-membered rings contain one interaction that is much weaker than the others \cite{Deeds:2012} -- hence, the stronger interactions form a spanning tree of the ring. It could be that could be that spanning trees underlie the assembly efficiency of a greater variety of biological assembly systems.

\begin{acknowledgments}
We thank Stephanie van Willigenburg and Lior Silberman for discussions. 
M.H.-C. and A. F.-P. acknowledge support from the Natural Sciences and Engineering Research Council of Canada (NSERC), RGPIN-2023-04449 / Cette recherche a \'{e}t\'{e} financ\'{e}e par le Conseil de recherches en sciences naturelles et en g\'{e}nie du Canada (CRSNG). 
T.M. was partially supported by an NSERC Undergraduate Student Research Award (USRA). S.T. was partially supported by a Work Learn International Undergraduate Research Award (WLIURA) from UBC.
\end{acknowledgments}

\bibliography{Refs,HierarchicalAssembly} % Entries are in the Refs.bib file

\appendix

\onecolumngrid

%\nocite{*}


%%%%%%%%%%%%%%%%%%
%%%%%   Appendix         %%%%%
%%%%%%%%%%%%%%%%%%

%%%%%%  Square   %%%%%%
\newpage
\section{Reaction equations for a 2x2 square}\label{Asec:square}

Here is a list of all the reactions possible in the intermediate stages of forming a 2x2 square. 
%We denote monomers as \(1,2,3,4\), dimers as \((1,2),(1,3),(2,3),(2,4)\) etc. 

\begin{alignat*}{2}
  1.&&  1+2 &\react 12 \\
  2.&&  1+3 &\react 13 \\
  3.&&  2+4 &\react 24 \\
  4.&&  3+4 &\react 34\\
  5.&&  2+13 &\react 123 \\
  6.&&  3+12 &\react 123 \\
  7.&&  1+24 &\react 124 \\
  8.&&  4+12 &\react 124 \\
  9.&&  1+34 &\react 134 \\
  10.&&   4+13 &\react 134 \\
  11.&&  2+34 &\react 234 \\
  12.&&  3+24 &\react 234 \\
  13.&&  12+34 &\react 1234 \\
  14.&&  13+24 &\react 1234 \\
  15.&&  4+123 &\react 1234 \\
  16.&&  3+124 &\react 1234 \\
  17.&&  2+134 &\react 1234 \\
  18.&&\qquad  1+234 &\react 1234
\end{alignat*}

% \begin{align*}
%     1+2 &\react (1,2) \\
%     1+3 &\react (1,3) \\
%     2+4 &\react (2,4) \\
%     3+4 &\react (3,4) \\
%     2+(1,3) &\react (1,2,3) \\
%     3+(1,2) &\react (1,2,3) \\
%     1+(2,4) &\react (1,2,4) \\
%     4+(1,2) &\react (1,2,4) \\
%     1+(3,4) &\react (1,3,4) \\
%     4+(1,3) &\react (1,3,4) \\
%     2+(3,4) &\react (2,3,4) \\
%     3+(2,4) &\react (2,3,4) \\
%     (1,2)+(3,4) &\react (1,2,3,4) \\
%     (1,3)+(2,4) &\react (1,2,3,4) \\
%     4+(1,2,3) &\react (1,2,3,4) \\
%     3+(1,2,4) &\react (1,2,3,4) \\
%     2+(1,3,4) &\react (1,2,3,4) \\
%     1+(2,3,4) &\react (1,2,3,4)
% \end{align*}

Here is the full list of reaction equations for the 2x2 square.
The reaction rates are denoted as \(K_{\pm i}\), where
\(i\) is the equation number, and \(\pm\) refer to forwards (on) or backwards (off) respectively.

\begin{align*}
\frac{dc_{1}}{dt} &= K_{-1}c_{12} + K_{-2}c_{13} + K_{-7}c_{124} + K_{-9}c_{134} + 
    K_{-18}c_{1234}
        - (K_1c_2 + K_2c_3 + K_7c_{24} + K_9c_{34} + K_{18}c_{234})c_1 \\
\frac{dc_2}{dt} &= K_{-1}c_{12} + K_{-3}c_{24} + K_{-5}c_{123} + K_{-11}c_{234} + 
    K_{-17}c_{1234}
    - (K_1c_1 + K_3c_4 + K_5c_{13} + K_{11}c_{34} + K_{17}c_{134})c_2 \\
\frac{dc_3}{dt} &= K_{-2}c_{13} + K_{-4}c_{34} + K_{-6}c_{123} + K_{-12}c_{234} + 
    K_{-16}c_{1234}
    - (K_2c_1 + K_4c_4 + K_6c_{12} + K_{12}c_{24} + K_{16}c_{123})c_3 \\
\frac{dc_4}{dt} &= K_{-3}c_{24} + K_{-4}c_{34} + K_{-8}c_{124} + K_{-10}c_{134} + 
    K_{-15}c_{1234}
    - (K_3c_2 + K_4c_3 + K_8c_{12} + K_{10}c_{13} + K_{15}c_{123})c_4 \\
\frac{dc_{12}}{dt} &= K_1c_1c_2 + K_{-6}c_{123} + K_{-8}c_{124} + K_{-13}c_{1234}
    - (K_{-1}+ K_6c_3 + K_8c_4 + K_{13}c_{34})c_{12}\\
\frac{dc_{13}}{dt} &= K_2c_1c_3 + K_{-5}c_{123} + K_{-10}c_{134} + K_{-14}c_{1234}
    - (K_{-2} + K_5c_2 + K_{10}c_4 + K_{14}c_{24})c_{13}\\
\frac{dc_{24}}{dt} &= K_3c_2c_4 + K_{-7}c_{124} + K_{-12}c_{234} + K_{-14}c_{1234}
    - (K_{-3} + K_7c_1 + K_{12}c_3 + K_{14}c_{13})c_{24}\\
\frac{dc_{34}}{dt} &= K_4c_3c_4 + K_{-9}c_{134} + K_{-11}c_{234} + K_{-13}c_{1234}
    - (K_{-4} + K_9c_1 + K_{11}c_2 + K_{13}c_{12})c_{34}\\
\frac{dc_{123}}{dt} &= K_5c_2c_{13} + K_6c_3c_{12} + K_{-15}c_{1234}
    - (K_{-5} + K_{-6} + K_{15}c_4)c_{123}\\
\frac{dc_{124}}{dt} &= K_7c_1c_{24} + K_8c_4c_{12} + K_{-16}c_{1234}
    - (K_{-7} + K_{-8} + K_{16}c_3)c_{124}\\
\frac{dc_{134}}{dt} &= K_9c_1c_{34} + K_{10}c_4c_{13} + K_{-17}c_{1234}
    - (K_{-9} + K_{-10} + K_{-17}c_2)c_{134}\\
\frac{dc_{234}}{dt} &= K_{11}c_2c_{34} + K_{12}c_3c_{24} + K_{-18}c_{1234}
    - (K_{-11} + K_{-12} + K_{18}c_1)c_{234}\\
\frac{dc_{1234}}{dt} &= K_{13}c_{12}c_{34} + K_{14}c_{13}c_{24} + K_{15}c_4c_{123} 
    + K_{16}c_3c_{124} 
    + K_{17}c_2c_{134} + K_{18}c_1c_{234} \\
    &- (K_{-13} + K_{-14} + K_{-15} + K_{-16} + K_{-17} + K_{-18})c_{1234}
\end{align*}

\section{Algorithm to create reaction equations for a given adjacency graph}\label{Asec:alg}

 All of the ODE systems of equations for the examples in the paper (except the 2x2 squares) are too large to feasibly write out by hand or check for errors. Therefore, we have developed an algorithm to automatically generate a text file containing code, in the Julia programming language, which evaluates the reaction equations for a given target structure, specified by its adjacency graph $G$. 

Specifically, given some fragment $H\subset G$, we are interested in the evolution of its concentration $c_H$. 
Suppose each of the reactions with reactant \(H\) are \(H + D_i \react E_i\) with energy change
    \(\sum_{\ell=1}^{N_i} \delta_{k^i_\ell}\), where $\{\delta_{k^i_\ell}\}_{\ell=1}^{N_i}$ is the list of bonds that glue together $H$ and $D_i$ (in the language of graph theory, the \emph{cutset} between $H$ and $D_i$). Call this list of bonds the \emph{adjacent bonds}. Suppose each of the reactions with
    product \(H\) are \(A_j + B_j \react H\) each with adjacent bonds \(\{\delta_{k^j_\ell}\}_{\ell=1}^{N_j}\). Then the concentration evolves as 
    \begin{align} \label{eq:ch}
        &\frac{dC_H}{dt} = \sum_i \left(k^{\rm off}_{H,D_i}C_{E_i}  - k^{\rm on}_{H,D_i}C_H C_{D_i}\right)
        + \sum_j \left(k^{\rm on}_{A_j,B_j}C_{A_j}C_{B_j} - k^{\rm off}_{A_j,B_j} C_H \right).
    \end{align}
    The on and off rates satisfy detailed balance, \eqref{konoff}, with energy change equal to the sum of the energy parameters in the cutset between the two clusters involved. 
    For most of the text we choose $k^{\rm on}_{A,B}=1$, which we call the \emph{standard model of reaction rates}, so the equations we solve are
         \[
        \frac{dC_H}{dt} = \sum_i \left(C_{E_i} e^{-\sum_{\ell=1}^{N_i} \delta_{k^i_\ell}} - C_H C_{D_i}\right)
        + \sum_j \left(C_{A_j}C_{B_j} - e^{-\sum_{l=1}^{N_j}\delta_{k_l^j}} C_H \right).
    \]
    We also explore a model that chooses $k^{\rm on}_{A,B}=1/V(AB)$,  where $V(AB)$ is the total number of vertices in clusters $A,B$. We call this the \emph{fluid model of reaction rates}. 
For this model, the equations we solve are 
    \[
    \frac{dC_H}{dt} = \sum_i \frac{1}{V(E_i)}\left(C_{E_i} e^{-\sum_{\ell=1}^{N_i} \delta_{k^i_\ell}} - C_H C_{D_i}\right)
        + \frac{1}{V(H)}\sum_j C_{A_j}C_{B_j} - e^{-\sum_{\ell=1}^{N_j} \delta_{k^j_\ell}} C_H
\]
Our algorithm produces code that evaluates the RHS of \eqref{eq:ch} for all fragments $H$.

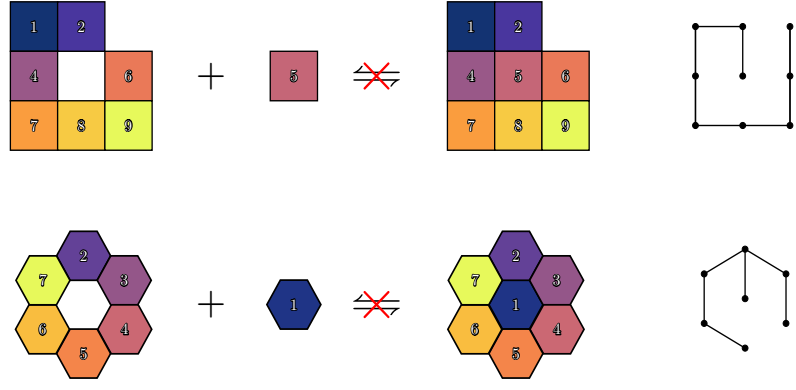
\begin{figure}
        \centering
\resizebox{10.5cm}{5cm}{
            \begin{tikzpicture}[scale=1]
                \filldraw[color=black, fill={rgb,256:red,19;green,51;blue,114}, thick] (0,2) rectangle ++(1,1);
                \node[] at (0.5,2.5) {\contour{black}{\(\color{white} 1\)}};
    
                \filldraw[color=black, fill={rgb,256:red,73;green,54;blue,159}, thick] (1,2) rectangle ++(1,1);
                \node[] at (1.5,2.5) {\contour{black}{\(\color{white} 2\)}};
    
                \filldraw[color=black, fill={rgb,256:red,154;green,88;blue,137}, thick] (0,1) rectangle ++(1,1);
                \node[] at (0.5,1.5) {\contour{black}{\(\color{white} 4\)}};
    
                \filldraw[color=black, fill={rgb,256:red,235;green,121;blue,88}, thick] (2,1) rectangle ++(1,1);
                \node[] at (2.5,1.5) {\contour{black}{\(\color{white} 6\)}};
    
                \filldraw[color=black, fill={rgb,256:red,251;green,158;blue,62}, thick] (0,0) rectangle ++(1,1);
                \node[] at (0.5,0.5) {\contour{black}{\(\color{white} 7\)}};
    
                \filldraw[color=black, fill={rgb,256:red,248;green,203;blue,67}, thick] (1,0) rectangle ++(1,1);
                \node[] at (1.5,0.5) {\contour{black}{\(\color{white} 8\)}};
    
                \filldraw[color=black, fill={rgb,256:red,232;green,250;blue,91}, thick] (2,0) rectangle ++(1,1);
                \node[] at (2.5,0.5) {\contour{black}{\(\color{white} 9\)}};
    
                \begin{scope}[shift={(4.25,0)}]
                    \node[] at (0,1.5) {\(\scalebox{2.5}{\(\mathbf{+}\)}\)};
    
                    \filldraw[color=black, fill={rgb,256:red,198;green,102;blue,119},thick] (1.25,1) rectangle ++(1,1);
                    \node[] at (1.75,1.5) {\contour{black}{\(\color{white} 5\)}};
    
                    \node[] at (3.5,1.5) {\(\scalebox{3.5}[2.5]{\(\leftrightharpoons\)}\)};
    
                    \node[] at (3.5,1.5) {\(\scalebox{3.5}{\(\color{red} \mathbf{\times}\)}\)};
                \end{scope}
    
                \begin{scope}[shift={(9.25,0)}]
                    \filldraw[color=black, fill={rgb,256:red,19;green,51;blue,114}, thick] (0,2) rectangle ++(1,1);
                    \node[] at (0.5,2.5) {\contour{black}{\(\color{white} 1\)}};
    
                    \filldraw[color=black, fill={rgb,256:red,73;green,54;blue,159}, thick] (1,2) rectangle ++(1,1);
                    \node[] at (1.5,2.5) {\contour{black}{\(\color{white} 2\)}};
    
                    \filldraw[color=black, fill={rgb,256:red,154;green,88;blue,137}, thick] (0,1) rectangle ++(1,1);
                    \node[] at (0.5,1.5) {\contour{black}{\(\color{white} 4\)}};
                    
                    \filldraw[color=black, fill={rgb,256:red,198;green,102;blue,119},thick] (1,1) rectangle ++(1,1);
                    \node[] at (1.5,1.5) {\contour{black}{\(\color{white} 5\)}};
    
                    \filldraw[color=black, fill={rgb,256:red,235;green,121;blue,88}, thick] (2,1) rectangle ++(1,1);
                    \node[] at (2.5,1.5) {\contour{black}{\(\color{white} 6\)}};
    
                    \filldraw[color=black, fill={rgb,256:red,251;green,158;blue,62}, thick] (0,0) rectangle ++(1,1);
                    \node[] at (0.5,0.5) {\contour{black}{\(\color{white} 7\)}};
    
                    \filldraw[color=black, fill={rgb,256:red,248;green,203;blue,67}, thick] (1,0) rectangle ++(1,1);
                    \node[] at (1.5,0.5) {\contour{black}{\(\color{white} 8\)}};
    
                    \filldraw[color=black, fill={rgb,256:red,232;green,250;blue,91}, thick] (2,0) rectangle ++(1,1);
                    \node[] at (2.5,0.5) {\contour{black}{\(\color{white} 9\)}};
                \end{scope}
    
                \begin{scope}[shift={(14,0)}]
                    \foreach \x in {0.5,1.5,2.5}
                    \foreach \y in {0.5,1.5,2.5}
                    {
                        \draw[fill] (\x,\y) circle (1.75pt);
                    }
                    \foreach \k in {0.5,1.5}
                    \foreach \l in {1.5,2.5}
                    \foreach \m in {0.5,2.5}
                    {
                        \draw[thick] (\m,\k)--(\m,\l);
                    }
                    \draw[thick] (0.5,0.5)--(1.5,0.5);
                    \draw[thick] (1.5,0.5)--(2.5,0.5);
                    \draw[thick] (0.5,2.5)--(1.5,2.5);
                    \draw[thick] (1.5,2.5)--(1.5,1.5);
                \end{scope}
    
                \begin{scope}[shift={(1.05,-3)}]
                    \node[regular polygon,draw,regular polygon sides = 6,minimum size = 0.45in,fill={rgb,256:red,99;green,65;blue,152},draw=black,line width=1pt] (h2) at (0.5,0.8660) {\contour{black}{\(\color{white} 2\)}};
    
                    \node[regular polygon,draw,regular polygon sides = 6,minimum size = 0.45in,fill={rgb,256:red,149;green,86;blue,138},draw=black,line width=1pt] (h3) at (1.36,0.37) {\contour{black}{\(\color{white} 3\)}};
    
                    \node[regular polygon,draw,regular polygon sides = 6,minimum size = 0.45in,fill={rgb,256:red,204;green,104;blue,115},draw=black,line width=1pt] (h4) at (1.37,-0.62) {\contour{black}{\(\color{white} 4\)}};
    
                    \node[regular polygon,draw,regular polygon sides = 6,minimum size = 0.45in,fill={rgb,256:red,245;green,134;blue,74},draw=black,line width=1pt] (h5) at (0.5,-1.11) {\contour{black}{\(\color{white} 5\)}};
    
                    \node[regular polygon,draw,regular polygon sides = 6,minimum size = 0.45in,fill={rgb,256:red,250;green,190;blue,63},draw=black,line width=1pt] (h6) at (-0.37,-0.62) {\contour{black}{\(\color{white} 6\)}};
    
                    \node[regular polygon,draw,regular polygon sides = 6,minimum size = 0.45in,fill={rgb,256:red,232;green,250;blue,91},draw=black,line width=1pt] (h7) at (-0.36,0.37) {\contour{black}{\(\color{white} 7\)}};
    
                    \begin{scope}[shift={(3.2,-0.12)}]
                        \node[] at (0,0) {\(\scalebox{2.5}{\(\mathbf{+}\)}\)};
        
                        \node[regular polygon,draw,regular polygon sides = 6,minimum size = 0.45in,fill={rgb,256:red,31;green,52;blue,135},draw=black,line width=1pt] (h1) at (1.75,0) {\contour{black}{\(\color{white} 1\)}};
        
                        \node[] at (3.5,0) {\(\scalebox{3.5}[2.5]{\(\leftrightharpoons\)}\)};
        
                        \node[] at (3.5,0) {\(\scalebox{3.5}{\(\color{red} \mathbf{\times}\)}\)};
                    \end{scope}
                    \begin{scope}[shift={(9.15,0)}]
                        \node[regular polygon,draw,regular polygon sides = 6,minimum size = 0.45in,fill={rgb,256:red,31;green,52;blue,135},draw=black,line width=1pt] (h1) at (0.5,-0.12) {\contour{black}{\(\color{white} 1\)}};
    
                        \node[regular polygon,draw,regular polygon sides = 6,minimum size = 0.45in,fill={rgb,256:red,99;green,65;blue,152},draw=black,line width=1pt] (h2) at (0.5,0.8660) {\contour{black}{\(\color{white} 2\)}};
    
                        \node[regular polygon,draw,regular polygon sides = 6,minimum size = 0.45in,fill={rgb,256:red,149;green,86;blue,138},draw=black,line width=1pt] (h3) at (1.36,0.37) {\contour{black}{\(\color{white} 3\)}};
    
                        \node[regular polygon,draw,regular polygon sides = 6,minimum size = 0.45in,fill={rgb,256:red,204;green,104;blue,115},draw=black,line width=1pt] (h4) at (1.37,-0.62) {\contour{black}{\(\color{white} 4\)}};
    
                        \node[regular polygon,draw,regular polygon sides = 6,minimum size = 0.45in,fill={rgb,256:red,245;green,134;blue,74},draw=black,line width=1pt] (h5) at (0.5,-1.11) {\contour{black}{\(\color{white} 5\)}};
    
                        \node[regular polygon,draw,regular polygon sides = 6,minimum size = 0.45in,fill={rgb,256:red,250;green,190;blue,63},draw=black,line width=1pt] (h6) at (-0.37,-0.62) {\contour{black}{\(\color{white} 6\)}};
    
                        \node[regular polygon,draw,regular polygon sides = 6,minimum size = 0.45in,fill={rgb,256:red,232;green,250;blue,91},draw=black,line width=1pt] (h7) at (-0.36,0.37) {\contour{black}{\(\color{white} 7\)}};
                    \end{scope}
                    \begin{scope}[shift={(14.5,0)}]
                        \coordinate (a) at (0,1);
                        \coordinate (b) at (0.8660,0.5);
                        \coordinate (c) at (0.8660,-0.5);
                        \coordinate (d) at (0,-1);
                        \coordinate (e) at (-0.8660,-0.5);
                        \coordinate (f) at (-0.8660,0.5);
                        \coordinate (g) at (0,0);
    
                        \draw[fill] (a) circle (1.75pt);
                        \draw[fill] (b) circle (1.75pt);
                        \draw[fill] (c) circle (1.75pt);
                        \draw[fill] (d) circle (1.75pt);
                        \draw[fill] (e) circle (1.75pt);
                        \draw[fill] (f) circle (1.75pt);
                        \draw[fill] (g) circle (1.75pt);
    
                        \draw[thick] (a)--(g);
                        \draw[thick] (a)--(b);
                        \draw[thick] (b)--(c);
                        \draw[thick] (d)--(e);
                        \draw[thick] (e)--(f);
                        \draw[thick] (f)--(a);
                    \end{scope}
                \end{scope}
            \end{tikzpicture}
}
\caption{Example of  reactions that we do not include in our list of reactions, because they would require particles to physically pass through each other when on a lattice. Top shows an example on a square lattice, bottom shows an example on a hexagonal lattice. The corresponding spanning trees of the target structure are on the right.}
    \label{Afig:blockages}
\end{figure}

Here is a description of  the steps of the algorithm. 
Let \(N\) denote the number of vertices or monomers of our target structure \(G\).  The program performs an initial set-up as follows. 
    \begin{itemize}
        \item Assign an index to each edge. This is done in lexicographical order of the numbers of the vertices. For example, in the case of the \(2 \times 2\) square with vertices labeled \(1,2,3,4\) we have the following indexing of edges
        \begin{align*}
            (1,2) \longleftrightarrow 1,\quad (1,3) \longleftrightarrow 2, \quad (1,4) \longleftrightarrow 3, \quad (2,4) \longleftrightarrow 4\,.
        \end{align*}

        \item Instantiate a list of all substructures of our completed structure, equivalently the list of all clusters in our system. This is done by taking all elements of \(V \in \mathcal{P}(\set{1,\hdots,N})\) (where $\mathcal P$ denotes the power set operator), and checking if the induced subgraph with vertex set \(V\) is connected. Finally, a hashmap structure is used to assign indexing to the clusters.

        \item Determine the adjacent edges $E_{AB}$ between each pair of clusters $A,B$. This is necessary in order to later compute the adjacent bonds \(\set{\delta_i}_{i\in E_{AB}}\) between each pair of clusters. To do this, a list of shared edges between the two clusters is instantiated, which can be accessed using the indices assigned in the previous step, by the pair \((\text{index(A)},\text{index}(B))\). %This is necessary for when the binding energy between the two is assigned, so that the incident energy between the two substructures can be assigned as \(e^{\sum_i \delta_i}\), where \(\set{\delta_i}_1^m\) are the set of binding energies of their shared edges (here we can ignore the scaling factor by dividing it out of the entire system of equations).

        \item
        Determine, for each cluster $P$, a list of 
        all the reactions it is involved in. This is done in two steps. 
        \begin{enumerate}
            \item First, determine a list of clusters from which $P$ can form, by merging clusters together. %This is done individually for each polymer, by iterating over all of the substructures. 
        This is similar to the initialization of the list of clusters, where given the initial cluster \(P\), we iterate over the power set of its vertices, to find all of the vertex sets \(V\), where the induced subgraph \(G_V\) of \(V\) is connected, as well as the induced subgraph \(G_{P\setminus V}\) of \(P \setminus V\). In this case, we encode interactions for \(G_V\) and \(G_{P \setminus V}\) reacting to form \(P\), as well as \(P\) breaking apart to form \(G_V\) and \(G_{P \setminus V}\).
        \item Second, find all the reactions that \(P\) is involved in to make a bigger cluster. To do this, we find all of the clusters (connected subgraphs) \(H \subset G\) containing \(P\), such that the induced subgraph \(K\) of \(H \setminus P\) is also a cluster. We then record the interactions for \(H\) breaking apart to form \(P\) and \(K\), as well as \(P\) and \(K\) binding to form \(H\).

        \end{enumerate}   

        For the second step, we disallow certain reactions 
        for graphs which can be physically embedded in lattices, where the reaction would require particles physically passing through each other. See Figure \ref{Afig:blockages} for examples. We call such reactions ``blockages''. 
        For this, we manually check for each of the disallowed cases at this step, and we do not add interactions to subgraphs corresponding to these cases. For example in the case of the \(3 \times 3\) square, we do not encode interactions for \(P = 5\), when \(\set{2,4,6,8} \subset V_K\), where \(V_K\) is the vertex set of \(K\).

    \end{itemize}

    After these initial steps, the program writes a text file corresponding to the reaction equations of the system, noting the following steps.

    \begin{itemize}
        \item We pre-compute coefficients involved in the reaction rates, to increase efficiency when solving the ODEs and performing parameter optimization. 
        For example, if clusters \(P_1\) and \(P_2\) are connected by edges with associated energies \(\delta_1, \delta_2, \delta_6\), then \(e_{1\_2\_6} = \exp(\delta_1+ \delta_2 + \delta_6)\) will be written to the file.
        To do this, we iterate over all of the interactions from the set up phase, and label the associated energies. 

        \item We compute the right-hand side of the reaction equations \eqref{eq:ch} for each species, by iterating over all of the reactions associated to a species \(P\), and writing the reactions to a string.
    \end{itemize}

\medskip

The computational complexity of producing, solving and optimizing the reaction diffusion equation scales very quickly with the number of vertices. Several metrics for this scaling are explored in Table \ref{tbl:complexity}. 
We note that optimizing is much more computationally challenging than producing the equations, which in turn is much more challenging than solving them. 

%This complexity makes the production, computation and optimization of the reaction diffusion equations quickly infeasible in more complicated or larger systems. %Due to these difficulties, parameter optimization using reaction diffusion equations was not possible for larger systems and even the optimization provided in C.f. Figure \ref{fig:examples} has memory blowups for many initial conditions, making a large sample of the optimizations difficult.

\begin{table}
\centering % used for centering table
\begin{tabular}{c c c c c c} % centered columns (4 columns)
\hline\hline %inserts double horizontal lines
Species & \# Vertices & \#  Edges & \# clusters & \#  reactions & File Size\\ % inserts table
%heading
\hline % inserts single horizontal line
2x2 Square &4 & 4 & 13 & 18 & 2 Kb \\
\hline
E1 &9 & 12 & 218 & 1376 & 149 Kb \\
\hline
E2 &7 &12 &95 & 448 & 55 Kb \\% [1ex] adds vertical space
\hline
E3 &8 &12 &167 & 1079 & 132 Kb \\
\hline
E4 &6 &15 &63 & 301 & 59 Kb \\
\hline
4x4 Square &16 & 24 & 11506 & 224,305 & 27.6 Mb \\
\hline %inserts single line
\end{tabular}
\caption{
%\mhc{e1, e2 are with blockages. number is slightly higher without}
Complexity of the reaction equations for various examples. For each example, we show, in order: the number of vertices, the number of interaction energy parameters or, equivalently, edges (which is the number of parameters we are optimizing over), the number of distinct clusters / connected subgraphs, the number of rate constants, and the file size of the code which specifies the reaction equations. }
\label{tbl:complexity} % is used to refer this table in the text
\end{table}

\section{Additional parameter studies}

\begin{figure}
\centering
\includegraphics[width=1\linewidth]{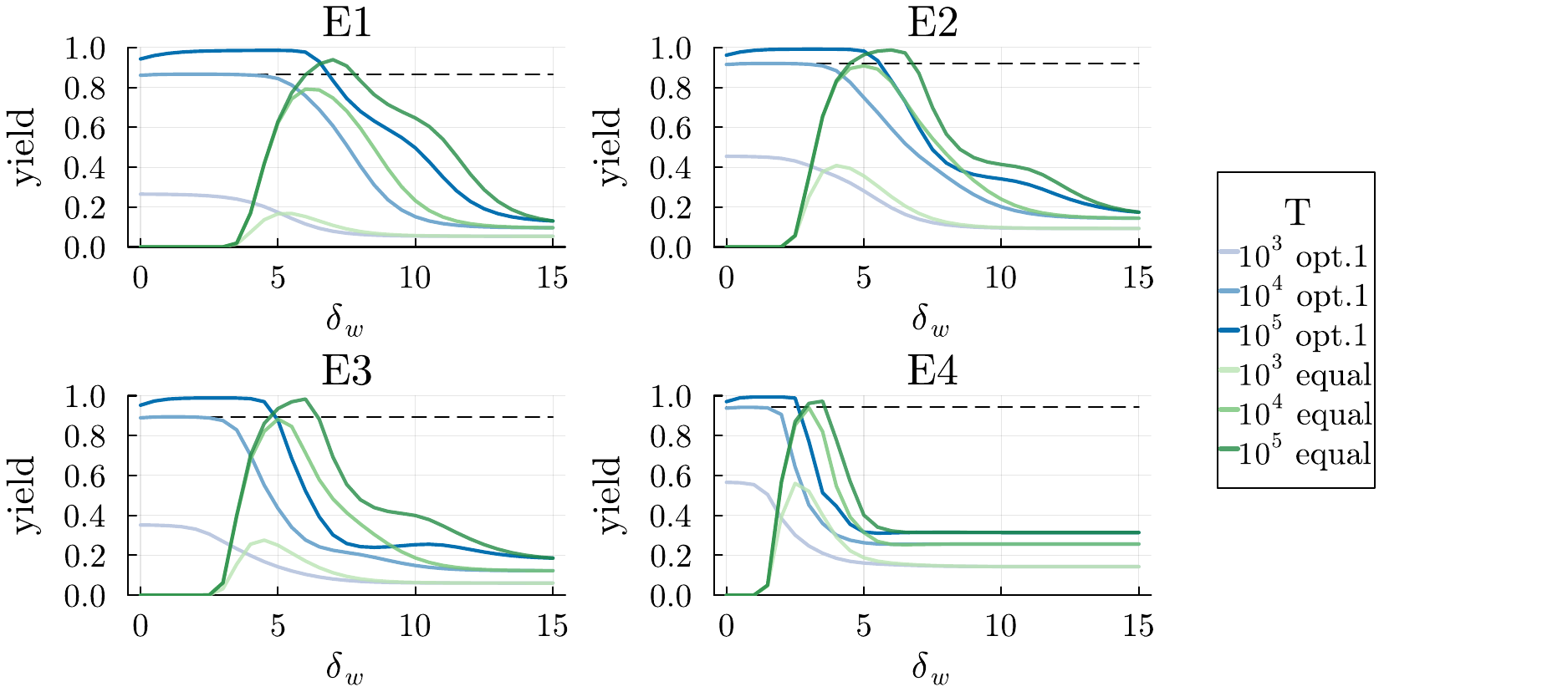}
\caption{Yields for each example E1,E2,E3,E4 as a function of weak bond strength $\delta_w$, computed by solving the reaction equations %with the standard rate model ($k_{A,B}^{\rm on}=1$), 
at different times $T=10^3,10^4,10^5$ as shown in the legend. The interactions are set as follows:  we take the optimal interactions for $T=10^4$ (shown in Figure \ref{fig:examples}), and keeps the strong bonds fixed at strength $\delta_s=15$, and set the weak bonds to strength $\delta_w$.
The pink dashed line labelled ``equal'' shows the yield when all interaction parameters have the same strength $\delta_w$. 
}\label{Afig:weakvary}
\end{figure}

%\begin{figure}
%\centering
%\includegraphics[width=1\linewidth]{Include/weakvary_diffusion.pdf}
%\caption{Yields for each example E1, E2, E3, E4, computed by solving the reaction equations with the fluid rate model ($k_{A,B}^{\rm on}=1/V(AB)$), at different times $T=10^4,4{\cdot}10^4,10^5$ as shown in the legend, and for different structures. 
%Opt. 1 uses the optimal interactions from the \emph{standard} rate model with $T=10^4$, as given in Figure \ref{fig:examples}, and varies the strength of the weak bonds $\delta_w$.  Opt. 2 starts with the optimal interactions from the \emph{fluid} rate model, shown in \mhc{***ref***}, but varies them as follows: it keeps the strong bonds fixed at strength $\delta_s=15$, sets the ``medium'' bonds (as described in the text) to strength $\delta_w$ which is varied, and sets the ``weak'' bonds to strength $0$.
%``Equal'' shows the yield when all interaction parameters have the same strength $\delta_w$.
%}\label{Afig:weakvarydiffusion}
%\end{figure}

% \begin{figure}
% \includegraphics[width=0.9\linewidth]{Plots/strongvary.pdf}
% \caption{Yields at $T=10^4$ for each example E1, E2, E3, E4, as a function of strong bond strength $\delta_s$, computed by solving the reaction equations with the with the standard rate model ($k_{A,B}^{\rm on}=1$) with different weak bond strengths $\delta_w$ as  shown in the legend. \mhc{say which graphs}
% }\label{Afig:rhotrongvary}
% \end{figure}

\begin{figure}
\centering
\includegraphics[width=0.85\linewidth]{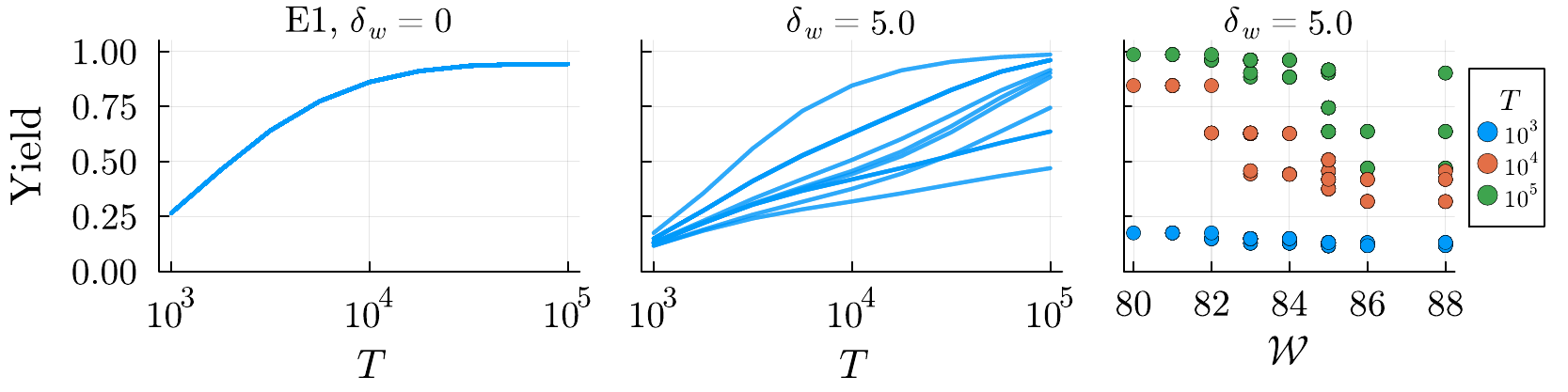}
\includegraphics[width=0.85\linewidth]{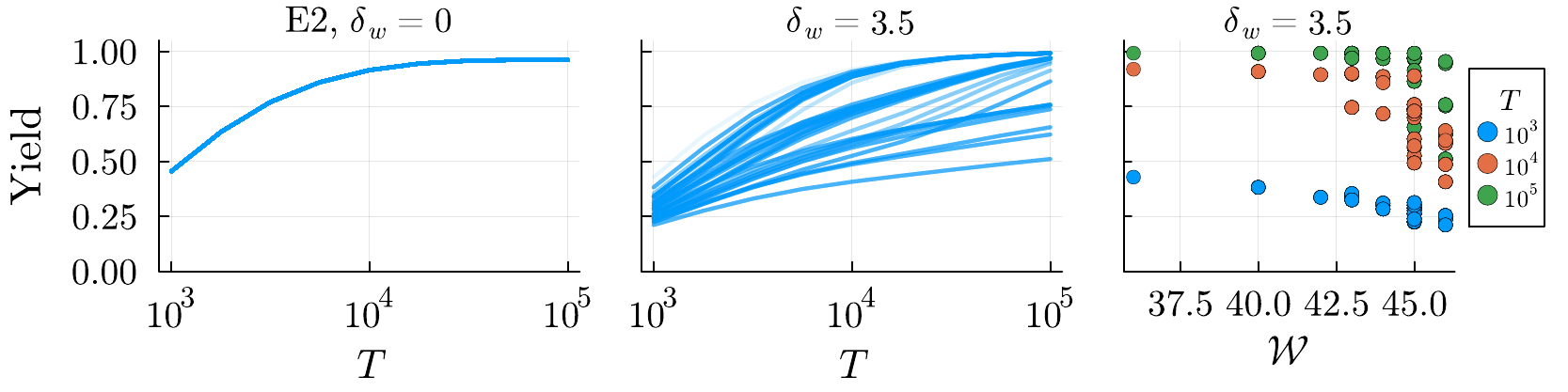}
\includegraphics[width=0.85\linewidth]{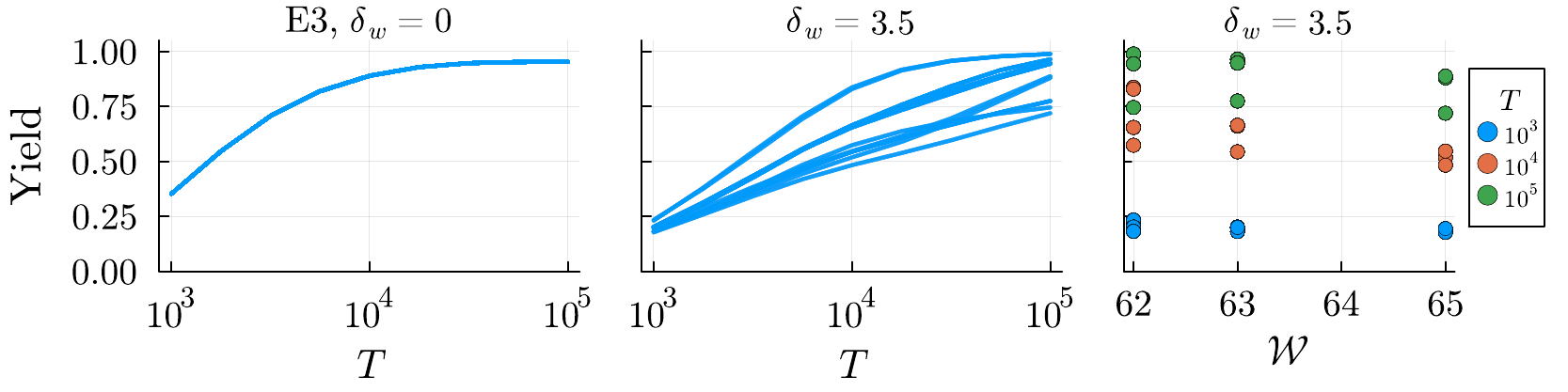}
\caption{Comparing spanning trees for examples (E1,E2,E3), with $\delta_s=15$. Left column: yield versus time for all spanning trees, with $\delta_w=0$. Middle column: same, but with $\delta_w>0$ as shown in the title. Right column: yield versus Weiner index $\mathcal W$ for all spanning trees, at times $T=10^3,10^4,10^5$ as shown in the legend. 
Correlation coefficients for the yield versus $\mathcal W$ at time $T=10^4$ are (E1) -0.83, (E2) -0.68, (E3) -0.68. 
These calculations do not remove ``blockages''; removing them would make a small number of yield curves when $\delta_w=0$ nonequal. 
}\label{Afig:yields}
\end{figure}

\begin{figure}
\centering
\includegraphics[width=0.65\linewidth]{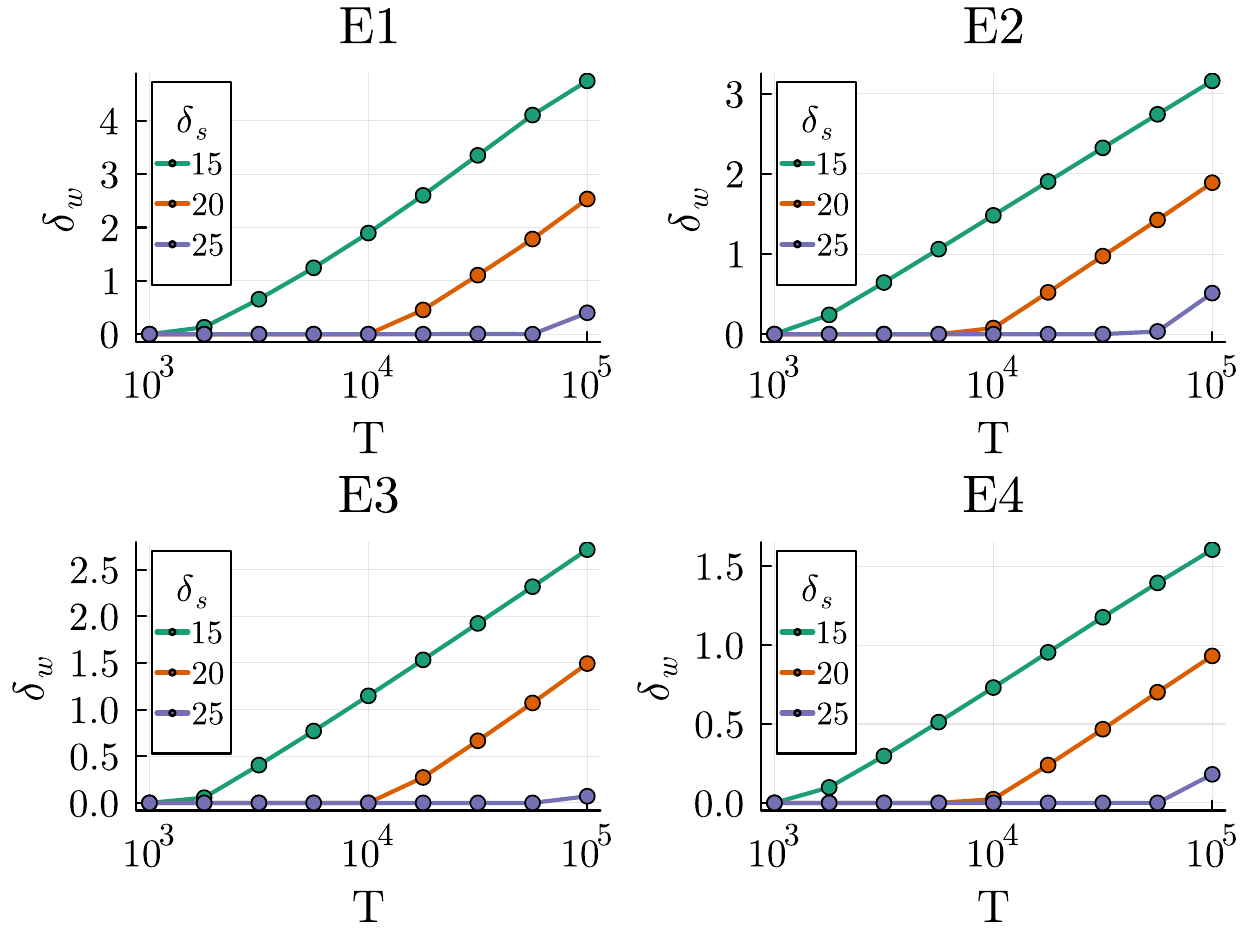}
\caption{Weak bond strength $\delta_w$ which optimizes the yield at time $T$, as a function of $T$, and for 3 different values of strong bond strength $\delta_s$ as shown in the legend. For each example, we started with the network of strong bonds obtained as an optimum for $T=10^4$ in Fig.~\ref{fig:examples}, and set the remaining (weak) bonds to have common energy $\delta_w$. We then optimised $\delta_w$ for a given choice of $T,\delta_s$. 
}\label{Afig:dwopt}
\end{figure}

\begin{figure}
\centering
\includegraphics[width=0.7\linewidth]{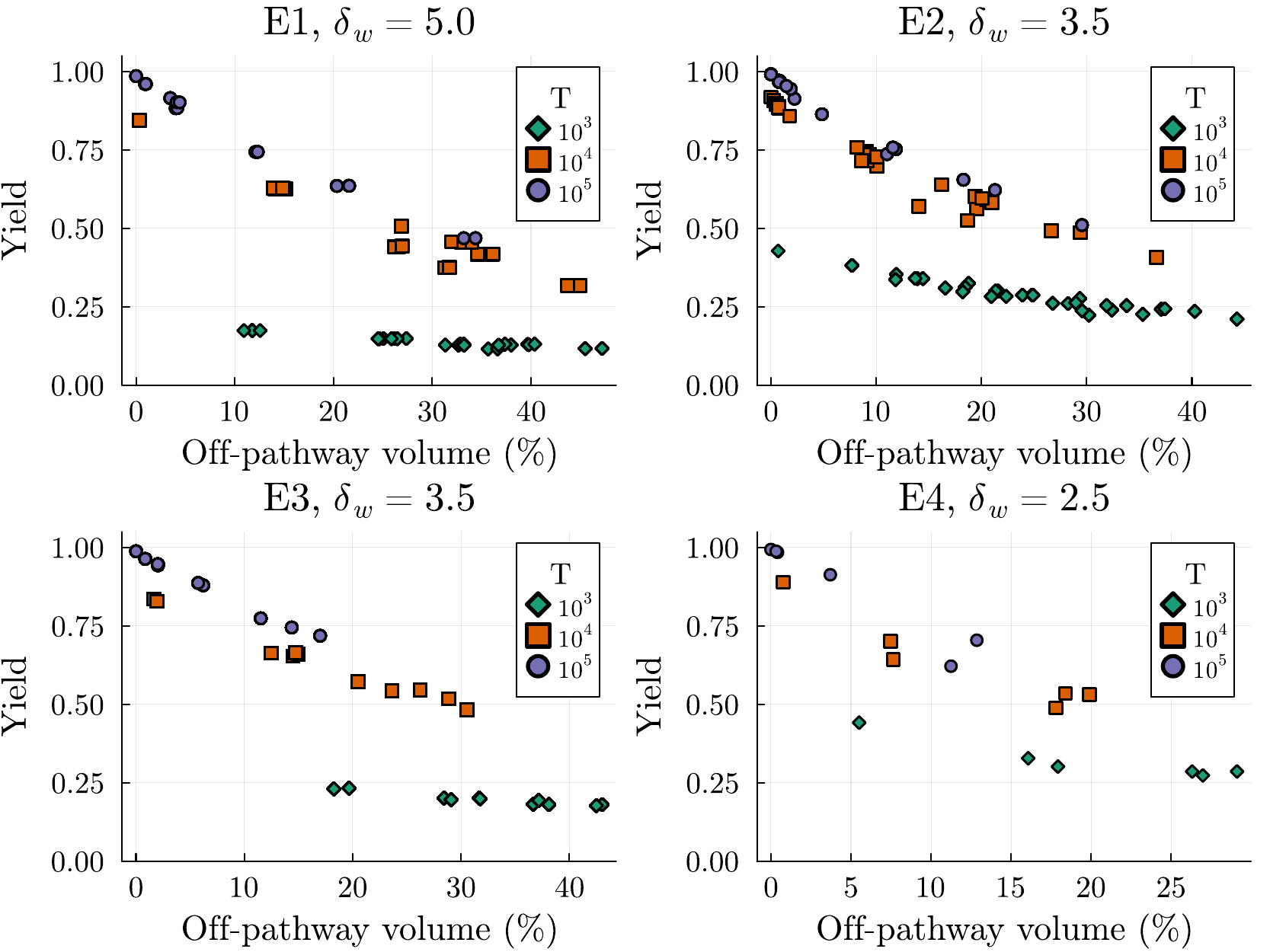}
\caption{Correlation of the yield at times $T=10^3,10^4,10^5$ with the off-pathway volume (the fraction of concentration in ``off-pathway'' clusters as described in the text). Interaction parameters were the same as in Fig. \ref{Afig:yields}. 
Correlation coefficients at time $T=10^4$ are (E1) -0.97, (E2) -0.98, (E3) -0.99, (E4) -0.95.
}\label{Afig:offpathway}
\end{figure}

\begin{figure}
\centering
\includegraphics[width=0.65\linewidth]{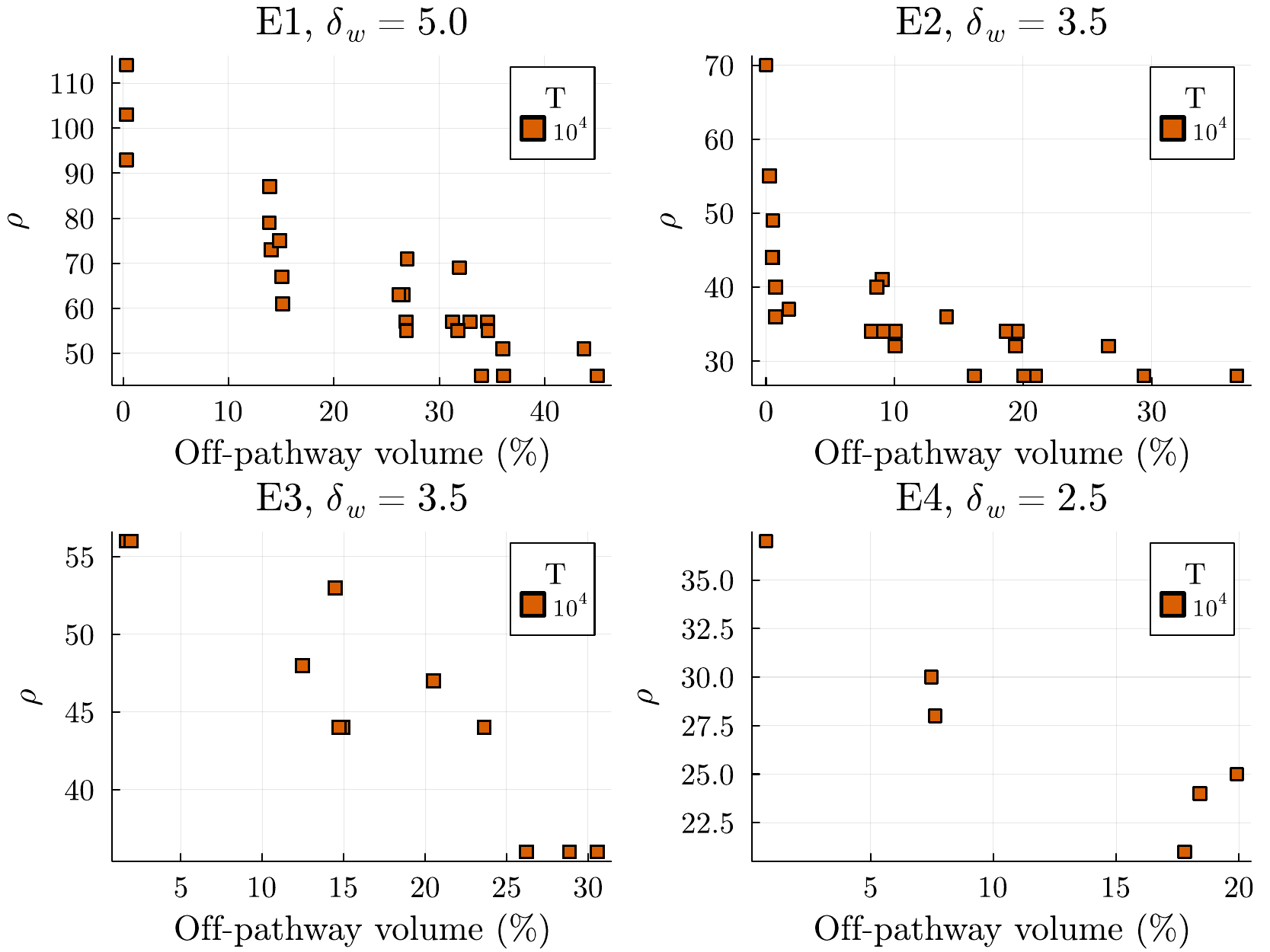}
\caption{Correlation of the $\rho$-index with the off-pathway volume for all spanning trees, for each example, at time $T=10^4$. Interaction parameters were the same as in Fig. \ref{Afig:yields}. 
Correlation coefficients at time $T=10^4$ are (E1) -0.91, (E2) -0.75, (E3) -0.87, (E4) -0.92.
}\label{Afig:rho}
\end{figure}

\begin{figure}
\centering
\includegraphics[width=0.65\linewidth]{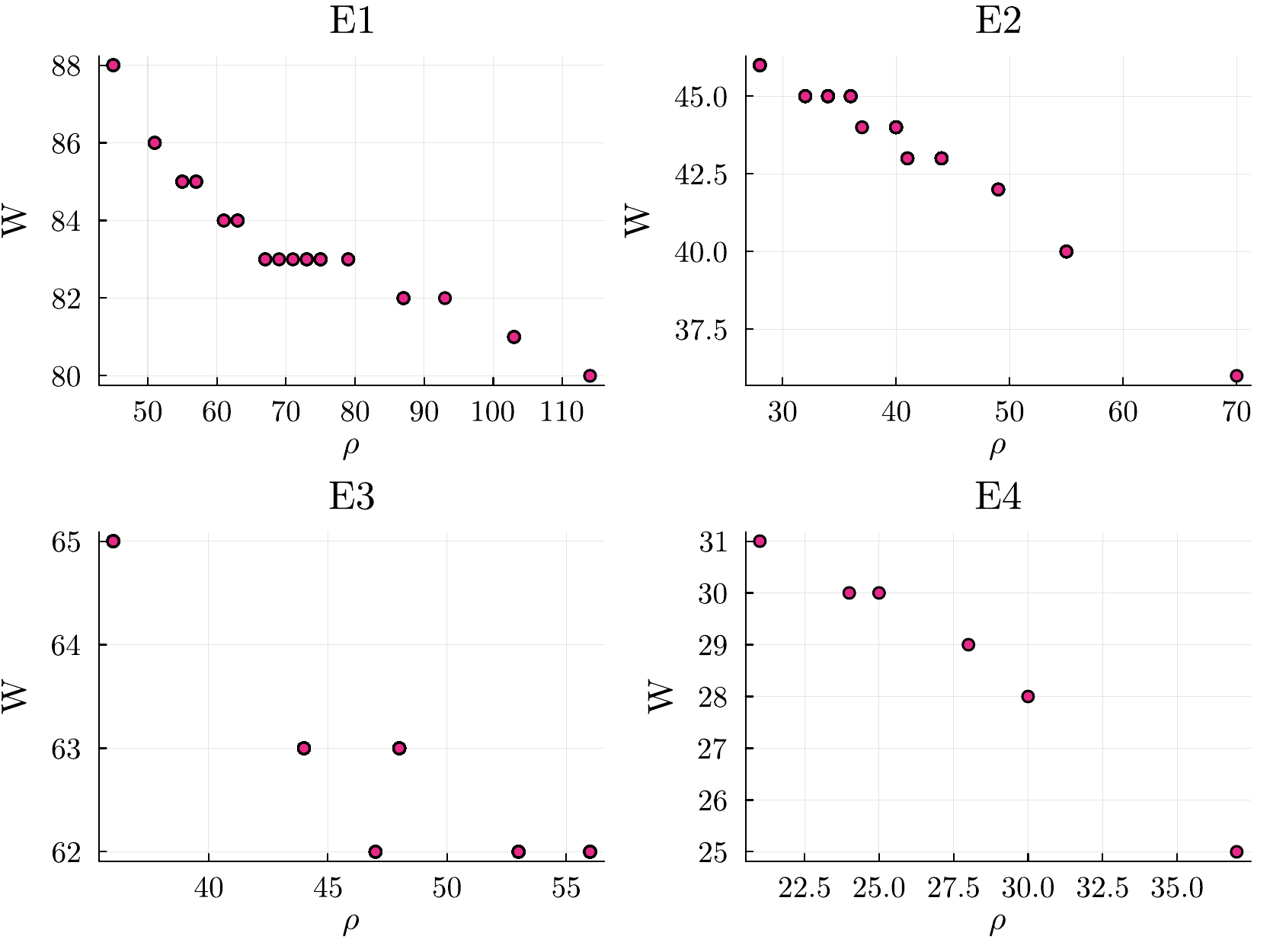}
\caption{Correlation of the $\rho$-index with the Weiner index for all spanning trees, for each example.
Correlation coefficients  are (E1) -0.93, (E2) -0.98, (E3) -0.92, (E4) -0.99.
}\label{Afig:Wrho}
\end{figure}

\begin{figure}
\centering
\includegraphics[width=0.65\linewidth]{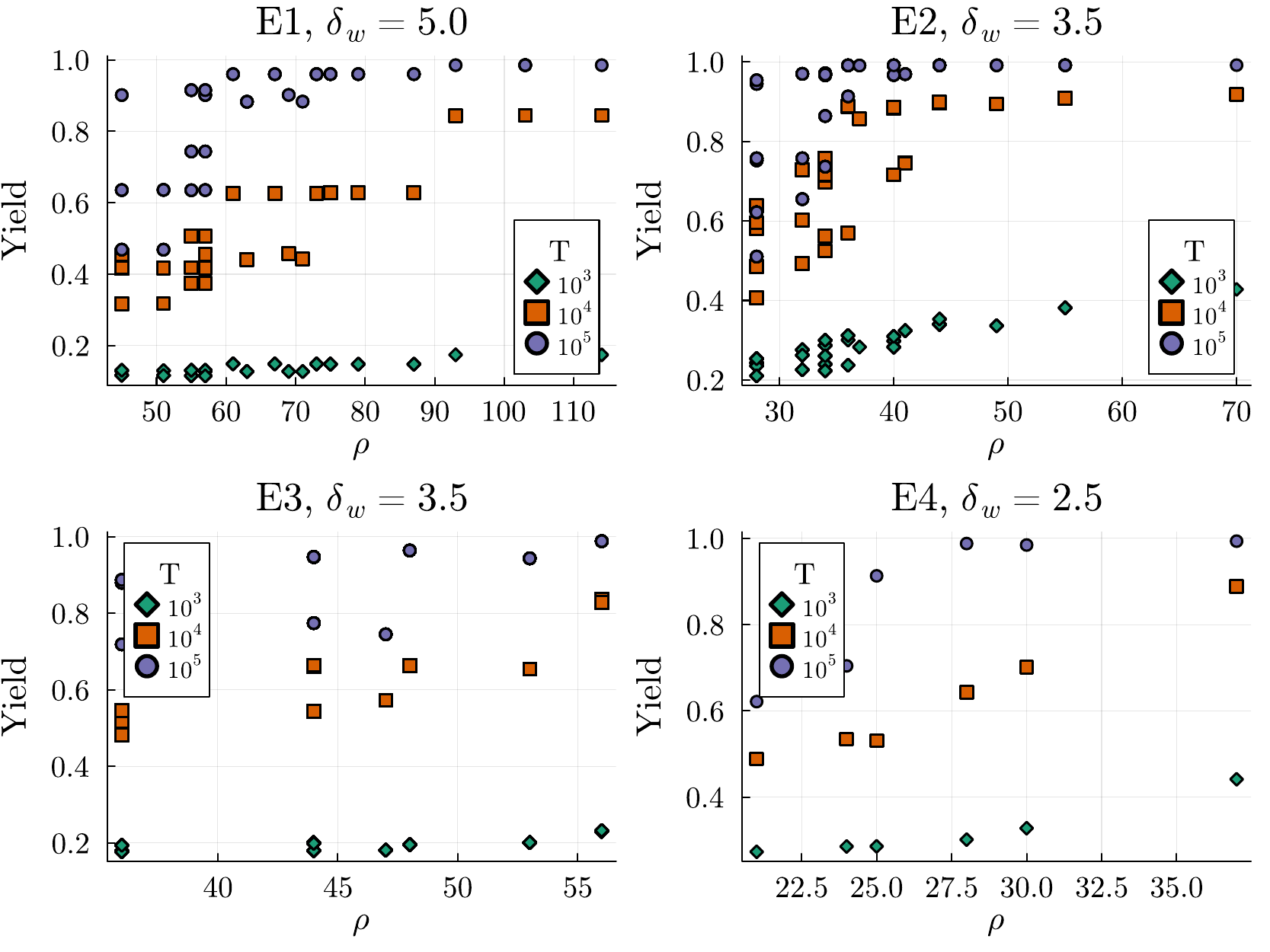}
\caption{Yield versus $\rho$-index for all spanning trees, for each example, at times $T=10^3,10^4,10^5$ as shown in the legend. 
Correlation coefficients for the yield at $T=10^4$ versus $\rho$ are (E1) 0.90, (E2) 0.76, (E3) 0.83, (E4) 0.99.
}\label{Afig:rhoYield}
\end{figure}

\begin{figure}
\centering
\includegraphics[width=0.8\linewidth]{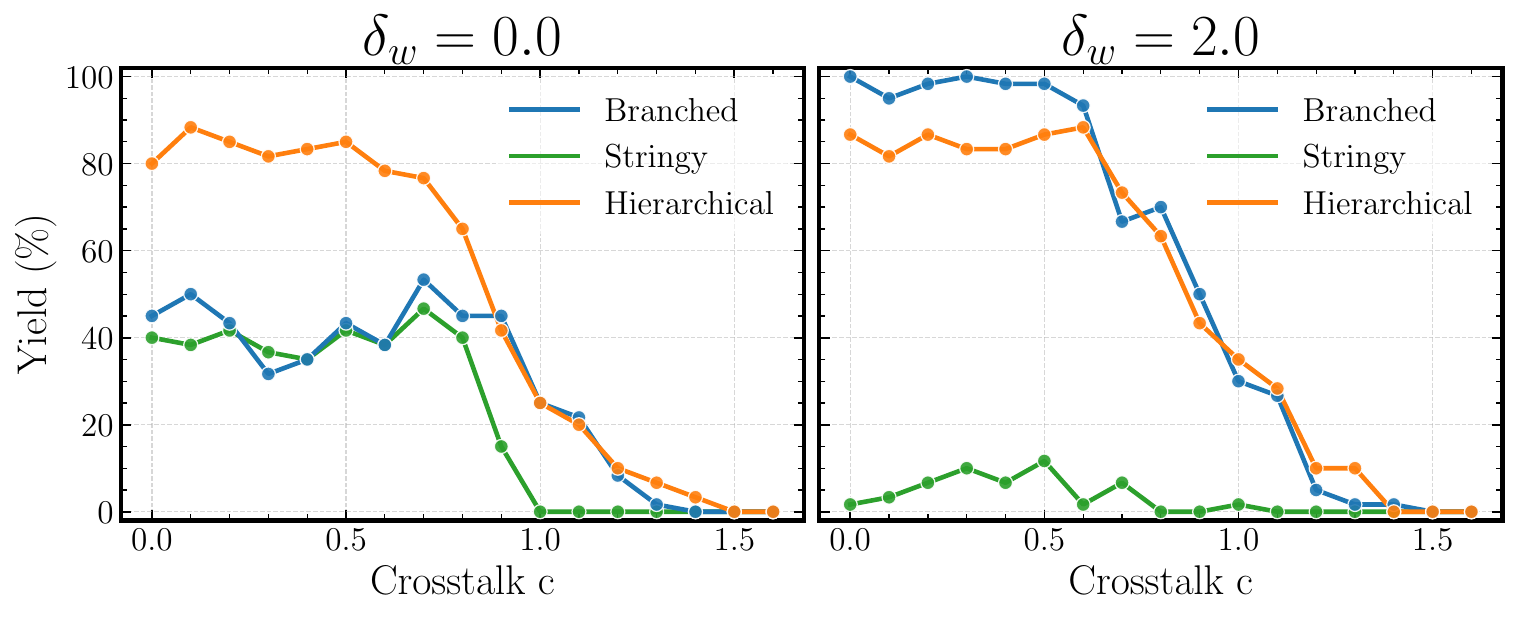}
\caption{Yield of target structures (squares formed from 64 particles) at time $64\times 10^5$ MC sweeps, as a function of the strength of the crosstalk energy $c$. Simulations considered Branched and Stringy spanning trees as well as a Hierarchical strategy as described in Fig.~\ref{fig:sims}. Spanning tree simulations used strong bond energy $\delta_s=15$ and weak bond energy $\delta_w=0$ (left) and $\delta_w=2$ (right). All simulations considered $N_c=60$ copies at volume fraction $0.05$.
}\label{Afig:crosstalk}
\end{figure}

%
%\begin{figure}
%\centering
%\include{TikzPics/tikzoptimalexamples.tex}
%\caption{Spanning trees with the smallest Weiner index, for each example. \tm{is this accurate?}
%% Yield at time $T=10^4$ for each example for all spanning trees with strong bond strength $\delta_s=15$, and with weak bond strength (a) $\delta_w=0$ and (b) $\delta_w=2$ (right). The spanning trees are sorted by decreasing yield. The yield is plotted with small markers. The total number of spanning trees are as follows: E1: 192, E2: 320, E3: 384, E4: 324. \tm{do these numbers account for graph isomorphisms?\tighe{No, because yield depends on the weak bond structure as well as the strong bond structure so is not an isomorphism invariant, in the section on complete graphs the symmetry makes the weak bond structures the same, so everything is filtered by isomorphism class}}
%%(c) shows the optimal spanning trees for each example for $\delta_w=2$. 
%}\label{Afig:best}
%\end{figure}

This section collects figures exploring changes in the model to various parameters, including the sensitivity of the optimal interaction parameters to changes in the strengths of the weak and strong bonds, and correlations of the yield with various graph invariants. %Below, we refer to the model where binding rates are all unity ($k_{A,B}^{\rm on}=1$) and the \emph{standard} rate model, and the model where binding rates decrease with fragment size ($k_{A,B}^{\rm on}=1/V(AB)$) as the \emph{fluid} rate model. 

\begin{itemize}
    \item Fig. \ref{Afig:weakvary} shows the yield as a function of the strength of the weak bonds, for the optimal structures shown in Fig.~\ref{fig:examples}, at $T=10^4$. %, for the standard rate model. 
   % \item Fig. \ref{Afig:weakvarydiffusion} shows the yield as a function of the strength of the ``medium'' bonds, for the fluid rate model. 
    \item Fig. \ref{Afig:yields} shows the yield for all spanning trees, for examples (E1,E2,E3), with no weak bonds, and with some weak bonds. It also shows the correlation of the Weiner index with the yields.
        \item Fig. \ref{Afig:dwopt} shows the optimal value of the weak bond strength at different times, for different values of strong bond interaction. 
    \item Fig. \ref{Afig:offpathway} shows the correlation of the yield with the off-pathway volume, for all spanning trees for each example.
    \item Fig. \ref{Afig:rho} shows the correlation of the $\rho$-index with the off-pathway volume, for all spanning trees for each example.
    \item Fig. \ref{Afig:Wrho} shows the correlation of the $\rho$-index with the Weiner index, for all spanning trees for each example.
    \item Fig. \ref{Afig:rhoYield} shows the correlation of the $\rho$-index with the yield at various times, for all spanning trees for each example.
    %\item Fig. \ref{Afig:best} 
    \item Fig. \ref{Afig:crosstalk} shows the yield of target structures formed from 64 monomers, as a function of crosstalk strength. 

\end{itemize}

%Correlation data:\\
%
%wc = 1\\
% corr(badvol, yield) = -0.973345326792357\\
% corr(badvol, S) = -0.9051338204208625\\
% corr(S,W) = -0.930787635327742\\
%wc = 2\\
% corr(badvol, yield) = -0.9770356646290673\\
% corr(badvol, S) = -0.748789071623088\\
% corr(S,W) = -0.9801238966718406\\
%wc = 3
% corr(badvol, yield) = -0.9868723070098228\\
% corr(badvol, S) = -0.8748641588736513\\
% corr(S,W) = -0.9175292645280151\\
%wc = 4
% corr(badvol, yield) = -0.9483847270386749\\
% corr(badvol, S) = -0.9161528051630061\\
% corr(S,W) = -0.9921835635830106
%
% \medskip
% 
%wc = 1, corr(W,yield) = -0.8327841515797286\\
%wc = 1, corr(S,yield) = 0.8953745990099113\\
%wc = 2, corr(W,yield) = -0.6837543000588273\\
%wc = 2, corr(S,yield) = 0.7615266582750597\\
%wc = 3, corr(W,yield) = -0.6847679847653284\\
%wc = 3, corr(S,yield) = 0.8259687965086054\\
%wc = 4, corr(W,yield) = -0.9934409833149727\\
%wc = 4, corr(S,yield) = 0.9906016791794565\\

%%%%%%   Proof   %%%%%%
\section{Proof of the Spanning Tree Theorem}\label{Asec:proof}

In this section we state and prove  the Spanning Tree Theorem more precisely. 

Consider a graph $H=(V,E)$ where $V$ is the set of vertices, and $E$ is the set of edges. We write $V(H)$ to mean the vertex set of $H$ and $E(H)$ to mean the edge set of $H$. We write $N\cdot V(H)$ to denote the multiset where each element of $V(H)$ has multiplicity $N$. 
The symbol  \(\sqcup\) will denote a disjoint union of elements.\\

  \textbf{\textit{Spanning Tree Theorem.}}
    Let \(H\) be a connected, acyclic graph. Suppose we have connected subgraphs $\set{G_i}_{i=1}^s$, such that each $G_i \subset H$, and 
    \[ \bigsqcup_{1 \leq i \leq s}V(G_i) = N\cdot V(H).\] 
    Then there exists a partition  \(\set{S_j}_{j=1}^{N}\) of  \(\set{G_i}_{i=1}^s\), and edge sets \(\set{E_j}_{j = 1}^{N}\) with $E_j\subset E$, such that, for each $j=1,\ldots,N$, 
    \begin{equation}\label{eq:thm}
     \bigsqcup_{G_i\in S_j} V(G_i) = V(H), \qquad E_j \sqcup \Big\{\bigsqcup_{G_i\in S_j} E(G_i)\Big\} = E(H).
    \end{equation}
    %\(E_j\bigcup_{G_i\in S_j} G_i = H\). 
    \newline
    
 In the language of our problem, \(H\) is the target structure,  and each \(G_i\) represents an intermediate state in the assembly process with $\set{G_i}_{i=1}^s$ being the whole collection of $s$ current fragments. Each \(S_j\) is a collection of intermediate states, i.e. a collection of \(G_i\), such that they may be glued together with edges/bonds in edge set  \(E_j\) to form one copy of \(H\). 
 
 The key idea of the proof below, is that if we have a connected subgraph $G'$, and if we consider an edge $(u,v)\subset E(H)$ between vertices $u\in G', v\notin G'$, then the subgraphs $G_i$ which contain $v$ and which have some overlap with $G'$ (have common vertices) must all contain $u$. By counting the number of copies of vertices $u$ and $v$, we determine this cannot always happen, so there must exist a subgraph $G''$ not yet in our collection which may be glued to $G'$ to form a larger connected subgraph. We continue this process until we complete one copy of $H$, and then repeat.  
    
%    \tm{is there any way to give a, high-level overview, in words, of the key idea of the proof below? \tighe{ANS: Given an incomplete cluster \(C\), such that for some edge \((u,v) \in T\) we have \(u \in C, \; v \not \in C\), then the graphs containing \(v\) that "overlap" with \(C\) (i.e. cannot bond with \(C\)) all contain \(u\)}}\\

    \textbf{Proof}
    We proceed by induction on \(N\). If \( N=1\), then \(\bigsqcup_i V(G_i) = V(H)\). Since this union of vetices is disjoint, we must also have 
    that  \(\bigsqcup_i E(G_i) \subset E(H)\). The existence of some edge set \(E_1\), such that \(E_1\sqcup\{\bigsqcup_i E(G_i) \}= H\) is immediate.

    Now suppose $N>1$, and suppose the statement of the theorem holds up to $N-1$. Notice that if there is some sub-collection \(\mathcal{U} \subset \set{G_i}_{i=1}^s\), such that \(\bigsqcup_\mathcal{U} V(G_i) = V(H)\), then the set of graphs \(\set{G_i}_{i=1}^s \setminus \mathcal{U}\) satisfies the original hypotheses on \(N-1\) copies of the vertex set, and so by the induction hypothesis there exists a partition $\set{S_j}_{j=1}^{m_1}$ such that \eqref{eq:thm} holds for this partition. Thus, the problem is reduced 
    to being able to construct one copy of \(H\). 
    
    The core of the argument lies in showing the following statement: for any collection \(\mathcal{V}\subset \set{G_i}_{i=1}^s\) of subgraphs  such that (i) the vertices in $\{G_i\}_{G_i\in \mathcal V}$ are disjoint, but their union does not equal $V(H)$, and (ii) the subgraphs in $\mathcal V$ may be connected by edges in $E'\subset E(H)$, then there exists a subgraph \(K \in \set{G_i}_{i=1}^s \setminus \mathcal{V}\), 
    such that the set of graphs $V\cup\{K\}$ may be connected by edges in $E(H)$. 
    
    Once we have shown this statement, then we could replace \(\mathcal{V}\) with \(\mathcal{V} \cup \set{K}\) and repeat this process until the vertex set in $\mathcal V$ equals $V(H)$. Then we will have constructed our single copy of $H$.

Now we  show existence of such a \(K \in \set{G_i}_{i=1}^s\setminus \mathcal{V}\). 
    Assume for the sake of contradiction that no such \(K\) exists. We may number the vertices of \(H\) as \(v_1,\hdots,v_\ell\) where without loss of generality we may choose the ordering such that \(\set{v_1,v_2,\hdots,v_r} = \bigcup_{G_i \in \mathcal{V}}V(G_i)\). 
    Consider the edge set $\bar E = E'\sqcup \bigsqcup_{G_i \in \mathcal{V}}E(G_i)$, and let $C$ denote a maximal connected component of the graph complement of $\bar E$ with respect to $H$. There must be exactly one edge in $E(H)$ connecting $C$ to $E'$, because $H$ is connected and acyclic. 
Without loss of generality, we may assume the label of the vertex this edge is attached to is \(v_{r+1}\). 

It follows that each \(G_i \in \set{G_i}_{i=1}^s\setminus \mathcal{V}\) containing \(v_{r+1}\) is such that \(G_i \not \subset C\), or else we could choose \(K = G_i\). But this implies that \(G_i\) must contain \(v_r\), since this is the only vertex in \(H\) connected to \(C\). 
This implies that the set of graphs in $\set{G_i}_{i=1}^s \setminus \mathcal{V}$ which contain $v_{r+1}$, is the same as the set of graphs in $\set{G_i}_{i=1}^s \setminus \mathcal{V}$ which contain $v_r$: 
    \begin{align*}
        \set{G_i \in \set{G_i}_{i=1}^s \setminus \mathcal{V} \mid v_{r+1} \in G_i} = 
        \set{G_i \in \set{G_i}_{i=1}^s \setminus \mathcal{V} \mid v_r \in G_i}.
    \end{align*}
    But now let's count the number of copies of $v_r$ and $v_{r+1}$: there are $N$ copies of $v_{r+1}$ in $\set{G_i}_{i=1}^s \setminus \mathcal{V}$, so by the identity above, there are also $N$ copies of $v_r$ in $\set{G_i}_{i=1}^s \setminus \mathcal{V}$. However there is one more copy of $v_r$ in $\mathcal V$, leading to $N+1$ total copies of $v_r$, a contradiction. 
% \qed

%%%%%%   Graph stuff   %%%%%%
\section{Properties of graph invariants}

In this section we discuss some properties of the graph invariants we have introduced. In this section, $T$ denotes a spanning tree, and $G$ denotes the graph it is embedded in.

%\tm{I got rid of $I(T)$ in the text, because it didn't seem necessary, and it made the story more complicated in a way that didn't seem to add anything. But this section uses it. Do you need to introduce it here? Or do you think it should be put back in the text? Or is there a way to write this section without $I(T)$? 
%
%If you put it back in the main text, then you'll have to convince me that it adds something beyond $N(T)$.} 
%
%\tm{feel free to edit text as you see fit -- there are lots of explanations that could be made more clear, by using more direct / less passive language, and by reordering the sentences, and by making them shorter. }

\subsection{Greedy Algorithm Optimizing \(\rho(T)\)}\label{Asec:greedy}
Consider a vertex \(u \in T\), and suppose that \(u\) is connected to the vertices \(v_1,\hdots,v_k\). If \(f(v)\) denotes the number of subtrees containing \(v\), and suppose that \(f(u) > f(w)\), then (here \(T \cup_{(\ell,v)}\) denotes the resulting tree by adding a leaf to vertex \(v\) of \(T\))
    \begin{align*}
        \rho(T \cup_{(\ell,u)}) = 1 + f(u) + \rho(T) > \rho(T \cup_{(\ell,w)}) = 1 + f(w) + \rho(T)
    \end{align*}
    This provides a straightforward greedy algorithm specifying removal of a leaf, then where the leaf can be moved to in order to maximize \(\rho(T)\). Note in particular, this implies that for any graph admitting a star shaped spanning tree \(K_{1,n-1} \subset G\) we have that \(K_{1,n-1}\) minimizes \(\rho(T)\) with respect to \(G\).

    Similar formation processes to this greedy algorithm have been observed to occur in the natural self assembly process of crystal growth. 
    The formation exhibited in by the greedy algorithm is layer by layer formation (LBL), the same as experimentally exhibited crystal formation on smooth surfaces in \cite{Luo:2023}. It was also found that crystal growth exhibits the minimization of the Wiener index, an invariant which is correlated to \(\rho(T)\) as explained in the main text. Bonchev, Mekenyan, and Fritsche \cite{Bonchev:1980} exhibited that a crystal formation model (for ideal high-energy surfaces) based on minimizing the Wiener Index was consistent with Kossel—Stransky theory and experimental observations.

\subsection{Star shaped trees are optimal}\label{Asec:starenergetic}
If a graph \(G\) contains a star shaped spanning tree \(T\) and a non-star shaped spanning tree \(T'\) then \(G\) must contain an off-pathway cluster (with respect to \(T'\)) with a strong bond. To see this let \(v\) be the center vertex of \(T\). Then, since \(T'\) is not a star, there is some vertex \(u\) which is not connected to \(v\) in \(T'\). If \(w\) is a vertex that is adjacent to \(v\) but not \(u\) in \(T'\), then \((wvu)\) is the desired off-pathway subgraph of \(G\). In the case that no such \(w\) exists, since \(T'\) is a tree we can conclude that \(v\) is a leaf of \(T'\) in this case since \(T'\) is not a star, there must be a vertex \(s\) with \(d(v,s) \geq 3\); in this case let \(P = (v,s_1,\hdots,s_{d-1},s)\) be a shortest path between \(u\) and \(s\). Then \((u,s_1,s)\) is the desired off-pathway subgraph of \(G\). This proves that if the star-shaped graph \(K_{1,n-1} \subset G\) then there are no off-pathway clusters with respect to this spanning tree. 
\section{Data -- Optimal interactions}\label{sec:data}

The following sections show the data obtained from optimizing the concentration of a target structure for examples E1, E2, E3, E4. %, with two different choices of reaction rates. 

The optimal parameters are shown in tables, sorted by yield, with the optimal parameter values  reported to 1 decimal place and the yield to 4 decimal places. The corresponding optimal graphs are sketched below the table, with the following conventions: black corresponds to the largest bond allowed ($\delta_{\rm max}=15$), no edge corresponds to a 0 energy bond, and grey bonds with labels show the values of that bond. An exception is E4 where we didn't label any bonds as the graphs would be too cluttered. 

%Some details specific to each choice of reaction rate are given below. 

%\paragraph{Standard model: Reaction rate $k_{A,B}^{\rm on}=1$.}
%For each example, the optimization was run 6 times for each of times $T=10^3,10^4$ with random initial conditions. 

%\paragraph{Fluid model: Reaction rate $k_{A,B}^{\rm on}=1/V(AB)$.}
%For each example, we show output from the optimization for 2 runs for each of times $T=10^3,10^4$, and 4 runs for $T=4\cdot 10^4$. (We sometimes ran the optimization more than this, but we selected certain data to highlight various qualitative and quantitative features.) We chose a longer timescale than for the previous case, because the rate constants are smaller on average in this case. 

\setlength{\tabcolsep}{6pt}

% !TEX root = main.tex
\subsubsection{Example E1}%, Standard model}

\begin{tabular}{c | l | l | c c c c c c c c c c c c |}
run & time & yield & $\delta_{12}$&$\delta_{14}$&$\delta_{23}$&$\delta_{25}$&$\delta_{36}$&$\delta_{45}$&$ \delta_{47}$&$\delta_{56}$&$\delta_{58}$&$\delta_{69}$&$\delta_{78}$&$\delta_{89}$   \\\hline
1& $10^3$& 0.2650& 0.0&15.0&15.0&15.0&0.0&15.0&0.0&15.0&15.0&0.0&15.0&15.0\\
2 & $10^3$& 0.2650& 15.0&0.0&15.0&15.0&0.0&15.0&0.0&15.0&15.0&0.0&15.0&15.0\\
3 & $10^3$& 0.2650& 15.0&0.0&15.0&15.0&0.0&15.0&0.0&15.0&15.0&0.0&15.0&15.0\\
4 & $10^3$& 0.2649& 0.0&15.0&15.0&0.0&15.0&15.0&15.0&15.0&15.0&15.0&0.0&0.0\\
5 & $10^3$& 0.2649&  0.0&15.0&15.0&15.0&0.0&15.0&0.0&15.0&0.0&15.0&15.0&15.0\\
6 & $10^3$& 0.2283& 15.0&15.0&15.0&0.0&15.0&15.0&0.0&0.0&0.0&15.0&15.0&15.0\\
\hline
7& $10^4$& 0.8657& 1.9&15.0&15.0&15.0&1.9&15.0&1.9&15.0&15.0&1.9&15.0&15.0\\
8 &$10^4$& 0.8657& 1.9&15.0&15.0&15.0&1.9&15.0&1.9&15.0&15.0&1.9&15.0&15.0\\
9 & $10^4$& 0.8655& 1.9&15.0&1.9&15.0&15.0&15.0&1.5&15.0&0.7&15.0&15.0&15.0\\
10& $10^4$& 0.8655& 15.0&15.0&15.0&0.7&1.5&15.0&1.9&15.0&15.0&15.0&15.0&1.9\\
11 & $10^4$& 0.8655& 1.9&15.0&15.0&15.0&1.5&15.0&1.9&0.7&15.0&15.0&15.0&15.0\\
12 & $10^4$& 0.8655& 15.0&1.9&15.0&15.0&1.5&15.0&1.9&0.7&15.0&15.0&15.0&15.0\\\hline
\end{tabular}\bigskip

\begin{tikzpicture}
% item 1
\begin{scope}[shift={(0,0)}]
%\node at (0,2.6) {(1)};  %[anchor=west, align=left]
\node[align=center] at (1,2.7) {(1) $T{=}10^3$, 26.50\%};
%vertex coordinates
\node[vertex] (a) at (0,2) {1};
\node[vertex] (b) at (1,2) {2};
\node[vertex] (c) at (2,2) {3};
\node[vertex] (d) at (0,1) {4};
\node[vertex] (e) at (1,1) {5};
\node[vertex] (f) at (2,1) {6};
\node[vertex] (g) at (0,0) {7};
\node[vertex] (h) at (1,0) {8};
\node[vertex] (i) at (2,0) {9};
% bonds
\draw[zero] (a)--(b); %12
\draw[strong] (a)--(d); %14
\draw[strong] (b)--(c); %23
\draw[strong] (b)--(e); %25
\draw[zero] (c)--(f); %36
\draw[strong] (d)--(e); %45
\draw[zero] (d)--(g); %47
\draw[strong] (e)--(f); %56
\draw[strong] (e)--(h); %58
\draw[zero] (f)--(i); %69
\draw[strong] (g)--(h); %78
\draw[strong] (h)--(i); %89
\end{scope}
% item 2
\begin{scope}[shift={(4,0)}]
%node at (0,2.6) {(2)};
\node[align=center] at (1,2.7) {(2) $T{=}10^3$, 26.50\%};
%vertex coordinates
\node[vertex] (a) at (0,2) {1};
\node[vertex] (b) at (1,2) {2};
\node[vertex] (c) at (2,2) {3};
\node[vertex] (d) at (0,1) {4};
\node[vertex] (e) at (1,1) {5};
\node[vertex] (f) at (2,1) {6};
\node[vertex] (g) at (0,0) {7};
\node[vertex] (h) at (1,0) {8};
\node[vertex] (i) at (2,0) {9};
% bonds
\draw[strong] (a)--(b); %12
\draw[zero] (a)--(d); %14
\draw[strong] (b)--(c); %23
\draw[strong] (b)--(e); %25
\draw[zero] (c)--(f); %36
\draw[strong] (d)--(e); %45
\draw[zero] (d)--(g); %47
\draw[strong] (e)--(f); %56
\draw[strong] (e)--(h); %58
\draw[zero] (f)--(i); %69
\draw[strong] (g)--(h); %78
\draw[strong] (h)--(i); %89
\end{scope}
% item 3
\begin{scope}[shift={(8,0)}]
%\node at (0,2.6) {(3)};
\node[align=center] at (1,2.7) {(3) $T{=}10^3$, 26.50\%};
%vertex coordinates
\node[vertex] (a) at (0,2) {1};
\node[vertex] (b) at (1,2) {2};
\node[vertex] (c) at (2,2) {3};
\node[vertex] (d) at (0,1) {4};
\node[vertex] (e) at (1,1) {5};
\node[vertex] (f) at (2,1) {6};
\node[vertex] (g) at (0,0) {7};
\node[vertex] (h) at (1,0) {8};
\node[vertex] (i) at (2,0) {9};
% bonds
\draw[strong] (a)--(b); %12
\draw[zero] (a)--(d); %14
\draw[strong] (b)--(c); %23
\draw[strong] (b)--(e); %25
\draw[zero] (c)--(f); %36
\draw[strong] (d)--(e); %45
\draw[zero] (d)--(g); %47
\draw[strong] (e)--(f); %56
\draw[strong] (e)--(h); %58
\draw[zero] (f)--(i); %69
\draw[strong] (g)--(h); %78
\draw[strong] (h)--(i); %89
\end{scope}
% item 4
\begin{scope}[shift={(12,0)}]
%\node at (0,2.6) {(4)};
\node[align=center] at (1,2.7) {(4) $T{=}10^3$, 26.49\%};
%vertex coordinates
\node[vertex] (a) at (0,2) {1};
\node[vertex] (b) at (1,2) {2};
\node[vertex] (c) at (2,2) {3};
\node[vertex] (d) at (0,1) {4};
\node[vertex] (e) at (1,1) {5};
\node[vertex] (f) at (2,1) {6};
\node[vertex] (g) at (0,0) {7};
\node[vertex] (h) at (1,0) {8};
\node[vertex] (i) at (2,0) {9};
% bonds
\draw[zero] (a)--(b); %12
\draw[strong] (a)--(d); %14
\draw[strong] (b)--(c); %23
\draw[zero] (b)--(e); %25
\draw[strong] (c)--(f); %36
\draw[strong] (d)--(e); %45
\draw[strong] (d)--(g); %47
\draw[strong] (e)--(f); %56
\draw[strong] (e)--(h); %58
\draw[strong] (f)--(i); %69
\draw[zero] (g)--(h); %78
\draw[zero] (h)--(i); %89
\end{scope}
% item 5
\begin{scope}[shift={(0,-4)}]
%\node at (0,2.6) {(5)};
\node[align=center] at (1,2.7) {(5) $T{=}10^3$, 26.49\%};
%vertex coordinates
\node[vertex] (a) at (0,2) {1};
\node[vertex] (b) at (1,2) {2};
\node[vertex] (c) at (2,2) {3};
\node[vertex] (d) at (0,1) {4};
\node[vertex] (e) at (1,1) {5};
\node[vertex] (f) at (2,1) {6};
\node[vertex] (g) at (0,0) {7};
\node[vertex] (h) at (1,0) {8};
\node[vertex] (i) at (2,0) {9};
% bonds
\draw[zero] (a)--(b); %12
\draw[strong] (a)--(d); %14
\draw[strong] (b)--(c); %23
\draw[strong] (b)--(e); %25
\draw[zero] (c)--(f); %36
\draw[strong] (d)--(e); %45
\draw[zero] (d)--(g); %47
\draw[strong] (e)--(f); %56
\draw[zero] (e)--(h); %58
\draw[strong] (f)--(i); %69
\draw[strong] (g)--(h); %78
\draw[strong] (h)--(i); %89
\end{scope}
% item 6
\begin{scope}[shift={(4,-4)}]
%\node at (0,2.6) {(6)};
\node[align=center] at (1,2.7) {(6) $T{=}10^3$, 22.83\%};
%vertex coordinates
\node[vertex] (a) at (0,2) {1};
\node[vertex] (b) at (1,2) {2};
\node[vertex] (c) at (2,2) {3};
\node[vertex] (d) at (0,1) {4};
\node[vertex] (e) at (1,1) {5};
\node[vertex] (f) at (2,1) {6};
\node[vertex] (g) at (0,0) {7};
\node[vertex] (h) at (1,0) {8};
\node[vertex] (i) at (2,0) {9};
% bonds
\draw[strong] (a)--(b); %12
\draw[strong] (a)--(d); %14
\draw[strong] (b)--(c); %23
\draw[zero] (b)--(e); %25
\draw[strong] (c)--(f); %36
\draw[strong] (d)--(e); %45
\draw[zero] (d)--(g); %47
\draw[zero] (e)--(f); %56
\draw[zero] (e)--(h); %58
\draw[strong] (f)--(i); %69
\draw[strong] (g)--(h); %78
\draw[strong] (h)--(i); %89
\end{scope}

% item 7
\begin{scope}[shift={(8,-4)}]
%\node at (0,2.6) {(7)};
\node[align=center] at (1,2.7) {(7) $T{=}10^4$, 86.57\%};
%vertex coordinates
\node[vertex] (a) at (0,2) {1};
\node[vertex] (b) at (1,2) {2};
\node[vertex] (c) at (2,2) {3};
\node[vertex] (d) at (0,1) {4};
\node[vertex] (e) at (1,1) {5};
\node[vertex] (f) at (2,1) {6};
\node[vertex] (g) at (0,0) {7};
\node[vertex] (h) at (1,0) {8};
\node[vertex] (i) at (2,0) {9};
% bonds
\draw[weak] (a)--(b) node[midway, above] {1.9}; %12
\draw[strong] (a)--(d); %14
\draw[strong] (b)--(c); %23
\draw[strong] (b)--(e); %25
\draw[weak] (c)--(f) node[midway,right] {1.9}; %36
\draw[strong] (d)--(e); %45
\draw[weak] (d)--(g) node[midway,left] {1.9}; %47
\draw[strong] (e)--(f); %56
\draw[strong] (e)--(h); %58
\draw[weak] (f)--(i) node[midway,right] {1.9}; %69
\draw[strong] (g)--(h); %78
\draw[strong] (h)--(i); %89
\end{scope}
% item 8
\begin{scope}[shift={(12,-4)}]
%\node at (0,2.6) {(8)};
\node[align=center] at (1,2.7) {(8) $T{=}10^4$, 86.57\%};
%vertex coordinates
\node[vertex] (a) at (0,2) {1};
\node[vertex] (b) at (1,2) {2};
\node[vertex] (c) at (2,2) {3};
\node[vertex] (d) at (0,1) {4};
\node[vertex] (e) at (1,1) {5};
\node[vertex] (f) at (2,1) {6};
\node[vertex] (g) at (0,0) {7};
\node[vertex] (h) at (1,0) {8};
\node[vertex] (i) at (2,0) {9};
% bonds
\draw[weak] (a)--(b) node[midway,above] {1.9}; %12
\draw[strong] (a)--(d); %14
\draw[strong] (b)--(c); %23
\draw[strong] (b)--(e); %25
\draw[weak] (c)--(f) node[midway,right] {1.9}; %36
\draw[strong] (d)--(e); %45
\draw[weak] (d)--(g) node[midway,left] {1.9}; %47
\draw[strong] (e)--(f); %56
\draw[strong] (e)--(h); %58
\draw[weak] (f)--(i) node[midway,right] {1.9}; %69
\draw[strong] (g)--(h); %78
\draw[strong] (h)--(i); %89
\end{scope}
% item 9
\begin{scope}[shift={(0,-8)}]
%\node at (0,2.6) {(9)};
\node[align=center] at (1,2.7) {(9) $T{=}10^4$, 86.55\%};
%vertex coordinates
\node[vertex] (a) at (0,2) {1};
\node[vertex] (b) at (1,2) {2};
\node[vertex] (c) at (2,2) {3};
\node[vertex] (d) at (0,1) {4};
\node[vertex] (e) at (1,1) {5};
\node[vertex] (f) at (2,1) {6};
\node[vertex] (g) at (0,0) {7};
\node[vertex] (h) at (1,0) {8};
\node[vertex] (i) at (2,0) {9};
% bonds
\draw[weak] (a)--(b) node[midway,above] {1.9}; %12
\draw[strong] (a)--(d); %14
\draw[weak] (b)--(c) node[midway,above] {1.9}; %23
\draw[strong] (b)--(e); %25
\draw[strong] (c)--(f); %36
\draw[strong] (d)--(e); %45
\draw[weak] (d)--(g) node[midway,left] {1.5}; %47
\draw[strong] (e)--(f); %56
\draw[weak] (e)--(h) node[midway,right] {0.7}; %58
\draw[strong] (f)--(i); %69
\draw[strong] (g)--(h); %78
\draw[strong] (h)--(i); %89
\end{scope}

% item 10
\begin{scope}[shift={(4,-8)}]
%\node at (0,2.6) {(10)};
\node[align=center] at (1,2.7) {(10) $T{=}10^4$, 86.55\%};
%vertex coordinates
\node[vertex] (a) at (0,2) {1};
\node[vertex] (b) at (1,2) {2};
\node[vertex] (c) at (2,2) {3};
\node[vertex] (d) at (0,1) {4};
\node[vertex] (e) at (1,1) {5};
\node[vertex] (f) at (2,1) {6};
\node[vertex] (g) at (0,0) {7};
\node[vertex] (h) at (1,0) {8};
\node[vertex] (i) at (2,0) {9};
% bonds
\draw[strong] (a)--(b); %12
\draw[strong] (a)--(d); %14
\draw[strong] (b)--(c); %23
\draw[weak] (b)--(e) node[midway,right] {0.7}; %25
\draw[weak] (c)--(f) node[midway,right] {1.5}; %36
\draw[strong] (d)--(e); %45
\draw[weak] (d)--(g) node[midway,left] {1.9}; %47
\draw[strong] (e)--(f); %56
\draw[strong] (e)--(h); %58
\draw[strong] (f)--(i); %69
\draw[strong] (g)--(h); %78
\draw[weak] (h)--(i) node[midway,below] {1.9}; %89
\end{scope}
% item 11
\begin{scope}[shift={(8,-8)}]
%\node at (0,2.6) {(11)};
\node[align=center] at (1,2.7) {(11) $T{=}10^4$, 86.55\%};
%vertex coordinates
\node[vertex] (a) at (0,2) {1};
\node[vertex] (b) at (1,2) {2};
\node[vertex] (c) at (2,2) {3};
\node[vertex] (d) at (0,1) {4};
\node[vertex] (e) at (1,1) {5};
\node[vertex] (f) at (2,1) {6};
\node[vertex] (g) at (0,0) {7};
\node[vertex] (h) at (1,0) {8};
\node[vertex] (i) at (2,0) {9};
% bonds
\draw[weak] (a)--(b) node[midway,above] {1.9}; %12
\draw[strong] (a)--(d); %14
\draw[strong] (b)--(c); %23
\draw[strong] (b)--(e); %25
\draw[weak] (c)--(f) node[midway,right] {1.5}; %36
\draw[strong] (d)--(e); %45
\draw[weak] (d)--(g) node[midway,left] {1.9}; %47
\draw[weak] (e)--(f) node[midway,above] {0.7}; %56
\draw[strong] (e)--(h); %58
\draw[strong] (f)--(i); %69
\draw[strong] (g)--(h); %78
\draw[strong] (h)--(i); %89
\end{scope}
% item 12
\begin{scope}[shift={(12,-8)}]
%\node at (0,2.6) {(12)};
\node[align=center] at (1,2.7) {(12) $T{=}10^4$, 86.55\%};
%vertex coordinates
\node[vertex] (a) at (0,2) {1};
\node[vertex] (b) at (1,2) {2};
\node[vertex] (c) at (2,2) {3};
\node[vertex] (d) at (0,1) {4};
\node[vertex] (e) at (1,1) {5};
\node[vertex] (f) at (2,1) {6};
\node[vertex] (g) at (0,0) {7};
\node[vertex] (h) at (1,0) {8};
\node[vertex] (i) at (2,0) {9};
% bonds
\draw[strong] (a)--(b); %12
\draw[weak] (a)--(d) node[midway,left] {1.9}; %14
\draw[strong] (b)--(c); %23
\draw[strong] (b)--(e); %25
\draw[weak] (c)--(f) node[midway, right] {1.5}; %36
\draw[strong] (d)--(e); %45
\draw[weak] (d)--(g) node[midway,left] {1.9}; %47
\draw[weak] (e)--(f) node[midway,above] {0.7}; %56
\draw[strong] (e)--(h); %58
\draw[strong] (f)--(i); %69
\draw[strong] (g)--(h); %78
\draw[strong] (h)--(i); %89
\end{scope}
\end{tikzpicture}

% Notes
% \begin{itemize}[nosep]
% \item 2=3, 7=8
% \item Run 6 has lower yield than the others. This corresponds to a ``blocked'' tree. The yield is lower even though the weak bonds are 0. 
% \item Runs 9,10,11,12, which have off-pathway clusters with 2 weak bonds, have slightly lower yield than runs 7,8. The weak bonds in the former are weaker than in the latter.
% \end{itemize}

% !TEX root = main.tex
\subsubsection{Example E2}%, Standard model}

\begin{tabular}{c | l | l | c c c c c c c c c c c c | }
run & time & yield & $\delta_{12}$&$\delta_{13}$&$\delta_{14}$&$\delta_{15}$&$\delta_{16}$&$\delta_{17}$&$\delta_{23}$&$\delta_{27}$&$\delta_{34}$&$\delta_{45}$&$\delta_{56}$&$\delta_{67}$   \\\hline
1 & $10^3$& 0.4544&15.0&15.0&15.0&15.0&0.0&15.0&0.0&0.0&0.0&0.0&0.0&15.0\\
2 & $10^3$& 0.4543&0.0&15.0&15.0&0.0&15.0&15.0&0.0&15.0&0.0&0.0&15.0&0.0\\
3& $10^3$&0.4542&0.0&0.0&15.0&15.0&15.0&15.0&15.0&0.0&15.0&0.0&0.0&0.0\\
4 & $10^3$& 0.4541&15.0&0.0&15.0&15.0&0.0&0.0&0.0&15.0&15.0&0.0&15.0&0.0\\
5 & $10^3$&0.4541&0.0&0.0&15.0&0.0&15.0&15.0&0.0&15.0&15.0&15.0&0.0&0.0 \\
6 & $10^3$& 0.4538&0.0&15.0&15.0&0.0&0.0&0.0&15.0&15.0&0.0&15.0&15.0&0.0\\
\hline
7& $10^4$& 0.9194&1.4&15.0&15.0&15.0&15.0&15.0&1.1&15.0&2.0&1.7&1.9&1.7\\
8 & $10^4$& 0.9194&15.0&15.0&15.0&15.0&1.4&15.0&1.7&2.0&1.9&1.6&15.0&1.1\\
9& $10^4$&0.9193&15.0&15.0&15.0&15.0&2.4&2.4&1.6&15.0&1.9&1.7&15.0&0.04 \\
10 &$10^4$&0.9193&1.1&15.0&1.1&15.0&15.0&15.0&1.4&15.0&1.4&15.0&1.7&1.7 \\
11 & $10^4$&0.9191&15.0&0.5&0.9&15.0&1.0&15.0&15.0&1.4&15.0&1.2&1.5&15.0 \\
12 & $10^4$&0.9190&15.0&0.9&0.3&0.8&1.1&0.8&15.0&15.0&15.0&15.0&1.2&15.0 \\
\hline
\end{tabular}\bigskip

% Notes
% \begin{itemize}[nosep]
% \item 7,8 have same strong-weak topology. Not exactly the same values for the weak bonds, but close. Both have the highest yield (but differ from others only at 4th decimal point).
% \item 6 has lowest yield, and it is a stringy structure. This is even when the weak bonds are 0.
% \end{itemize}

\begin{tikzpicture}
\begin{scope}[scale=1.3,shift={(0,0)}]
% item 1
%\node at (-1,1.4) {(1)};
\node[align=center] at (0,1.4) {(1) $T{=}10^3$, 45.44\%};
% vertex coordinates
 \node[vertex] (a) at (0,0) {1};
 \node[vertex] (b) at (0.5,0.8660) {2};
\node[vertex] (c) at (1,0) {3};
\node[vertex] (d) at (0.5,-0.8660) {4};
\node[vertex] (e) at (-0.5,-0.8660) {5};
\node[vertex] (f) at (-1,0) {6};
\node[vertex] (g) at (-0.5,0.8660) {7};
% bonds
\draw[strong] (a)--(b); %12
\draw[strong] (a)--(c); %13
\draw[strong] (a)--(d); %14
\draw[strong] (a)--(e); %15
\draw[zero] (a)--(f); %16
\draw[strong] (a)--(g); %17
\draw[zero] (b)--(c); %23
\draw[zero] (b)--(g); %27
\draw[zero] (c)--(d); %34
\draw[zero] (d)--(e); %45
\draw[zero] (e)--(f); %56
\draw[strong] (f)--(g); %67
\end{scope}
\begin{scope}[scale=1.3,shift={(3,0)}]
% item 2
%\node at (-1,1.4) {(2)};
\node[align=center] at (0,1.4) {(2) $T{=}10^3$, 45.43\%};
% vertex coordinates
 \node[vertex] (a) at (0,0) {1};
 \node[vertex] (b) at (0.5,0.8660) {2};
\node[vertex] (c) at (1,0) {3};
\node[vertex] (d) at (0.5,-0.8660) {4};
\node[vertex] (e) at (-0.5,-0.8660) {5};
\node[vertex] (f) at (-1,0) {6};
\node[vertex] (g) at (-0.5,0.8660) {7};
% bonds
\draw[zero] (a)--(b); %12
\draw[strong] (a)--(c); %13
\draw[strong] (a)--(d); %14
\draw[zero] (a)--(e); %15
\draw[strong] (a)--(f); %16
\draw[strong] (a)--(g); %17
\draw[zero] (b)--(c); %23
\draw[strong] (b)--(g); %27
\draw[zero] (c)--(d); %34
\draw[zero] (d)--(e); %45
\draw[strong] (e)--(f); %56
\draw[zero] (f)--(g); %67
\end{scope}
\begin{scope}[scale=1.3,shift={(6,0)}]
% item 3
%\node at (-1,1.4) {(3)};
\node[align=center] at (0,1.4) {(3) $T{=}10^3$, 45.42\%};
% vertex coordinates
 \node[vertex] (a) at (0,0) {1};
 \node[vertex] (b) at (0.5,0.8660) {2};
\node[vertex] (c) at (1,0) {3};
\node[vertex] (d) at (0.5,-0.8660) {4};
\node[vertex] (e) at (-0.5,-0.8660) {5};
\node[vertex] (f) at (-1,0) {6};
\node[vertex] (g) at (-0.5,0.8660) {7};
% bonds
\draw[zero] (a)--(b); %12
\draw[zero] (a)--(c); %13
\draw[strong] (a)--(d); %14
\draw[strong] (a)--(e); %15
\draw[strong] (a)--(f); %16
\draw[strong] (a)--(g); %17
\draw[strong] (b)--(c); %23
\draw[zero] (b)--(g); %27
\draw[strong] (c)--(d); %34
\draw[zero] (d)--(e); %45
\draw[zero] (e)--(f); %56
\draw[zero] (f)--(g); %67
\end{scope}
\begin{scope}[scale=1.3,shift={(9,0)}]
% item 4
%\node at (-1,1.4) {(4)};
\node[align=center] at (0,1.4) {(4) $T{=}10^3$, 45.41\%};
% vertex coordinates
 \node[vertex] (a) at (0,0) {1};
 \node[vertex] (b) at (0.5,0.8660) {2};
\node[vertex] (c) at (1,0) {3};
\node[vertex] (d) at (0.5,-0.8660) {4};
\node[vertex] (e) at (-0.5,-0.8660) {5};
\node[vertex] (f) at (-1,0) {6};
\node[vertex] (g) at (-0.5,0.8660) {7};
% bonds
\draw[strong] (a)--(b); %12
\draw[zero] (a)--(c); %13
\draw[strong] (a)--(d); %14
\draw[strong] (a)--(e); %15
\draw[zero] (a)--(f); %16
\draw[zero] (a)--(g); %17
\draw[zero] (b)--(c); %23
\draw[strong] (b)--(g); %27
\draw[strong] (c)--(d); %34
\draw[zero] (d)--(e); %45
\draw[strong] (e)--(f); %56
\draw[zero] (f)--(g); %67
\end{scope}
\begin{scope}[scale=1.3,shift={(0,-3)}]
% item 5
%\node at (-1,1.4) {(5)};
\node[align=center] at (0,1.4) {(5) $T{=}10^3$, 45.41\%};
% vertex coordinates
 \node[vertex] (a) at (0,0) {1};
 \node[vertex] (b) at (0.5,0.8660) {2};
\node[vertex] (c) at (1,0) {3};
\node[vertex] (d) at (0.5,-0.8660) {4};
\node[vertex] (e) at (-0.5,-0.8660) {5};
\node[vertex] (f) at (-1,0) {6};
\node[vertex] (g) at (-0.5,0.8660) {7};
% bonds
\draw[zero] (a)--(b); %12
\draw[zero] (a)--(c); %13
\draw[strong] (a)--(d); %14
\draw[zero] (a)--(e); %15
\draw[strong] (a)--(f); %16
\draw[strong] (a)--(g); %17
\draw[zero] (b)--(c); %23
\draw[strong] (b)--(g); %27
\draw[strong] (c)--(d); %34
\draw[strong] (d)--(e); %45
\draw[zero] (e)--(f); %56
\draw[zero] (f)--(g); %67
\end{scope}
\begin{scope}[scale=1.3,shift={(3,-3)}]
% item 6
%\node at (-1,1.4) {(6)};
\node[align=center] at (0,1.4) {(6) $T{=}10^3$, 45.38\%};
% vertex coordinates
 \node[vertex] (a) at (0,0) {1};
 \node[vertex] (b) at (0.5,0.8660) {2};
\node[vertex] (c) at (1,0) {3};
\node[vertex] (d) at (0.5,-0.8660) {4};
\node[vertex] (e) at (-0.5,-0.8660) {5};
\node[vertex] (f) at (-1,0) {6};
\node[vertex] (g) at (-0.5,0.8660) {7};
% bonds
\draw[zero] (a)--(b); %12
\draw[strong] (a)--(c); %13
\draw[strong] (a)--(d); %14
\draw[zero] (a)--(e); %15
\draw[zero] (a)--(f); %16
\draw[zero] (a)--(g); %17
\draw[strong] (b)--(c); %23
\draw[strong] (b)--(g); %27
\draw[zero] (c)--(d); %34
\draw[strong] (d)--(e); %45
\draw[strong] (e)--(f); %56
\draw[zero] (f)--(g); %67
\end{scope}

\begin{scope}[scale=1.3,shift={(6,-3)}]
% item 7
%\node at (-1,1.4) {(7)};
\node[align=center] at (0,1.4) {(7) $T{=}10^4$, 91.94\%};
% vertex coordinates
 \node[vertex] (a) at (0,0) {1};
 \node[vertex] (b) at (0.5,0.8660) {2};
\node[vertex] (c) at (1,0) {3};
\node[vertex] (d) at (0.5,-0.8660) {4};
\node[vertex] (e) at (-0.5,-0.8660) {5};
\node[vertex] (f) at (-1,0) {6};
\node[vertex] (g) at (-0.5,0.8660) {7};
% bonds
\draw[weak] (a)--(b) node[pos=0.7, left] {1.4}; %12
\draw[strong] (a)--(c); %13
\draw[strong] (a)--(d); %14
\draw[strong] (a)--(e); %15
\draw[strong] (a)--(f); %16
\draw[strong] (a)--(g); %17
\draw[weak] (b)--(c) node[midway,right] {1.1}; %23
\draw[strong] (b)--(g); %27
\draw[weak] (c)--(d) node[midway,right] {2.0}; %34
\draw[weak] (d)--(e) node[midway,below] {1.7}; %45
\draw[weak] (e)--(f) node[midway,left] {1.9}; %56
\draw[weak] (f)--(g) node[midway,left] {1.7}; %67
\end{scope}
\begin{scope}[scale=1.3,shift={(9,-3)}]
% item 8
%\node at (-1,1.4) {(8)};
\node[align=center] at (0,1.4) {(8) $T{=}10^4$, 91.94\%};
% vertex coordinates
 \node[vertex] (a) at (0,0) {1};
 \node[vertex] (b) at (0.5,0.8660) {2};
\node[vertex] (c) at (1,0) {3};
\node[vertex] (d) at (0.5,-0.8660) {4};
\node[vertex] (e) at (-0.5,-0.8660) {5};
\node[vertex] (f) at (-1,0) {6};
\node[vertex] (g) at (-0.5,0.8660) {7};
% bonds
\draw[strong] (a)--(b); %12
\draw[strong] (a)--(c); %13
\draw[weak] (a)--(d) node[pos=0.3,right] {1.4}; %14
\draw[strong] (a)--(e); %15
\draw[strong] (a)--(f); %16
\draw[strong] (a)--(g); %17
\draw[weak] (b)--(c) node[midway,right] {1.7}; %23
\draw[weak] (b)--(g) node[midway,above] {1.4}; %27
\draw[strong] (c)--(d); %34
\draw[weak] (d)--(e) node[midway,below] {1.2}; %45
\draw[weak] (e)--(f) node[midway,left] {1.5}; %56
\draw[weak] (f)--(g) node[pos=0.7,left] {1.1}; %67
\end{scope}
\begin{scope}[scale=1.3,shift={(0,-6)}]
% item 9
%\node at (-1,1.4) {(9)};
\node[align=center] at (0,1.4) {(9) $T{=}10^4$, 91.93\%};
% vertex coordinates
 \node[vertex] (a) at (0,0) {1};
 \node[vertex] (b) at (0.5,0.8660) {2};
\node[vertex] (c) at (1,0) {3};
\node[vertex] (d) at (0.5,-0.8660) {4};
\node[vertex] (e) at (-0.5,-0.8660) {5};
\node[vertex] (f) at (-1,0) {6};
\node[vertex] (g) at (-0.5,0.8660) {7};
% bonds
\draw[strong] (a)--(b); %12
\draw[strong] (a)--(c); %13
\draw[strong] (a)--(d); %14
\draw[strong] (a)--(e); %15
\draw[weak] (a)--(f) node[midway,below] {2.4}; %16
\draw[weak] (a)--(g) node[pos=0.7,right] {2.4}; %17
\draw[weak] (b)--(c) node[midway,right] {1.6}; %23
\draw[strong] (b)--(g); %27
\draw[weak] (c)--(d) node[midway,right] {1.9}; %34
\draw[weak] (d)--(e) node[midway,below] {1.7}; %45
\draw[strong] (e)--(f); %56
\draw[weak] (f)--(g) node[midway,left] {0.04}; %67
\end{scope}
\begin{scope}[scale=1.3,shift={(3,-6)}]
% item 10
%\node at (-1,1.4) {(10)};
\node[align=center] at (0,1.4) {(10) $T{=}10^4$, 91.93\%};
% vertex coordinates
 \node[vertex] (a) at (0,0) {1};
 \node[vertex] (b) at (0.5,0.8660) {2};
\node[vertex] (c) at (1,0) {3};
\node[vertex] (d) at (0.5,-0.8660) {4};
\node[vertex] (e) at (-0.5,-0.8660) {5};
\node[vertex] (f) at (-1,0) {6};
\node[vertex] (g) at (-0.5,0.8660) {7};
% bonds
\draw[weak] (a)--(b) node[pos=0.7,left] {1.1}; %12
\draw[strong] (a)--(c); %13
\draw[weak] (a)--(d) node[pos=0.7,left] {1.1}; %14
\draw[strong] (a)--(e); %15
\draw[strong] (a)--(f); %16
\draw[strong] (a)--(g); %17
\draw[weak] (b)--(c) node[midway,right] {1.4}; %23
\draw[strong] (b)--(g); %27
\draw[weak] (c)--(d) node[midway,right] {1.4}; %34
\draw[strong] (d)--(e); %45
\draw[weak] (e)--(f) node[midway,left] {1.7}; %56
\draw[weak] (f)--(g) node[midway,left] {1.7}; %67
\end{scope}

\begin{scope}[scale=1.3,shift={(6,-6)}]
% item 11
%\node at (-1,1.4) {(11)};
\node[align=center] at (0,1.4) {(11) $T{=}10^4$, 91.91\%};
% vertex coordinates
 \node[vertex] (a) at (0,0) {1};
 \node[vertex] (b) at (0.5,0.8660) {2};
\node[vertex] (c) at (1,0) {3};
\node[vertex] (d) at (0.5,-0.8660) {4};
\node[vertex] (e) at (-0.5,-0.8660) {5};
\node[vertex] (f) at (-1,0) {6};
\node[vertex] (g) at (-0.5,0.8660) {7};
% bonds
\draw[strong] (a)--(b); %12
\draw[weak] (a)--(c) node[midway,above] {0.5}; %13
\draw[weak] (a)--(d) node[pos=0.3,right] {0.9}; %14
\draw[strong] (a)--(e); %15
\draw[weak] (a)--(f) node[midway,above] {1.0}; %16
\draw[strong] (a)--(g); %17
\draw[strong] (b)--(c); %23
\draw[weak] (b)--(g) node[midway,above] {1.4}; %27
\draw[strong] (c)--(d); %34
\draw[weak] (d)--(e) node[midway,above] {1.2}; %45
\draw[weak] (e)--(f) node[midway,left] {1.5}; %56
\draw[strong] (f)--(g); %67
\end{scope}
\begin{scope}[scale=1.3,shift={(9,-6)}]
% item 12
%\node at (-1,1.4) {(12)};
\node[align=center] at (0,1.4) {(12) $T{=}10^4$, 91.90\%};
% vertex coordinates
 \node[vertex] (a) at (0,0) {1};
 \node[vertex] (b) at (0.5,0.8660) {2};
\node[vertex] (c) at (1,0) {3};
\node[vertex] (d) at (0.5,-0.8660) {4};
\node[vertex] (e) at (-0.5,-0.8660) {5};
\node[vertex] (f) at (-1,0) {6};
\node[vertex] (g) at (-0.5,0.8660) {7};
% bonds
\draw[strong] (a)--(b); %12
\draw[weak] (a)--(c) node[midway,above] {0.9}; %13
\draw[weak] (a)--(d) node[pos=0.3,right] {0.3}; %14
\draw[weak] (a)--(e) node[pos=0.7,right] {0.8}; %15
\draw[weak] (a)--(f) node[midway,above] {1.1}; %16
\draw[weak] (a)--(g) node[pos=0.7,right] {0.8}; %17
\draw[strong] (b)--(c); %23
\draw[strong] (b)--(g); %27
\draw[strong] (c)--(d); %34
\draw[strong] (d)--(e); %45
\draw[weak] (e)--(f) node[midway,left] {1.2}; %56
\draw[strong] (f)--(g); %67
\end{scope}

\end{tikzpicture}

% !TEX root = main.tex
\subsubsection{Example E3}%, Standard model}

\begin{tabular}{l | l | l | c c c c c c c c c c c c |}
\# & time & yield & $\delta_{12}$&$\delta_{13}$&$\delta_{15}$&$\delta_{24}$&$\delta_{26}$&$\delta_{34}$&$ \delta_{37}$&$\delta_{48}$&$\delta_{56}$&$\delta_{57}$&$\delta_{68}$&$\delta_{78}$  \\\hline
1 & $10^3$ & 0.3521& 15.0& 15.0& 15.0& 0.0& 15.0& 0.0& 0.0& 15.0& 0.0& 0.0& 15.0& 15.0  \\
2 & $10^3$ &  0.3521&15.0&0.0&0.0&0.0&15.0&15.0&15.0&0.0&0.0&15.0&15.0&15.0 \\
3 & $10^3$ & 0.3521& 15.0& 15.0& 15.0& 0.0& 0.0& 15.0& 15.0& 0.0& 15.0& 0.0& 15.0& 0.0  \\
4 & $10^3$ & 0.3521& 15.0& 15.0& 15.0& 15.0& 15.0& 0.0& 0.0& 15.0& 0.0& 0.0& 0.0& 15.0  \\
5 & $10^3$ & 0.3520& 0.0& 15.0& 15.0& 15.0& 0.0& 15.0& 15.0& 15.0& 0.0& 0.0& 15.0& 0.0  \\
6 & $10^3$ & 0.3520& 15.0& 0.0& 15.0& 15.0& 15.0& 15.0& 0.0& 15.0& 0.0& 0.0& 0.0& 15.0  \\
\hline
7 & $10^4$ & 0.8938& 1.7& 15.0& 1.5& 15.0& 0.6& 15.0& 15.0& 1.0& 1.7& 15.0& 15.0& 15.0  \\
8 & $10^4$ & 0.8938& 1.7& 0.5& 15.0& 1.3& 15.0& 15.0& 1.7& 15.0& 15.0& 1.3& 15.0& 15.0  \\
9 & $10^4$  &0.8936&15.0&15.0&15.0&15.0&1.1&0.6&15.0&15.0&15.0&1.5&0.9&0.9  \\
10 & $10^4$ &0.8920&1.2&15.0&15.0&15.0&15.0&2.9&15.0&2.7&3.5&0.0&2.3&15.0\\
11 & $10^4$ & 0.8909&1.8&15.0&15.0&4.9&15.0&4.5&0.0&15.0&1.8&15.0&4.9&4.5  \\
12 &$10^4$ & 0.8907&  15.0& 4.7& 5.0& 5.1& 4.8& 15.0& 15.0& 0.4& 15.0& 1.4& 2.0& 15.0 \\\hline
\end{tabular}\bigskip

% Notes
% \begin{itemize}[nosep]
% \item 7,8 are similar, and have highest yield. Not quite topologically the same, but almost.
% \item 10,11,12 are not connected by strong bonds. Weak bonds are fairly strong.
% \end{itemize}

\begin{tikzpicture}
% item 1
\begin{scope}[scale=0.8, shift={(0,0)}]
%\node at (-1,3.2) {(1)};
\node[align=center] at (0.5,3.5) {(1) $T{=}10^3$, 35.21\%};
 %vertex coordinates
\node[vertex] (a) at (-1,2.5) {1};
\node[vertex] (b) at (1,2.5) {2};
\node[vertex] (c) at (0,2) {3};
\node[vertex] (d) at (2,2) {4};
\node[vertex] (e) at (-1,0.5) {5};
\node[vertex] (f) at (1,0.5) {6};
\node[vertex] (g) at (0,0) {7};
\node[vertex] (h) at (2,0) {8};
%bonds
\draw[strong] (a)--(b);  % 12
\draw[strong] (a)--(c);  % 13
\draw[strong] (a)--(e);  % 15
\draw[zero] (b)--(d);  % 24
\draw[strong] (b)--(f);  % 26
\draw[zero] (c)--(d);  % 34
\draw[zero] (c)--(g);  % 37
\draw[strong] (d)--(h);  % 48
\draw[zero] (e)--(f);  % 56
\draw[zero] (e)--(g);  % 57
\draw[strong] (f)--(h);  % 68
\draw[strong] (g)--(h);  % 78
\end{scope}
% item 2
\begin{scope}[scale=0.8, shift={(5,0)}]
%\node at (-1,3.2) {(2)};
\node[align=center] at (0.5,3.5) {(2) $T{=}10^3$, 35.21\%};
 %vertex coordinates
\node[vertex] (a) at (-1,2.5) {1};
\node[vertex] (b) at (1,2.5) {2};
\node[vertex] (c) at (0,2) {3};
\node[vertex] (d) at (2,2) {4};
\node[vertex] (e) at (-1,0.5) {5};
\node[vertex] (f) at (1,0.5) {6};
\node[vertex] (g) at (0,0) {7};
\node[vertex] (h) at (2,0) {8};
%bonds
\draw[strong] (a)--(b);  % 12
\draw[zero] (a)--(c);  % 13
\draw[zero] (a)--(e);  % 15
\draw[zero] (b)--(d);  % 24
\draw[strong] (b)--(f);  % 26
\draw[strong] (c)--(d);  % 34
\draw[strong] (c)--(g);  % 37
\draw[zero] (d)--(h);  % 48
\draw[zero] (e)--(f);  % 56
\draw[strong] (e)--(g);  % 57
\draw[strong] (f)--(h);  % 68
\draw[strong] (g)--(h);  % 78
\end{scope}
% item 3
\begin{scope}[scale=0.8,shift={(10,0)}]
%\node at (-1,3.2) {(3)};
\node[align=center] at (0.5,3.5) {(3) $T{=}10^3$, 35.21\%};
 %vertex coordinates
\node[vertex] (a) at (-1,2.5) {1};
\node[vertex] (b) at (1,2.5) {2};
\node[vertex] (c) at (0,2) {3};
\node[vertex] (d) at (2,2) {4};
\node[vertex] (e) at (-1,0.5) {5};
\node[vertex] (f) at (1,0.5) {6};
\node[vertex] (g) at (0,0) {7};
\node[vertex] (h) at (2,0) {8};
%bonds
\draw[strong] (a)--(b);  % 12
\draw[strong] (a)--(c);  % 13
\draw[strong] (a)--(e);  % 15
\draw[zero] (b)--(d);  % 24
\draw[zero] (b)--(f);  % 26
\draw[strong] (c)--(d);  % 34
\draw[strong] (c)--(g);  % 37
\draw[zero] (d)--(h);  % 48
\draw[strong] (e)--(f);  % 56
\draw[zero] (e)--(g);  % 57
\draw[strong] (f)--(h);  % 68
\draw[zero] (g)--(h);  % 78
\end{scope}
% item 4
\begin{scope}[scale=0.8, shift={(15,0)}]
%\node at (-1,3.2) {(4)};
\node[align=center] at (0.5,3.5) {(4) $T{=}10^3$, 35.21\%};
 %vertex coordinates
\node[vertex] (a) at (-1,2.5) {1};
\node[vertex] (b) at (1,2.5) {2};
\node[vertex] (c) at (0,2) {3};
\node[vertex] (d) at (2,2) {4};
\node[vertex] (e) at (-1,0.5) {5};
\node[vertex] (f) at (1,0.5) {6};
\node[vertex] (g) at (0,0) {7};
\node[vertex] (h) at (2,0) {8};
%bonds
\draw[strong] (a)--(b);  % 12
\draw[strong] (a)--(c);  % 13
\draw[strong] (a)--(e);  % 15
\draw[strong] (b)--(d);  % 24
\draw[strong] (b)--(f);  % 26
\draw[zero] (c)--(d);  % 34
\draw[zero] (c)--(g);  % 37
\draw[strong] (d)--(h);  % 48
\draw[zero] (e)--(f);  % 56
\draw[zero] (e)--(g);  % 57
\draw[zero] (f)--(h);  % 68
\draw[strong] (g)--(h);  % 78
\end{scope}
% item 5
\begin{scope}[scale=0.8, shift={(0,-5)}]
%\node at (-1,3.2) {(5)};
\node[align=center] at (0.5,3.5) {(5) $T{=}10^3$, 35.20\%};
 %vertex coordinates
\node[vertex] (a) at (-1,2.5) {1};
\node[vertex] (b) at (1,2.5) {2};
\node[vertex] (c) at (0,2) {3};
\node[vertex] (d) at (2,2) {4};
\node[vertex] (e) at (-1,0.5) {5};
\node[vertex] (f) at (1,0.5) {6};
\node[vertex] (g) at (0,0) {7};
\node[vertex] (h) at (2,0) {8};
%bonds
\draw[zero] (a)--(b);  % 12
\draw[strong] (a)--(c);  % 13
\draw[strong] (a)--(e);  % 15
\draw[strong] (b)--(d);  % 24
\draw[zero] (b)--(f);  % 26
\draw[strong] (c)--(d);  % 34
\draw[strong] (c)--(g);  % 37
\draw[strong] (d)--(h);  % 48
\draw[zero] (e)--(f);  % 56
\draw[zero] (e)--(g);  % 57
\draw[strong] (f)--(h);  % 68
\draw[zero] (g)--(h);  % 78
\end{scope}
% item 6
\begin{scope}[scale=0.8, shift={(5,-5)}]
%\node at (-1,3.2) {(6)};
\node[align=center] at (0.5,3.5) {(6) $T{=}10^3$, 35.20\%};
 %vertex coordinates
\node[vertex] (a) at (-1,2.5) {1};
\node[vertex] (b) at (1,2.5) {2};
\node[vertex] (c) at (0,2) {3};
\node[vertex] (d) at (2,2) {4};
\node[vertex] (e) at (-1,0.5) {5};
\node[vertex] (f) at (1,0.5) {6};
\node[vertex] (g) at (0,0) {7};
\node[vertex] (h) at (2,0) {8};
%bonds
\draw[strong] (a)--(b);  % 12
\draw[zero] (a)--(c);  % 13
\draw[strong] (a)--(e);  % 15
\draw[strong] (b)--(d);  % 24
\draw[strong] (b)--(f);  % 26
\draw[strong] (c)--(d);  % 34
\draw[zero] (c)--(g);  % 37
\draw[strong] (d)--(h);  % 48
\draw[zero] (e)--(f);  % 56
\draw[zero] (e)--(g);  % 57
\draw[zero] (f)--(h);  % 68
\draw[strong] (g)--(h);  % 78
\end{scope}

% item 7
\begin{scope}[scale=0.8, shift={(10,-5)}]
%\node at (-1,3.2) {(7)};
\node[align=center] at (0.5,3.5) {(7) $T{=}10^4$, 89.38\%};
%vertex coordinates
\node[vertex] (a) at (-1,2.5) {1};
\node[vertex] (b) at (1,2.5) {2};
\node[vertex] (c) at (0,2) {3};
\node[vertex] (d) at (2,2) {4};
\node[vertex] (e) at (-1,0.5) {5};
\node[vertex] (f) at (1,0.5) {6};
\node[vertex] (g) at (0,0) {7};
\node[vertex] (h) at (2,0) {8};
%bonds
\draw[weak] (a)--(b) node[midway,above] {1.7};  % 12
\draw[strong] (a)--(c);  % 13
\draw[weak] (a)--(e) node[midway,left] {1.5};  % 15
\draw[strong] (b)--(d);  % 24
\draw[weak] (b)--(f) node[midway,right] {0.6};  % 26
\draw[strong] (c)--(d);  % 34
\draw[strong] (c)--(g);  % 37
\draw[weak] (d)--(h) node[midway,right] {1.0};  % 48
\draw[weak] (e)--(f) node[midway,above right] {1.7};  % 56
\draw[strong] (e)--(g);  % 57
\draw[strong] (f)--(h);  % 68
\draw[strong] (g)--(h);  % 78
\end{scope}
% item 8
\begin{scope}[scale=0.8, shift={(15,-5)}]
%\node at (-1,3.2) {(8)};
\node[align=center] at (0.5,3.5) {(8) $T{=}10^4$, 89.38\%};
 %vertex coordinates
\node[vertex] (a) at (-1,2.5) {1};
\node[vertex] (b) at (1,2.5) {2};
\node[vertex] (c) at (0,2) {3};
\node[vertex] (d) at (2,2) {4};
\node[vertex] (e) at (-1,0.5) {5};
\node[vertex] (f) at (1,0.5) {6};
\node[vertex] (g) at (0,0) {7};
\node[vertex] (h) at (2,0) {8};
%bonds
\draw[weak] (a)--(b) node[midway,above] {1.7};  % 12
\draw[weak] (a)--(c) node[pos=0.3,below] {0.5};  % 13
\draw[strong] (a)--(e);  % 15
\draw[weak] (b)--(d) node[midway,above right] {1.3};  % 24
\draw[strong] (b)--(f);  % 26
\draw[strong] (c)--(d);  % 34
\draw[weak] (c)--(g) node[midway,right] {1.7};  % 37
\draw[strong] (d)--(h);  % 48
\draw[strong] (e)--(f);  % 56
\draw[weak] (e)--(g) node[midway,below left] {1.3};  % 57
\draw[strong] (f)--(h);  % 68
\draw[strong] (g)--(h);  % 78
\end{scope}
% item 9
\begin{scope}[scale=0.8, shift={(0,-10)}]
%\node at (-1,3.2) {(9)};
\node[align=center] at (0.5,3.5) {(9) $T{=}10^4$, 89.36\%};
 %vertex coordinates
\node[vertex] (a) at (-1,2.5) {1};
\node[vertex] (b) at (1,2.5) {2};
\node[vertex] (c) at (0,2) {3};
\node[vertex] (d) at (2,2) {4};
\node[vertex] (e) at (-1,0.5) {5};
\node[vertex] (f) at (1,0.5) {6};
\node[vertex] (g) at (0,0) {7};
\node[vertex] (h) at (2,0) {8};
%bonds
\draw[strong] (a)--(b);  % 12
\draw[strong] (a)--(c);  % 13
\draw[strong] (a)--(e);  % 15
\draw[strong] (b)--(d);  % 24
\draw[weak] (b)--(f) node[midway,right] {1.1};  % 26
\draw[weak] (c)--(d) node[pos=0.25,below] {0.6};  % 34
\draw[strong] (c)--(g);  % 37
\draw[strong] (d)--(h);  % 48
\draw[strong] (e)--(f);  % 56
\draw[weak] (e)--(g) node[pos=0.2,below] {1.5};  % 57
\draw[weak] (f)--(h) node[pos=0.7,above] {0.9};  % 68
\draw[weak] (g)--(h) node[midway,below] {0.9};  % 78
\end{scope}
% item 10
\begin{scope}[scale=0.8, shift={(5,-10)}]
%\node at (-1,3.2) {(10)};
\node[align=center] at (0.5,3.5) {(10) $T{=}4{\cdot}10^4$, 89.20\%};
 %vertex coordinates
\node[vertex] (a) at (-1,2.5) {1};
\node[vertex] (b) at (1,2.5) {2};
\node[vertex] (c) at (0,2) {3};
\node[vertex] (d) at (2,2) {4};
\node[vertex] (e) at (-1,0.5) {5};
\node[vertex] (f) at (1,0.5) {6};
\node[vertex] (g) at (0,0) {7};
\node[vertex] (h) at (2,0) {8};
%bonds
\draw[weak] (a)--(b) node[midway,above] {1.2};  % 12
\draw[strong] (a)--(c);  % 13
\draw[strong] (a)--(e);  % 15
\draw[strong] (b)--(d);  % 24
\draw[strong] (b)--(f);  % 26
\draw[weak] (c)--(d) node[midway,below left] {2.9};  % 34
\draw[strong] (c)--(g);  % 37
\draw[weak] (d)--(h) node[midway,right] {2.7};  % 48
\draw[weak] (e)--(f) node[midway,above right] {3.5};  % 56
\draw[weak] (e)--(g) node[midway,below left] {0.0};  % 57
\draw[weak] (f)--(h) node[pos=0.7,above] {2.3};  % 68
\draw[strong] (g)--(h);  % 78
\end{scope}
% item 11
\begin{scope}[scale=0.8, shift={(10,-10)}]
%\node at (-1,3.2) {(11)};
\node[align=center] at (0.5,3.5) {(11) $T{=}10^4$, 89.09\%};
 %vertex coordinates
\node[vertex] (a) at (-1,2.5) {1};
\node[vertex] (b) at (1,2.5) {2};
\node[vertex] (c) at (0,2) {3};
\node[vertex] (d) at (2,2) {4};
\node[vertex] (e) at (-1,0.5) {5};
\node[vertex] (f) at (1,0.5) {6};
\node[vertex] (g) at (0,0) {7};
\node[vertex] (h) at (2,0) {8};
%bonds
\draw[weak] (a)--(b) node[midway,above] {1.8};  % 12
\draw[strong] (a)--(c);  % 13
\draw[strong] (a)--(e);  % 15
\draw[weak] (b)--(d) node[pos=0.7,above] {4.9};  % 24
\draw[strong] (b)--(f);  % 26
\draw[weak] (c)--(d) node[pos=0.25,below] {4.5};  % 34
\draw[weak] (c)--(g) node[midway,left] {0.0};  % 37
\draw[strong] (d)--(h);  % 48
\draw[weak] (e)--(f) node[pos=0.75,above] {1.8};  % 56
\draw[strong] (e)--(g);  % 57
\draw[weak] (f)--(h) node[pos=0.7,above] {4.9};  % 68
\draw[weak] (g)--(h) node[midway,below] {4.5};  % 78
\end{scope}
% item 12
\begin{scope}[scale=0.8, shift={(15,-10)}]
%\node at (-1,3.2) {(12)};
\node[align=center] at (0.5,3.5) {(12) $T{=}10^4$, 89.07\%};
 %vertex coordinates
\node[vertex] (a) at (-1,2.5) {1};
\node[vertex] (b) at (1,2.5) {2};
\node[vertex] (c) at (0,2) {3};
\node[vertex] (d) at (2,2) {4};
\node[vertex] (e) at (-1,0.5) {5};
\node[vertex] (f) at (1,0.5) {6};
\node[vertex] (g) at (0,0) {7};
\node[vertex] (h) at (2,0) {8};
%bonds
\draw[strong] (a)--(b);  % 12
\draw[weak] (a)--(c) node[pos=0.3,below] {4.7};  % 13
\draw[weak] (a)--(e) node[midway,left] {5.0};  % 15
\draw[weak] (b)--(d) node[pos=0.7,above] {5.1};  % 24
\draw[weak] (b)--(f) node[midway,right] {4.8};  % 26
\draw[strong] (c)--(d);  % 34
\draw[strong] (c)--(g);  % 37
\draw[weak] (d)--(h) node[midway,right] {0.4};  % 48
\draw[strong] (e)--(f);  % 56
\draw[weak] (e)--(g) node[pos=0.3,below] {1.4};  % 57
\draw[weak] (f)--(h) node[pos=0.7,above] {2.0};  % 68
\draw[strong] (g)--(h);  % 78
\end{scope}
\end{tikzpicture}

% !TEX root = main.tex
\subsubsection{Example E4}%, Standard model}

\addtolength{\tabcolsep}{-0.2em}
\begin{tabular}{c | l | l | c c c c c c c c c c c c c c c | }
run & time & yield & $\delta_{12}$&$\delta_{13}$&$\delta_{14}$&$\delta_{15}$&$\delta_{16}$&$\delta_{23}$&$\delta_{24}$&$ \delta_{25}$&$\delta_{26}$&$\delta_{34}$&$\delta_{35}$&$\delta_{36}$&$\delta_{45}$&$\delta_{46}$&$\delta_{56}$   \\\hline
1 & $10^3$& 0.5656& 0.0&15.0&0.0&15.0&0.0&15.0&0.0&0.0&0.0&15.0&0.0&0.0&0.0&0.0&15.0\\
2 & $10^3$& 0.5656& 15.0&0.0&15.0&0.0&0.0&0.0&0.0&0.0&0.0&15.0&15.0&0.0&0.0&15.0&0.0 \\
3& $10^3$& 0.5656& 15.0&15.0&0.0&0.0&0.0&0.0&15.0&0.0&15.0&0.0&0.0&0.0&15.0&0.0&0.0\\
4 & $10^3$& 0.5656&0.0&0.0&15.0&0.0&0.0&15.0&0.0&0.0&0.0&0.0&15.0&0.0&15.0&0.0&15.0 \\
5 & $10^3$& 0.5656& 0.0&15.0&0.0&0.0&0.0&15.0&0.0&0.0&0.0&15.0&0.0&15.0&15.0&0.0&0.0\\
6 & $10^3$& 0.5656& 0.0&0.0&15.0&0.0&0.0&0.0&15.0&0.0&0.0&0.0&0.0&15.0&15.0&15.0&0.0 \\\hline
7 & $10^4$& 0.9417&15.0&1.4&15.0&15.0&1.4&0.0&1.4&2.0&0.0&15.0&0.0&2.0&1.4&15.0&0.0\\
8 & $10^4$& 0.9417&15.0&2.0&1.4&0.0&0.0&15.0&15.0&1.4&1.4&1.4&0.0&0.0&15.0&15.0&2.0\\
9 & $10^4$& 0.9417&1.0&15.0&15.0&1.0&1.0&0.0&15.0&1.5&1.5&2.4&0.0&0.0&15.0&15.0&1.5 \\
10 &$10^4$& 0.9417&1.5&1.5&0.0&15.0&1.0&1.5&0.0&15.0&1.0&0.0&15.0&1.0&2.4&15.0&15.0\\
11& $10^4$& 0.9417& 0.0&0.0&2.4&0.0&15.0&1.5&15.0&1.5&1.0&15.0&1.5&1.0&15.0&15.0&1.0\\
12& $10^4$& 0.9416& 0.0&0.0&2.4&15.0&0.0&15.0&2.4&0.0&0.0&15.0&0.7&1.7&15.0&15.0&1.7 \\
\hline
\end{tabular}\bigskip

\begin{tikzpicture}
% item 1
\begin{scope}[scale=1.4,shift={(0,0)}]
 %\node at (-1,1) {(1)};
 \node[align=center] at (0.095,1.4) {(1) $T{=}10^3$, 56.56\%};
%vertex coordinates
\node[vertex] (a) at (0,0) {1};
\node[vertex] (b) at (0.31,0.95) {2};
\node[vertex] (c) at (1,0) {3};
\node[vertex] (d) at (0.31,-0.95) {4};
\node[vertex] (e) at (-0.81,0.59) {5};
\node[vertex] (f) at (-0.81,-0.59) {6};
% bonds
\draw[zero] (a)--(b); %12
\draw[strong] (a)--(c); %13
\draw[zero] (a)--(d); %14
\draw[strong] (a)--(e); %15
\draw[zero] (a)--(f);  %16
\draw[strong] (b)--(c); %23
\draw[zero] (b)--(d); %24
\draw[zero] (b)--(e); %25
\draw[zero] (b)--(f); %26
\draw[strong] (c)--(d); %34
\draw[zero] (c)--(e);  %35
\draw[zero] (c)--(f); %36
\draw[zero] (d)--(e); %45
\draw[zero] (d)--(f); %46
\draw[strong] (e)--(f); %56
\end{scope}
% item 2
\begin{scope}[scale=1.4,shift={(3,0)}]
 %\node at (-1,1) {(2)};
  \node[align=center] at (0.095,1.4) {(2) $T{=}10^3$, 56.56\%};
%vertex coordinates
\node[vertex] (a) at (0,0) {1};
\node[vertex] (b) at (0.31,0.95) {2};
\node[vertex] (c) at (1,0) {3};
\node[vertex] (d) at (0.31,-0.95) {4};
\node[vertex] (e) at (-0.81,0.59) {5};
\node[vertex] (f) at (-0.81,-0.59) {6};
% bonds
\draw[strong] (a)--(b); %12
\draw[zero] (a)--(c); %13
\draw[strong] (a)--(d); %14
\draw[zero] (a)--(e); %15
\draw[zero] (a)--(f);  %16
\draw[zero] (b)--(c); %23
\draw[zero] (b)--(d); %24
\draw[zero] (b)--(e); %25
\draw[zero] (b)--(f); %26
\draw[strong] (c)--(d); %34
\draw[strong] (c)--(e);  %35
\draw[zero] (c)--(f); %36
\draw[zero] (d)--(e); %45
\draw[strong] (d)--(f); %46
\draw[zero] (e)--(f); %56
\end{scope}
% item 3
\begin{scope}[scale=1.4,shift={(6,0)}]
 %\node at (-1,1) {(3)};
  \node[align=center] at (0.095,1.4) {(3) $T{=}10^3$, 56.56\%};
%vertex coordinates
\node[vertex] (a) at (0,0) {1};
\node[vertex] (b) at (0.31,0.95) {2};
\node[vertex] (c) at (1,0) {3};
\node[vertex] (d) at (0.31,-0.95) {4};
\node[vertex] (e) at (-0.81,0.59) {5};
\node[vertex] (f) at (-0.81,-0.59) {6};
% bonds
\draw[strong] (a)--(b); %12
\draw[strong] (a)--(c); %13
\draw[zero] (a)--(d); %14
\draw[zero] (a)--(e); %15
\draw[zero] (a)--(f);  %16
\draw[zero] (b)--(c); %23
\draw[strong] (b)--(d); %24
\draw[zero] (b)--(e); %25
\draw[strong] (b)--(f); %26
\draw[zero] (c)--(d); %34
\draw[zero] (c)--(e);  %35
\draw[zero] (c)--(f); %36
\draw[strong] (d)--(e); %45
\draw[zero] (d)--(f); %46
\draw[zero] (e)--(f); %56
\end{scope}
% item 4
\begin{scope}[scale=1.4,shift={(9,0)}]
 %\node at (-1,1) {(4)};
  \node[align=center] at (0.095,1.4) {(4) $T{=}10^3$, 56.56\%};
%vertex coordinates
\node[vertex] (a) at (0,0) {1};
\node[vertex] (b) at (0.31,0.95) {2};
\node[vertex] (c) at (1,0) {3};
\node[vertex] (d) at (0.31,-0.95) {4};
\node[vertex] (e) at (-0.81,0.59) {5};
\node[vertex] (f) at (-0.81,-0.59) {6};
% bonds
\draw[zero] (a)--(b); %12
\draw[zero] (a)--(c); %13
\draw[strong] (a)--(d); %14
\draw[zero] (a)--(e); %15
\draw[zero] (a)--(f);  %16
\draw[strong] (b)--(c); %23
\draw[zero] (b)--(d); %24
\draw[zero] (b)--(e); %25
\draw[zero] (b)--(f); %26
\draw[zero] (c)--(d); %34
\draw[strong] (c)--(e);  %35
\draw[zero] (c)--(f); %36
\draw[strong] (d)--(e); %45
\draw[zero] (d)--(f); %46
\draw[strong] (e)--(f); %56
\end{scope}
% item 5
\begin{scope}[scale=1.4,shift={(0,-3)}]
% \node at (-1,1) {(5)};
 \node[align=center] at (0.095,1.4) {(5) $T{=}10^3$, 56.56\%};
%vertex coordinates
\node[vertex] (a) at (0,0) {1};
\node[vertex] (b) at (0.31,0.95) {2};
\node[vertex] (c) at (1,0) {3};
\node[vertex] (d) at (0.31,-0.95) {4};
\node[vertex] (e) at (-0.81,0.59) {5};
\node[vertex] (f) at (-0.81,-0.59) {6};
% bonds
\draw[zero] (a)--(b); %12
\draw[strong] (a)--(c); %13
\draw[zero] (a)--(d); %14
\draw[zero] (a)--(e); %15
\draw[zero] (a)--(f);  %16
\draw[strong] (b)--(c); %23
\draw[zero] (b)--(d); %24
\draw[zero] (b)--(e); %25
\draw[zero] (b)--(f); %26
\draw[strong] (c)--(d); %34
\draw[zero] (c)--(e);  %35
\draw[strong] (c)--(f); %36
\draw[strong] (d)--(e); %45
\draw[zero] (d)--(f); %46
\draw[zero] (e)--(f); %56
\end{scope}
% item 6
\begin{scope}[scale=1.4,shift={(3,-3)}]
 %\node at (-1,1) {(6)};
  \node[align=center] at (0.095,1.4) {(6) $T{=}10^3$, 56.56\%};
%vertex coordinates
\node[vertex] (a) at (0,0) {1};
\node[vertex] (b) at (0.31,0.95) {2};
\node[vertex] (c) at (1,0) {3};
\node[vertex] (d) at (0.31,-0.95) {4};
\node[vertex] (e) at (-0.81,0.59) {5};
\node[vertex] (f) at (-0.81,-0.59) {6};
% bonds
\draw[zero] (a)--(b); %12
\draw[zero] (a)--(c); %13
\draw[strong] (a)--(d); %14
\draw[zero] (a)--(e); %15
\draw[zero] (a)--(f);  %16
\draw[zero] (b)--(c); %23
\draw[strong] (b)--(d); %24
\draw[zero] (b)--(e); %25
\draw[zero] (b)--(f); %26
\draw[zero] (c)--(d); %34
\draw[zero] (c)--(e);  %35
\draw[strong] (c)--(f); %36
\draw[strong] (d)--(e); %45
\draw[strong] (d)--(f); %46
\draw[zero] (e)--(f); %56
\end{scope}

% item 7
\begin{scope}[scale=1.4,shift={(6,-3)}]
 %\node at (-1,1) {(7)};
  \node[align=center] at (0.095,1.4) {(7) $T{=}10^4$, 94.17\%};
%vertex coordinates
\node[vertex] (a) at (0,0) {1};
\node[vertex] (b) at (0.31,0.95) {2};
\node[vertex] (c) at (1,0) {3};
\node[vertex] (d) at (0.31,-0.95) {4};
\node[vertex] (e) at (-0.81,0.59) {5};
\node[vertex] (f) at (-0.81,-0.59) {6};
% bonds
\draw[strong] (a)--(b); %12
\draw[weak] (a)--(c); %13
\draw[strong] (a)--(d); %14
\draw[strong] (a)--(e); %15
\draw[weak] (a)--(f);  %16
\draw[zero] (b)--(c); %23
\draw[weak] (b)--(d); %24
\draw[weak] (b)--(e); %25
\draw[zero] (b)--(f); %26
\draw[strong] (c)--(d); %34
\draw[zero] (c)--(e);  %35
\draw[weak] (c)--(f); %36
\draw[weak] (d)--(e); %45
\draw[strong] (d)--(f); %46
\draw[zero] (e)--(f); %56
\end{scope}
% item 8
\begin{scope}[scale=1.4,shift={(9,-3)}]
 %\node at (-1,1) {(8)};
 \node[align=center] at (0.095,1.4) {(8) $T{=}10^4$, 94.17\%};
%vertex coordinates
\node[vertex] (a) at (0,0) {1};
\node[vertex] (b) at (0.31,0.95) {2};
\node[vertex] (c) at (1,0) {3};
\node[vertex] (d) at (0.31,-0.95) {4};
\node[vertex] (e) at (-0.81,0.59) {5};
\node[vertex] (f) at (-0.81,-0.59) {6};
% bonds
\draw[strong] (a)--(b); %12
\draw[weak] (a)--(c); %13
\draw[weak] (a)--(d); %14
\draw[zero] (a)--(e); %15
\draw[zero] (a)--(f);  %16
\draw[strong] (b)--(c); %23
\draw[strong] (b)--(d); %24
\draw[weak] (b)--(e); %25
\draw[weak] (b)--(f); %26
\draw[weak] (c)--(d); %34
\draw[zero] (c)--(e);  %35
\draw[zero] (c)--(f); %36
\draw[strong] (d)--(e); %45
\draw[strong] (d)--(f); %46
\draw[weak] (e)--(f); %56
\end{scope}
% item 9
\begin{scope}[scale=1.4,shift={(0,-6)}]
 %\node at (-1,1) {(9)};
 \node[align=center] at (0.095,1.4) {(9) $T{=}10^4$, 94.17\%};
%vertex coordinates
\node[vertex] (a) at (0,0) {1};
\node[vertex] (b) at (0.31,0.95) {2};
\node[vertex] (c) at (1,0) {3};
\node[vertex] (d) at (0.31,-0.95) {4};
\node[vertex] (e) at (-0.81,0.59) {5};
\node[vertex] (f) at (-0.81,-0.59) {6};
% bonds
\draw[weak] (a)--(b); %12
\draw[strong] (a)--(c); %13
\draw[strong] (a)--(d); %14
\draw[weak] (a)--(e); %15
\draw[weak] (a)--(f);  %16
\draw[zero] (b)--(c); %23
\draw[strong] (b)--(d); %24
\draw[weak] (b)--(e); %25
\draw[weak] (b)--(f); %26
\draw[weak] (c)--(d); %34
\draw[zero] (c)--(e);  %35
\draw[zero] (c)--(f); %36
\draw[strong] (d)--(e); %45
\draw[strong] (d)--(f); %46
\draw[weak] (e)--(f); %56
\end{scope}
% item 10
\begin{scope}[scale=1.4,shift={(3,-6)}]
 %\node at (-1,1) {(10)};
 \node[align=center] at (0.095,1.4) {(10) $T{=}10^4$, 94.17\%};
%vertex coordinates
\node[vertex] (a) at (0,0) {1};
\node[vertex] (b) at (0.31,0.95) {2};
\node[vertex] (c) at (1,0) {3};
\node[vertex] (d) at (0.31,-0.95) {4};
\node[vertex] (e) at (-0.81,0.59) {5};
\node[vertex] (f) at (-0.81,-0.59) {6};
% bonds
\draw[weak] (a)--(b); %12
\draw[weak] (a)--(c); %13
\draw[zero] (a)--(d); %14
\draw[strong] (a)--(e); %15
\draw[weak] (a)--(f);  %16
\draw[weak] (b)--(c); %23
\draw[zero] (b)--(d); %24
\draw[strong] (b)--(e); %25
\draw[weak] (b)--(f); %26
\draw[zero] (c)--(d); %34
\draw[strong] (c)--(e);  %35
\draw[weak] (c)--(f); %36
\draw[weak] (d)--(e); %45
\draw[strong] (d)--(f); %46
\draw[strong] (e)--(f); %56
\end{scope}
% item 11
\begin{scope}[scale=1.4,shift={(6,-6)}]
% \node at (-1,1) {(11)};
\node[align=center] at (0.095,1.4) {(11) $T{=}10^4$, 94.17\%};
%vertex coordinates
\node[vertex] (a) at (0,0) {1};
\node[vertex] (b) at (0.31,0.95) {2};
\node[vertex] (c) at (1,0) {3};
\node[vertex] (d) at (0.31,-0.95) {4};
\node[vertex] (e) at (-0.81,0.59) {5};
\node[vertex] (f) at (-0.81,-0.59) {6};
% bonds
\draw[zero] (a)--(b); %12
\draw[zero] (a)--(c); %13
\draw[weak] (a)--(d); %14
\draw[zero] (a)--(e); %15
\draw[strong] (a)--(f);  %16
\draw[weak] (b)--(c); %23
\draw[strong] (b)--(d); %24
\draw[weak] (b)--(e); %25
\draw[weak] (b)--(f); %26
\draw[strong] (c)--(d); %34
\draw[weak] (c)--(e);  %35
\draw[weak] (c)--(f); %36
\draw[strong] (d)--(e); %45
\draw[strong] (d)--(f); %46
\draw[weak] (e)--(f); %56
\end{scope}
% item 12
\begin{scope}[scale=1.4,shift={(9,-6)}]
 %\node at (-1,1) {(12)};
 \node[align=center] at (0.095,1.4) {(12) $T{=}10^4$, 94.16\%};
%vertex coordinates
\node[vertex] (a) at (0,0) {1};
\node[vertex] (b) at (0.31,0.95) {2};
\node[vertex] (c) at (1,0) {3};
\node[vertex] (d) at (0.31,-0.95) {4};
\node[vertex] (e) at (-0.81,0.59) {5};
\node[vertex] (f) at (-0.81,-0.59) {6};
% bonds
\draw[zero] (a)--(b); %12
\draw[zero] (a)--(c); %13
\draw[weak] (a)--(d) ;%node[midway,right] {2.4}; %14
\draw[strong] (a)--(e); %15
\draw[zero] (a)--(f);  %16
\draw[strong] (b)--(c); %23
\draw[weak] (b)--(d) ;%node[pos=0.1,right] {2.4}; %24
\draw[weak] (b)--(e); %25
\draw[weak] (b)--(f); %26
\draw[strong] (c)--(d); %34
\draw[weak] (c)--(e) ;%node[pos=0.1,above] {0.7};  %35
\draw[weak] (c)--(f) ;%node[pos=0.1,below] {1.7}; %36
\draw[strong] (d)--(e); %45
\draw[strong] (d)--(f); %46
\draw[weak] (e)--(f) ;%node[midway,left] {1.7}; %56
\end{scope}
\end{tikzpicture}

\end{document}